\documentclass[a4paper,11pt]{article}
\usepackage{jheppub}
\usepackage{lineno}
\usepackage[utf8]{inputenc}
\usepackage[english]{babel}
\usepackage[T1]{fontenc}
\usepackage{amsmath}
\usepackage{hyperref}
\usepackage{physics}
\usepackage{bbold}
\usepackage[numbers]{natbib}
\usepackage{tikz}
\usepackage{lipsum}
\usepackage{ upgreek }
\usepackage{graphicx}

\graphicspath{{Figures/}}
\title{Entanglement and particle production from cosmological perturbations: a quantum optical simulation approach}
\author[a,e]{Pramod Kamal Kharel}
\affiliation[a]{Tri-Chandra Multiple Campus, Tribhuvan University, Ghantaghar, 44600 Kathmandu, Nepal}
\author[b,e]{Mausam Ghimire}
\affiliation[b]{Butwal Multiple Campus, Tribhuvan University, 32907 Butwal, Nepal}
\author[a,e]{Ashish Khanal}
\author[c,e]{Samyam Pudasaini}
\affiliation[c]{St.Xavier's College, Tribhuvan University, Maitighar, 44600 Kathmandu, Nepal}
\author[a,e]{Nabaraj Khatri}
\author[d,e]{Sayujya Bhandari}
\affiliation[d]{Amrit Campus, Tribhuvan University, Thamel, 44600 Kathmandu, Nepal}
\author[a,e]{Divash Rai}
\affiliation[e]{Holographic Himalaya, Lambagar,  44600 Tarakeshwar,  Nepal}
\author[f]{Kiran Adhikari}
\affiliation[f]{Emmy Noether Group for Theoretical Quantum Systems Design, Technical University of Munich, Arcisstraße 21, 80333 München}
\emailAdd{kiran.adhikari@tum.de}
\author[g]{Rajeev Singh}
\affiliation[g]{Department of Physics, West University of Timisoara, Bulevardul Vasile P\^arvan 4, Timisoara 300223, Romania}
\emailAdd{rajeev.singh@e-uvt.ro}
\abstract{In this work, we develop a computational framework based on the Gaussian formalism and symplectic circuit representation to explore cosmological perturbations during inflation. These tools offer an efficient means to study entanglement generation and particle production, particularly when analytical methods become insufficient and numerical simulations are essential. By evolving an initial Bunch-Davies vacuum through a two-mode squeezer, we simulate the behavior of the von Neumann entropy and logarithmic negativity across a wide range of cosmological backgrounds, each characterized by a distinct equation of state. The von Neumann entropy obtained via QuGIT simulations is compared with analytic Rényi entropy bounds, thereby validating the accuracy of our circuit implementation of the cosmological squeezing Hamiltonian in both accelerating and decelerating scenarios. We further investigate the role of thermal noise and demonstrate how the von Neumann entropy and logarithmic negativity are affected by its presence.
}
\begin{document}
\maketitle
\flushbottom
\section{Introduction}
\label{sec:intro}
An interesting property of the two-mode squeezed state of the cosmological perturbations is that it belongs to the class of Gaussian states in continuous-variable (CV) quantum optics. Its Wigner function is Gaussian, and its entanglement and correlations are fully characterized by its covariance matrix~\cite{Ferraro:2005hen,Weedbrook:2011wxo,Serafini:2017rrn,Adhikari:2022oxr,Adhikari:2022whf,Adhikari:2023evu,Li:2023ekd,Li:2024kfm,Rabinovici:2025otw}. This parallelism makes it possible to apply methods created for optical systems, especially those that use continuous-variable Gaussian states, directly to inflationary cosmology. Each wave vector $k$ and the pair of modes $(k,-k)$ of cosmological perturbations evolve under a linear quadratic Hamiltonian, which evolves the quantum state through a symplectic transformation acting on the covariance matrix. Importantly, any two-mode covariance matrix can be diagonalized using a symplectic transformation. The symplectic eigenvalues of the diagonalized covariance matrix encode the quantum uncertainties and the entanglement of the state~\cite{Ferraro:2005hen,Plenio:2007zz}.

Likewise, non-Gaussian features in cosmological perturbations are naturally produced by non-linear gravitational effects, indicating that the universe is not completely Gaussian. This arises, for instance, in models with a single matter field when taking into account the full cubic action for density perturbations. In simple cases with a canonical scalar field, the speed of sound is $ c_s = 1 $; however, in more general models, a reduced $ c_s $ amplifies equilateral-type non-Gaussianity and strengthens cubic interactions~\cite{Brahma:2020zpk}. Such cubic terms modify the Hamiltonian beyond quadratic form, even in minimal models, making it increasingly difficult to analytically solve these non-Gaussian features classically. However, quantum optics platforms offer a natural toolkit for simulating non‑Gaussian dynamics of cosmological perturbations using cubic phase gates in inflationary cosmology. 

It is a widely accepted fact that the inflationary cosmology paradigm is the `standard model' of the hot big bang cosmology~\cite{Guth:1980zm}. Inflation not only solves the horizon and the flatness problem but also predicts the sets of perturbations on the Friedmann-Lemaitre-Robertson-Walker (FLRW) background metric. These perturbations are the results of the amplification of quantum fluctuations, during inflation, into macroscopic perturbations which later evolve into large-scale structures such as galaxies, clusters, and cosmic microwave background anisotropies~\cite{Mukhanov:1982nu}.

The usual approach in studying these perturbations is to separate the metric $g^{(0)}_{\mu \nu}$ and the scalar field $\varphi_0(\eta)$, setting for inflation, into a homogeneous classical background and the small perturbations as follows
\begin{align}
    g_{\mu \nu} &= g^{(0)}_{\mu \nu} + \delta g_{\mu \nu}\,, \\
\varphi(\mathbf{x}, \eta) &= \varphi_0(\eta) + \delta\varphi(\mathbf{x}, \eta)\,.
\end{align}
These small perturbations, $\delta g_{\mu \nu}$ and $\delta\varphi(\mathbf{x}, \eta)$, are treated quantum mechanically while the background obeys the classical Friedmann equation~\cite{Mukhanov:1990me,Dodelson:2003ft,Straumann:2005mz}. It is sensible to treat the metric perturbation itself quantum mechanically during inflation, as the energy density during this epoch is so enormous that we expect quantum-gravity effects to play a role.

For the purpose of studying cosmological perturbations, it turns out that both the scalar metric and the field fluctuations are encoded in a single degree of freedom, the comoving curvature perturbation $\mathcal{R}$, which is a gauge-invariant variable. In Refs.~\cite{Mukhanov:1985rz,Sasaki:1986hm}, Mukhanov and Sasaki introduced a variable $v$, that is defined as:
\begin{equation}
 v = a \left[ \delta\varphi + \left( \frac{\varphi_0'}{\mathcal{H}} \right)\phi \right] \,,
\end{equation}
and the co-moving curvature perturbation $R$ is simply proportional to $v/a$, where $a$ is scale factor, $\varphi$ is scalar field, and $\phi$ is the scalar metric. In practical calculations, $v$ is preferred over $\mathcal{R}$ because its action takes the form of a simple quadratic action, which makes it straightforward to quantize as a canonical field. In this work, we will also use $v$ for the  canonical quantization of the scalar perturbations, deriving its linear equation of motion from the perturbed action.

Moreover, the quantization of these perturbations during inflation turns their vacuum fluctuations into real particles. By canonically quantizing these modes and by imposing a Bunch-Davies vacuum state, so that each mode matches a positive‐frequency plane wave for $k\gg aH$ (sub-Hubble region), the subsequent evolution mixes positive and negative frequencies. This mode mixing is most easily described by Bogoliubov transformations, where the early time annihilation operators $\hat{c}_{\mathbf k}$ relate to the late time operators $b_{\mathbf k}$ via
\begin{equation}
b_{\mathbf k} \;=\;\alpha_k^*\,\hat{c}_{\mathbf k} \;-\;\beta_k^*\,\hat{c}_{-\mathbf k}^\dagger\,,
\end{equation}
with $|\alpha_k|^2-|\beta_k|^2=1$. A nonzero $\beta_k$ means that what started out as a vacuum state at early times ends up containing particles when viewed at late times, with each mode having an average particle number of $|\beta_k|^2$~\cite{Parker:1969au,Mukhanov:1990me, Jiang:2025ktt}.

An alternative, yet equivalent, description of particle production is given through squeezed quantum states. In~\cite{Mukhanov:1990me, Albrecht:1992kf} it was shown that the amplification of ground state quantum fluctuations during inflation is a process of quantum squeezing.
Within sub-Hubble scales, the Fourier modes of these perturbations undergo oscillations. However, during the accelerated expansion of inflation, the physical wavelengths of these modes grow faster than the Hubble radius $H^{-1}(t)$. As a result, the modes eventually exit the Hubble horizon. Once they cross this threshold, their quantum state evolves into a squeezed vacuum state~\cite{Grishchuk:1990bj,Albrecht:1992kf}. The squeezing of quantum field modes during inflation produces particle pairs with opposing momenta via amplified quantum fluctuations. Hence, the cosmological perturbation in the FLRW background metric can be described in the language of a two-mode squeezed state.

The hallmark of the two-mode squeezed state is that the perturbed modes $k$ and $-k$ are highly entangled. The strong correlations between the entangled modes imply that the reduced density matrix of the squeezed state, which is mixed through a subtle coarse-graining, gives the non-vanishing von Neumann entropy. In~\cite{Brahma:2020zpk} the coarse-graining entropy was done by averaging over the squeezing angle, which provides a stochastic part that lead the reduced density matrix's off diagonal elements to zero. This process gives rise to entanglement entropy, which can account for part of the large entropy observed in the post-inflationary universe~\cite{Gasperini:1992xv, Martin:2021qkg, Martin:2021xml, Boutivas:2023ksg}. Also, many of the papers have argued that the squeezing phenomenon in the context of cosmology is associated with the emergence of classical behavior of our early quantum universe~\cite{Brandenberger:1990bx,Grishchuk:1990bj,Albrecht:1992kf,Polarski:1995jg,Lesgourgues:1996jc,Kiefer:1998qe,Calzetta:1995ys,Perez:2005gh,Kiefer:2008ku,Colas:2021llj,Li:2024ljz,Li:2024iji}.

This work contributes to the growing efforts of applying quantum optics and quantum information theory to the study of inflationary cosmology. Recent studies have demonstrated that quantum optical methods like continuous-variable quantum circuits, Gaussian-state manipulation, and entanglement measures provide powerful frameworks to explore primordial quantum states~\cite{Liu:2020wtr,Boutivas:2023mfg}. Several works have explored simulating inflationary cosmology using quantum tools. For example, digital simulation of particle creation in expanding universes were done using quantum circuits and extended field theory algorithms~\cite{Lloyd:2013xba,Bao:2017iye,Yang:2019kbb,Maceda:2024rrd,Piotrak:2025zhy}.  The redistribution of Gaussian entanglement for continuous variable of scalar field in an expanding universe was also analyzed by modeling the expansion as a Gaussian channel \cite{Li:2023gtf}. Earlier studies showed that quantum fluctuations during inflation naturally form squeezed states~\cite{Parker:1969au,Grishchuk:1993ds}. 
 Continuous variable photonic systems have been proposed to simulate quantum fields using the Gaussian formalism~\cite{Marshall:2015mna}. Other works studied entropy and decoherence, linking squeezing to the emergence of classical structures~\cite{Albrecht:1992kf,Grishchuk:1993ds,Zhai:2024odw}. Together, these works provide valuable context and motivation for our approach.
 
 In addition to quantum-optical simulations, condensed-matter systems have long been studied as analog models of cosmology. Analog models like Bose-Einstein condensates and coupled mechanical oscillators have been used as laboratory analogues of the early universe mimicking inflation-like expansion and particle pair production~\cite{Fischer:2004bf,Wittemer:2019agm,Bhardwaj:2020ndh,Rhyno:2023kud,Agullo:2024lry}. Experiment with the quantum fluids of light have been shown that a sudden change in a $3$D photon fluid can cause spontaneous creation of particles, forming acoustic-peak patterns like those cosmic inflation \cite{Steinhauer:2021fhb}. Various systems from expanding superfluid rings to photonic condensates and optical fiber circuits have been explored as analogues for early universe studies \cite{Pal:2024qno}. Other works include  probing sonic analogues of Gibbons–Hawking radiation, trans-Planckian effects, and quantum backreaction, and exploring phenomena such as analogue gravitational lensing, black hole information scrambling, nonlinear quantum effects, and phase transitions in dipolar systems~\cite{Fischer:2001jz,Fedichev:2003id,Fedichev:2003bv,Fedichev:2003dj,Cha:2016esj,Tian:2020bze,Ribeiro:2021fpk,Ribeiro:2022gln,Baak:2022hum,Baak:2023zjf,Pal:2024qno,Chandran:2025azu}. Besides quantum-optical and condensed-matter settings, synthetic mechanical lattices have also been employed to study analogues of inflationary and FLRW cosmology. In Refs. \cite{Rhyno:2023kud} it is demonstrated how the bosonic dynamics of scalar-field fluctuations are mapped onto the motion of mechanical oscillators, and key inflationary quantities such as particle production and the power spectrum are simulated and measured. These above approaches from the divergent field in contrast to circuit based method we employed, creates the motivation to explore more on this subject matter and discover the unknown.
 
 In this work, we use tools and techniques from quantum optics to simulate the inflationary squeezing Hamiltonian. We use Gaussian states and unitaries to model a two-mode squeezed vacuum and draw a quantum optics circuit~\cite{Gerry_Knight_2004}. We use this circuit to compute entanglement measures like von Neumann entropy and logarithmic negativity of the system. We also analyze the effect of thermal noise on the quantum state and different entanglement measures~\cite{Vidal:2002zz,Gerry_Knight_2004,Plenio:2005cwa,Plenio:2007zz,Choudhury:2016cso,Choudhury:2016pfr,Brady:2022ffk,Kranas:2023aph}. Moreover, we simulate the entropy measures of cosmological perturbations across various cosmological backgrounds. In particular, we focus on the behavior of the entanglement entropy for backgrounds with an arbitrary, constant equation of state $w$, covering both expanding (accelerating and decelerating) and contracting (accelerating and decelerating) backgrounds. 

 In section \ref{section 2}, we review the description of cosmological perturbations as two-mode squeezed states. We then find the approximate analytical solution for the squeezing parameter for accelerating and decelerating backgrounds and thus plot their numerical solutions. In section \ref{section 3}, we give a brief review of continuous variable quantum computing (CVQC) and the concept of universality in CVQC. We also review the covariance formalism of Gaussian states and some common Gaussian unitaries, along with a description to compute commonly used entanglement measures utilising the covariance matrix. In section \ref{section 4}, we provide analytical calculations for von Neumann entropy, R\'enyi entropy, logarithmic negativity, and purity for cosmological perturbations. We also show how one could simulate these measures for cosmological perturbations using CV gates. In section \ref{section 5}, we briefly discuss the results of our simulation. 
Finally, we conclude our work with all the cumulative major findings in section \ref{section 6} followed by mentioning the potential future directions in section \ref{sec:outlook}.



\section{Theory of quantum cosmological perturbations}
\label{section 2}
\subsection{Review of the formalism of cosmological perturbations}

We adopt the gauge-invariant approach for the equation of motion of the metric and scalar field perturbation. It ensures physical predictions are free from dependency on coordinate transformations~\cite{Mukhanov:1990me}. We, therefore, first expand the FLRW metric around a background state and introduce scalar perturbations with  four degrees of freedom. We simplify the metric by choosing an appropriate gauge, such as the longitudinal gauge, which reduces the number of independent variables. Then the total action consisting of both gravitational and matter components is minimized to express the equation of motion of the perturbation of the scalar field in terms of a gauge-invariant quantity, which is, in our case, the Mukhanov-Sasaki variable. 

\subsubsection{Longitudinal gauge invariant metric}
The metric $ g_{\mu\nu} $ splits into a background part $ g_{\mu\nu}^{(0)} $ (the unperturbed FLRW metric) and a small perturbation or deviation due to density fluctuations, gravitational waves, etc. $ \delta g_{\mu\nu} $:
\begin{equation}
    g_{\mu \nu} = g^{(0)}_{\mu \nu} + \delta g_{\mu \nu} \,.
\end{equation}
There are four degrees of freedom in $\delta g_{\mu \nu}$ which correspond to scalar metric fluctuations. In cosmology, perturbations can be scalar, vector, or tensor. Here, we focus on scalar perturbations (related to density and pressure fluctuations), which have four independent scalar functions
\begin{equation}\delta g_{\mu\nu} = a^2 \begin{pmatrix} 2\phi & -B_{,i} \\ -B_{,i} & 2(\psi \delta_{ij} - E_{,ij}) \end{pmatrix}\,,
\end{equation}
with $\phi$, $B$, $\psi$, $E$ being the four fluctuating degrees of freedom, whereas $a(\eta)$ is the scale factor in conformal time, $\phi$ is the perturbation to the time-time component (gravitational potential), $B$ is the shift in the time-space component (spatial gradient), $B_{,i} = \partial_i B$, $\psi$ is the perturbation to the spatial isotropic part (another potential), and
$E$ is the anisotropic perturbation to the spatial metric (via second derivatives) $E_{,ij} = \partial_i \partial_j E$.

By performing a small-amplitude transformation of spacetime coordinates (gauge transformations), the ``fictitious'' fluctuations in a homogeneous and isotropic universe are introduced,
\begin{equation}
\Delta Q(p) = \tilde{\delta Q}(p) - \delta Q(p)\,,
\end{equation}
which is a gauge artifact and carries no physical significance~\cite{Brandenberger:2003vk}.

Let us consider a gauge transformation. Under a coordinate transformation:
\begin{equation}
    \eta \rightarrow \tilde{\eta} = \eta + \xi^0(\eta, x), \qquad x^i \rightarrow \tilde{x}^i = x^i + \gamma^{ij} \xi_{|j}(\eta, x),
\end{equation}
where $\xi$ and $\xi^0$ are scalar functions and $\gamma^{ij}$ is the spatial background metric.

The variation of $\phi, B, \psi, E$ determining the perturbed metric is given by:
\begin{equation}
    \tilde{\phi} = \phi - \frac{a'}{a} \xi^0 - \xi^{0'}, \quad
    \tilde{\psi} = \psi + \frac{a'}{a} \xi^0, \quad
    \tilde{B} = B + \xi^0 - \xi', \quad
    \tilde{E} = E - \xi.
\end{equation}
For the longitudinal gauge, choose $ \xi^0 $ and $ \xi $ such that $ \tilde{B} = 0 $ and $ \tilde{E} = 0 $. This simplifies the metric to:
\begin{equation}
ds^2 = a^2(\eta)\left[-(1+2\phi)d\eta^2 + 2B_{,i} d\eta dx^i + (1-2\phi)\delta_{ij}+2E_{,ij} dx^i dx^j\right] \,.\label{G}
\end{equation}
But the gauge artifact is still there, and the process involved in removing the gauge artifact involves the following steps.

Firstly set:\begin{align}
    \tilde{B} = B + \xi^0 - \xi'=0 \,,\label{F}\\
    \tilde{E} = E - \xi=0 \rightarrow E = \xi \,.
\end{align}
And Eq. \ref{F} further reduces to,
$$B =- \xi^0 + \xi',\quad \text{or}, \quad \xi^0 = \xi'-B.$$
So by choosing $\xi=E$ and $\xi^0=\xi'-B$, we can set $B'=0$ and $E'=0$.
Therefore, in the new gauge, the metric perturbations are:
\begin{align}
    \tilde{\Phi} &= \phi - ((E'-B)')-\frac{a'}{a}(E'-B)\,, \\
    \tilde{\Psi} &= \psi - \frac{a'}{a}(E'-B)\,.
\end{align}
In this basis $\tilde{\psi}=\psi$ and $\tilde{\phi}=\phi$.
Hence, Eq. \ref{G} reduces to the form:
\begin{equation}
ds^2 = a^2(\eta)\left[-(1+2\phi)d\eta^2 + (1-2\psi)\delta_{ij}dx^i dx^j\right] \,,
\end{equation}
which is a metric that specifically describes a perturbed FLRW universe. In this framework, $\phi$ represents the first-order scalar perturbation that encodes information about the deviations from homogeneity and isotropy.

\subsubsection{Evolution equations}
We utilize the Mukhanov-Sasaki formalism to derive the equation of motion of the scalar perturbation~\cite{Mukhanov:1990me}. 
The total action is given as:
\begin{equation}
    S = -\frac{1}{16\pi G} \int R\sqrt{-g} d^4x + \int \left[ \frac{1}{2} \varphi_{,\alpha} \varphi^{,\alpha} - V(\varphi) \right] \sqrt{-g} d^4x\,. 
\end{equation}
The first term of the equation is the Einstein action, and the second term of the equation is the matter action.

The Einstein action part is of the form:
\begin{equation}
S_{\mathcal{EH}} = -\frac{1}{16 \pi G} \int R \sqrt{-g} \, d^4x\,. \label{K}
\end{equation}
We expand the action up to second order in perturbations:
\begin{align}
\delta_2 S_{EH} = \frac{1}{16\pi G} \int \Bigl\{ &\,
a^2 \Bigl[ -6\psi'^2 - 12\mathcal{H} (\phi + \psi) \psi' - 9 \mathcal{H}^2 (\phi + \psi)^2 \Bigr] \notag \\
& - 2 \psi_{,i} (2\phi_{,i} - \psi_{,i}) - 4\mathcal{H} (\phi + \psi) (B - E')_{,ii} + 4\mathcal{H} \psi' E_{,ii} \notag \\
& - 4\psi' (B - E')_{,ii} - 4\mathcal{H} \psi_{,i} B_{,i} + 6\mathcal{H}^2 (\phi + \psi) E_{,ii} \notag \\
& - 4\mathcal{H} E_{,ii} (B - E')_{,ij} + 4\mathcal{H} E_{,ii} B_{,jj} + 3\mathcal{H}^2 E_{,ii}^2 + 3\mathcal{H}^2 B_{,i} B_{,i} \notag \\
& + \mathfrak{D}_1^{gr} + \mathfrak{D}_2^{gr} \Bigr\} \, d^4x.
\end{align}
where $\mathfrak{D}_1^{gr}$ and $\mathfrak{D}_2^{gr}$ is total derivative term and $ H = \frac{\dot{a}}{a} $ is the Hubble parameter.

The total derivative term vanishes after integrating and in the longitudinal gauge ($ B = E = 0 $, $ \phi = \psi $), and we will get:
\begin{equation}
\delta_2 S_{EH}= \frac{1}{16\pi G} \int  \left\{ a^2 \left[ -6\phi'^2 - 24\mathcal{H} \phi \phi' - 36\mathcal{H}^2 \phi^2 \right] - 2(\phi_{,i})^2 \right\}d^4x\,.
\end{equation}
We briefly introduce the background Friedmann equations here, which are commonly substituted into later perturbative calculations in cosmology, particularly when simplifying the second-order action in cosmological perturbation theory.

The Friedmann equation for the time evolution of the background in the  case of a scalar matter field is:
\begin{equation}
H^2 = \frac{8 \pi G}{3} \left( \frac{1}{2} \dot{\varphi}^2 + V(\varphi) \right) =l^2\left( \frac{1}{2} \dot{\varphi}^2 + V(\varphi) \right)\,. \label{H}
\end{equation}
We use the second equation, the scalar field equation of motion:
\begin{equation} \ddot{\varphi} + 3H\varphi+V_{,\varphi} = 0\,. \label{I}
\end{equation}
This equation can be derived by varying the matter action with respect to $\varphi$. Taking the derivative of the first Friedmann equation with respect to time and subtracting $\dot\varphi$ times Eq. \ref{I}, we obtain:
\begin{equation}
\dot{H}= - \frac{3}{2} l^2 \dot{\varphi}^2 \,.\label{J}
\end{equation}
By utilizing the formula of conformal time, $d\eta=\frac{dt}{a(t)}$, Eq. \ref{H} and \ref{I} takes the form,
\begin{align}
\mathcal{H}^2 &= l^2 \left[ \frac{1}{2} {\varphi_0'}^2 + V(\varphi_0) a^2 \right]\,, \\
2 \mathcal{H}' + \mathcal{H}^2 &= 3l^2 \left[ -\frac{1}{2} {\varphi_0'}^2 + V(\varphi_0) a^2 \right]\,.
\end{align}
These two equations can be combined to give
\begin{equation}
\mathcal{H}^2 - \mathcal{H}' = \frac{3}{2} l^2 {\varphi_0'}^2.
\end{equation}
The above Friedmann equations help in reducing the redundancy and simplifying the background coefficients.
Now, the second-order perturbation of the matter action is:
\begin{equation}
    \delta_2 S_m = \int d^4x (-g_0)^{1/2} \left( \frac{\delta_2(-g)^{1/2}}{(-g_0)^{1/2}} \mathcal{L}_0 + \frac{2 \delta_1 (-g)^{1/2}}{(-g_0)^{1/2}} \delta_1 \mathcal{L} + \delta_2 \mathcal{L} \right),
\end{equation}
$\delta_1{L}$ and $\delta_2{L}$ can be found by expanding
\begin{equation}
    \mathcal{L}(\varphi) = \frac{1}{2} \varphi_{,\alpha} \varphi^{,\alpha} - V(\varphi).
\end{equation}
Expanding about the background field $\varphi_0$ using a Taylor series, we get the first perturbation of the Lagrangian and the second perturbation of the Lagrangian, where $L_0$ is the Lagrangian density evaluated at the background metric and $g_0$ is the background metric.
Now, applying the background equations~\ref{H},~\ref{I},~\ref{J} and integrating by parts and combining the second-order variations of the matter action $\delta_2 S_m$ and the Einstein-Hilbert action $\delta_2 S_{EH}$ (\ref{K}) and utilizing the longitudinal gauge, we obtain
\begin{align}
\delta_2 S = \delta_2 S_m+\delta_2 S_{EH}=\frac{1}{6l^2} \int \Bigl\{ &\,
a^2 \left[ -6\phi'^2 - 12\mathcal{H} \phi \phi' - 2 (\phi_{,i})^2 - 2 (\mathcal{H}' + 2\mathcal{H}^2) \phi^2 \right] \notag \\
& + 3l^2 \left[ \delta\phi'^2 - \delta\phi_{,i} \delta\phi_{,i} - V_{,\phi\phi} a^2 \delta\phi^2 \right] \notag \\
& + 6l^2 \left[ \phi_0' (4\phi)' \delta\phi - 2V_{,\phi} a^2 \phi \delta\phi \right] \Bigr\} \, d^4x.
\end{align}
From the linearized Einstein equation (momentum constraint):
\begin{equation}
    \phi' + \mathcal{H}\phi = \frac{3}{2} l^2 \varphi_0' \delta\varphi.
\end{equation}
This equation relates the perturbations $ \phi $ to the perturbation of the scalar field $ \delta\varphi $. However, these variables are gauge-dependent, meaning their values depend on the choice of coordinate system. To eliminate this dependence, we define the gauge-invariant Mukhanov-Sasaki variable~\cite{Mukhanov:1985rz, Sasaki:1986hm}:
\begin{equation}
    v = a \left[ \delta\varphi + \left( \frac{\varphi_0'}{\mathcal{H}} \right)\sqrt{2}\epsilon \phi \right],
\end{equation}
where, $\epsilon = -\frac{\dot{H}}{H^2}$, that is first slow-roll inflation parameter. 
After transforming the action into terms of the gauge-invariant variable v, the second-order perturbed action~\cite{Mukhanov:1990me} simplifies to:
\begin{equation}
   \label{action}
    \delta_2 S = \frac{1}{2} \int \left( v'^2 - v_{,i} v_{,i} + \frac{z''}{z} v^2 \right) d^4x,
\end{equation}
where $z = a\sqrt{2}\epsilon\frac{\varphi_0'}{\mathcal{H}}$ in conformal time.

This action represents perturbations of a scalar field coupled to an external time-varying source, which leads to the equation of motion:
\begin{equation}
v'' - \nabla^2 v - \frac{z''}{z} v = 0.
\end{equation}
This equation is a harmonic oscillator equation with time-dependent mass given by ${z''}/{z} $. On the scale smaller than the Hubble radius, the mass term is negligible, and the mode function oscillates with a constant amplitude. When the scale of a mode becomes larger than the Hubble radius, it stops oscillating. The moment when modes exit the Hubble radius, the quantum modes effectively shift into the classical regime, with their amplitudes starting to grow significantly over time.

\subsection{Quantization and formalism of squeezed states }

The classical analysis of the fluctuation is sufficient to describe how the perturbations evolve. However, to study the origin of the perturbations and particle production in expanding background space times quantum treatment is necessary~\cite{Brandenberger:1993zc}. In this section, we investigate the quantum nature of the perturbations to see how the expanding background leads to particle production. We treat them as field that evolves in an expanding spacetime. Using the Heisenberg picture, we construct the Hamiltonian and the equation of motion of ladder operators by using the Heisenberg equation of motion. And we do the Bogoliubov transformation to see the evolution of the field modes from the initial time $\eta_0$ (vacuum).

\subsubsection{Hamiltonian}
Firstly, the Hamiltonian is constructed to track how the fields evolve. For that, we need the canonical momentum $\pi$, which is defined as the conjugate momentum to the field $v$. The canonical momentum is obtained by taking the functional derivative of the Lagrangian density with respect to the time derivative of the field, $v'$. Up to a total derivative term, action in \ref{action} is equivalent to~\cite{Albrecht:1992kf}:
\begin{equation}
\delta_2 S= \frac{1}{2} \int d^4x \left[ (v')^2 - c_s^2 (v_i')^2 - 2 \frac{z'}{z} v' v  + \left( \frac{z'}{z} \right)^2 v^{2}\right],
\end{equation}
which we find convenient to work with. The Lagrangian density $\mathcal{L}$ is the integrand of the action, which is given as follows,

\begin{equation}
\mathcal{L} = \frac{1}{2} \left[ (v')^2 - c_s^2 (v_{,i})^2 - 2 \frac{z'}{z} v' v + \left( \frac{z'}{z} \right)^2 v^2 \right]\,.
\label{eq:lagrangian}
\end{equation}
The canonical momentum conjugate to the field $v$ is defined as,
\begin{equation}
\pi = \frac{\partial \mathcal{L}}{\partial v'} = v' - \frac{z'}{z} v. 
\end{equation}
Substituting into the Lagrangian density \ref{eq:lagrangian}, we find
\begin{equation}
\mathcal{L} = \frac{1}{2} \left[ \pi^2 + 2\frac{z'}{z} \pi v + \left( \frac{z'}{z} \right)^2 v^2 - c_s^2 (v_{,i})^2 - 2 \frac{z'}{z} \pi v - 2 \left( \frac{z'}{z} \right)^2 v^2 + \left( \frac{z'}{z} \right)^2 v^2 \right]. \tag{2.32}
\end{equation}
Upon simplification, the cross-terms and higher-order terms cancel appropriately, yielding
\begin{equation}
\mathcal{L} = \frac{1}{2} \left[ \pi^2 - c_s^2 (v_{,i})^2 \right] \,.
\label{eq:reduced lagrangian}
\end{equation}
We now compute the Hamiltonian density using the standard definition $\mathcal{H} = \pi v' - \mathcal{L}$. Substituting $v' = \pi + \frac{z'}{z} v$ and the expression for $\mathcal{L}$ from \ref{eq:reduced lagrangian}, we arrive at the Hamiltonian density which reads,
\begin{equation}
\mathcal{H} = \frac{1}{2} \pi^2 + \frac{z'}{z} \pi v + \frac{1}{2} c_s^2 (v_{,i})^2\,.
\end{equation}
The Hamiltonian is the integral over spatial coordinates,
\begin{equation}
    H = \frac{1}{2} \int d^3 x \left[ \pi^2 + c_s^2 (v_{,i})^2 + 2 \frac{z'}{z} v \pi \right]\,.
    \label{eq:hamiltoninan 1}
\end{equation}
(Note that $\frac{z'}{z} \pi v = \frac{z'}{z} v \pi$, so the terms are equivalent up to ordering.)
We promote the fields to operators and express them in Fourier space as
\begin{align}
\hat{v} &= \int\frac{d^{3}k}{(2\pi)^{3/2}}\hat{v}_{\vec{k}}e^{i\vec{k}\cdot \vec{x}}\,, \\
\hat{\pi} &= \int\frac{d^{3}k}{(2\pi)^{3/2}}\hat{\pi}_{\vec{k}}e^{i\vec{k}\cdot \vec{x}}\,.
\end{align}
Substituting these Fourier decompositions into the Hamiltonian \ref{eq:hamiltoninan 1}, and integrating over spatial coordinates using $\int d^3x e^{i(\vec{k}+\vec{k}')\cdot\vec{x}} = (2\pi)^3 \delta^{(3)}(\vec{k}+\vec{k}')$ yields
\begin{equation}
\hat{H} = \frac{1}{2} \int {d^3k} \left[ \hat{\pi}_{\vec{k}} \hat{\pi}_{-\vec{k}} + c_s^2 k^2 \hat{v}_{\vec{k}} \hat{v}_{-\vec{k}} + \frac{z'}{z} (\hat{v}_{\vec{k}} \hat{\pi}_{-\vec{k}} + \hat{v}_{-\vec{k}} \hat{\pi}_{\vec{k}}) \right].
\label{eq:two mode hamiltonian 1}
\end{equation}
The integrand corresponds to the two-mode Hamiltonian for each $\vec{k}$. 

We can expand the operators $\hat{v}_k$ and $\hat{\pi}_k$ in terms of ladder operators $\hat{c}_k$ and $\hat{c}_k^\dagger$ where it follows the usual commutation relation $[\hat{c}_k(\eta), c^\dagger_p(\eta)] = \delta^{(3)}({k} - {p})
$ which can be expressed as follows,
\begin{equation}
\hat{v}_k = \frac{1}{\sqrt{2k}} \left( \hat{c}_k + \hat{c}^{\dagger}_{-k} \right), \quad 
\hat{\pi}_k = -i \sqrt{\frac{k}{2}} \left( \hat{c}_k - \hat{c}^{\dagger}_{-k} \right)\,.
\label{eq:fourier decomposition}
\end{equation}
Finally, we can compute the Hamiltonian in the operator form by substituting \ref{eq:fourier decomposition} into \ref{eq:two mode hamiltonian 1}, we will get,
\begin{equation}
\label{eq:Hamiltonian}
  \hat{H} = \frac{1}{2} \int \frac{d^3k}{(2\pi)^{3}} \left[ k \left( \hat{c}_k \hat{c}_k^\dagger + \hat{c}_{-k} \hat{c}_{-k}^\dagger \right) - i \frac{z'}{z} \left( \hat{c}_k \hat{c}_{-k} - \hat{c}_k^\dagger \hat{c}_{-k}^\dagger \right) \right]\,.
\end{equation}
This is the required Hamiltonian in operator form. The first term of this Hamiltonian represents the harmonic oscillators and the energy for free quantum fields. The second term is crucial and has important information. The state is not seen generally in a vacuum after the latter times. The term $i\frac{z'}{z}$ shows the expansion of space time and squeezing. Specifically, the presence of the creation operator product $\hat{c}_k^\dagger \hat{c}_{-k}^\dagger$ in the second term is evident that quantum pairs with opposite momenta in modes $k$ and $-k$ are generated as the universe expands. This pair-creation process arises because of the evolving background spacetime, leading to quantum correlations in cosmological perturbations.

\subsubsection{Evolution of field modes}

Now, the next job is to see the evolution of ladder operators or field modes to understand how the field evolves in an expanding background with conformal time.  The Heisenberg equation of motion is given by,
\begin{equation}
    \frac{d \hat{c}_{\mathbf{k}}}{d\eta} = -i [\hat{c}_{\mathbf{k}} , \hat{H}] = - i k \hat{c}_{\mathbf{k}} + \left( \frac{z'}{z} \right) \hat{c}_{\mathbf{k}}^\dagger\,.
    \label{eq:evolution mode 1}
\end{equation} 
Similarly, for $\hat{c}_{\mathbf{k}}^\dagger$,
\begin{equation}
    \frac{d \hat{c}_{\mathbf{k}}^\dagger}{d\eta} =  -i [\hat{c}_{\mathbf{k}}^\dagger , \hat{H}]  = \left( \frac{z'}{z} \right) \hat{c}_{\mathbf{k}} + i k \hat{c}_{\mathbf{k}}^\dagger\,.
    \label{eq:evolution mode 2}
\end{equation}
This system of differential equations can be solved using a Bogoliubov transformation, which is the linear transformation that mixes creation and annihilation operators to describe how the state changes under a new physical condition and relates the initial field modes to the evolved ones.  It is assumed that the system starts from vacuum state $\ket{0}$ during inflation defined by the condition $\hat{c}_{\mathbf{k}}(\eta_{\text{0}})|0\rangle = 0$. The creation and annihilation operator at time $\eta > \eta_{0}$ is related by the creation and annihilation operator at initial time $\eta_{0}$ as follows,
\begin{align}
    \hat{c}_k(\eta) &= \alpha_k(\eta) \hat{c}_k(\eta_{\text{0}}) + \beta_k(\eta) \hat{c}_{-k}^\dagger(\eta_{\text{0}})
      \label{eq:Bogoliubov transformation1} \,,\\
    \hat{c}_k^\dagger(\eta) &= \alpha_k^*(\eta) \hat{c}_k^\dagger(\eta_{\text{0}}) + \beta_k^*(\eta) \hat{c}_{-k}(\eta_{\text{0}}),
    \label{eq:Bogoliubov transformation}
\end{align} 
where $\alpha_k(\eta)$ and $\beta_k(\eta)$ are the Bogoliubov coefficients where they obey,
\begin{align}
i \frac{d\alpha_k(\eta)}{d\eta} &= k \alpha_k(\eta) + i \frac{z'}{z} \beta_k^*(\eta)\,, \\
i \frac{d\beta_k(\eta)}{d\eta} &= k \beta_k(\eta) + i \frac{z'}{z} \alpha_k^*(\eta)\,.
\end{align}
Furthermore, $\alpha_k(\eta)$ and $\beta_k(\eta)$ must satisfy the normalization condition as,
\begin{equation}
|\alpha_k|^2 - |\beta_k|^2 = 1\,,
\end{equation}
such that the commutation relation between the creation and annihilation operators is preserved in time. In addition, the initial condition for the coefficients is set as $\alpha_k(\eta_0) = 1$ and $\beta_k(\eta_0)=0$, which suggests that the initial state is vacuum at $\eta= \eta_0$ .
For the given normalization condition, we can always choose the value for $\alpha_k =  e^{i\theta_k} \cosh r_k$ and $\beta_k =  e^{-i\theta_k+2i\phi_k} \sinh r_k$. Plugging this into Eq. \ref{eq:Bogoliubov transformation1} ,
\begin{equation}
    \hat{c}_k(\eta) =  e^{i\theta_k} \cosh r_k \hat{c}_k(\eta_0) + e^{-i\theta_k + 2i\phi_k} \sinh r_k \hat{c}_{-k}^\dagger (\eta_0)\,,
\end{equation}
where the quantities $r_k$, $\theta_k$, and $\phi_k$ are called the squeezing parameter, rotation angle, and squeezing angle, respectively, which are the functions of the conformal time. The squeezing parameter \textbf{$r_k$} determines how much squeezing occurs. It is a quantitative measure of the amount of squeezing happening to the variables with a changing universe. The squeezing angle \textbf{$\phi_k$} determines which variable undergoes squeezing and which undergoes stretching~\cite{Albrecht:1992kf}. Moreover, the squeezing angle $\phi_k$ determines how the uncertainty is redistributed between different quadratures of a quantum state and sets the direction of squeezing in the phase space. The rotation operator is only to change the phase, which is of no consequence to our study, and we will drop it from here~\cite{Bhattacharyya:2020rpy}. For inflation, these parameters are given as~\cite{Albrecht:1992kf}
\begin{align}
    r_k (\eta) &= -\sinh^{-1} \left( \frac{1}{2 c_s k \eta} \right), \\  
    \phi_k (\eta) &= -\frac{\pi}{4} - \frac{1}{2} \tan^{-1} \left( \frac{1}{2 c_s k \eta} \right), \\  
    \theta_k (\eta) &= -k\eta - \tan^{-1} \left( \frac{1}{2 c_s k \eta} \right). 
\end{align}

\subsubsection{Squeezing and rotation evolution operators}

The evolution operator for each mode $k$ of the Hamiltonian \ref{eq:Hamiltonian} is given by:
\begin{equation}
\hat{\mathcal{U}}_{\mathbf{k}} = \hat{\mathcal{S}}_{\mathbf{k}}(r_{\mathbf{k}}, \phi_{\mathbf{k}}) \hat{\mathcal{R}}_{\mathbf{k}}(\theta_{\mathbf{k}}),
\end{equation}
where $\hat{\mathcal{R}}(\theta_{\mathbf{k}})$  and $ \hat{\mathcal{S}}(r_{\mathbf{k}}, \phi_{\mathbf{k}})$ is the rotation and squeezing operator respectively  defined as,
\begin{align}
\hat{\mathcal{R}}(\theta_{\mathbf{k}}) &= \exp\left[-i \theta_{\mathbf{k}} (\hat{c}_{\mathbf{k}}^{\dagger} \hat{c}_{\mathbf{k}} + \hat{c}_{\mathbf{-k}}^{\dagger} \hat{c}_{\mathbf{-k}} +1 ) \right]
\,, \\
\hat{\mathcal{S}}(r_{\mathbf{k}}, \phi_{\mathbf{k}}) &= \exp \left[ \frac{r_{\mathbf{k}}}{\alpha} \left(e^{-2i\phi_{\mathbf{k}}} \hat{c}_{\mathbf{-k}} \hat{c}_{\mathbf{k}} - \text{h.c.} \right) \right]\,.
\label{eq:squeezing_operator}
\end{align}
 Since the rotation operator only produces a global phase, we will ignore it forward. The two-mode squeezing operator applied on the vacuum state produces a squeezed vacuum state:~\cite{Albrecht:1992kf} 
\begin{align}
SQ(k, \eta) &= S_k(r_k, \phi_k) \, |0_k \, 0_{-k}\rangle  = \frac{1}{\cosh r_k} \sum_{n=0}^\infty 
    e^{-2i n \phi_k} \tanh^n r_k \, |n_k \, n_{-k}\rangle \,,
    \label{eq:Squeezed state}
\end{align}
where, $|n_k \, , n_{-k}\rangle = \frac{1}{n!} \, (\hat{c}_k^{\dagger } \, \hat{c}_{-k}^{\dagger } )^n\, |0_k \, 0_{-k}\rangle$. For a mode k, the state is normalized as: 
\begin{align}
\langle SQ(k, \eta) | SQ(k, \eta) \rangle 
&= \frac{1}{\cosh^2 r_k} \sum_{n=0}^\infty \sum_{m=0}^\infty e^{-2i(n - m)\phi_k} \tanh^{m+n} r_k \, \delta_{m,n} = 1 \,.
\end{align}
In general, the squeezed vacuum of all the modes is given by the tensor product state,
\begin{equation}
|SQ(\eta)\rangle = \bigotimes_k |SQ(k, \eta)\rangle.
\end{equation}

\subsection{Squeezing parameter for expanding and contracting backgrounds}

In this section, we will study the changes in the squeezing parameters $r_k$ and $\phi_k$  in FLRW universes with scalar curvature perturbations in different cosmological backgrounds (accelerating, decelerating, expanding, and contracting).

For simplicity, we will change the variable from the conformal time to physical time using $d\eta = \frac{dt}{a(t)}$. Then, the differential equations for the squeezing parameters $r_k$ and $\phi_k$ is given by~\cite{Bhattacharyya:2020kgu}
\begin{align}
    \frac{d r_k}{d a} &= -\frac{1}{a} \cos(2\phi_k)\,, \label{eq:squeezing_parameter} \\
    \frac{d \phi_k}{d a} &= \frac{k}{a \mathcal{H}} + \frac{1}{a} \coth(2r_k) \sin(2\phi_k)\,
    \label{eq:squeezing_angle}
\end{align}
We numerically solve the squeezing parameters $r_k$ and $\phi_k$  using the fourth-order Runge-Kutta (RK4) method. Next, we derive the relation between the comoving wavenumber $k$, the conformal time $\eta$, and the scale factor $a$ at the moment when the modes cross the horizon. A power-law expansion for a flat universe with an arbitrary equation of state parameter $w$ is given by $a(t) \propto t^{\frac{2}{3(1+w)}} \label{B}$. We relate the scale factor with conformal time $\eta$ as
\begin{equation}
a(\eta) \propto \eta^{\frac{2}{1+3w}} = \eta^{-\beta} \label{D}\,,
\end{equation}
where $\beta=\frac{2}{1+3w}$. The Hubble parameter in terms of conformal time is defined as $H = \frac{a'}{a^2}$. Thus, the comoving wavenumber $k$, at the moment when the mode exits the horizon ($k = aH$), becomes 
\begin{equation}
k = aH(\eta) = \frac{a'}{a} \approx \frac{2}{1+3w} \frac{1}{\eta}\,.
\end{equation}
For the special case of inflation driven by a scalar field, where $w = -1$, we obtain $|k| = \frac{1}{\eta}$.
Now, expressing this relation in terms of the scale factor $a$, we use Eq.~\eqref{D} to find the dependence of horizon-crossing modes on $a$: 
\begin{equation}
k \approx \frac{1}{a^{\frac{1+3w}{2}}}\,.
\label{k_horizon}
\end{equation}
In particular, for $w = -1$, this simplifies to $k \approx a$, which is the required relation between the comoving wavenumber $k$ and the scale factor $a$ for modes crossing the horizon during inflation.

\subsubsection{Expanding background}
The expression for the scale factor $a(\eta)$ for the expanding background is given by~\cite{Bhattacharyya:2020kgu}:
\begin{equation}
a(\eta)=
\begin{cases}
\displaystyle\left(\frac{\eta_{0}}{\eta}\right)^{\beta},
& -\infty<\eta<0,\;\eta_{0}<0,\;\beta>0\;(w<-1/3)\quad\text{(accelerating)}\,,\\[8pt]
\displaystyle\left(\frac{\eta}{\eta_{0}}\right)^{|\beta|},
& 0<\eta<\infty,\;\eta_{0}>0,\;\beta<0\;(w>-1/3)\quad\text{(decelerating)}\,,
\end{cases}
\label{eq:scale-factor-exp}
\end{equation}
where,  $\beta = -\frac{2}{1+3w}$. Then, the equations of motion, Eqs. \ref{eq:squeezing_parameter} and \ref{eq:squeezing_angle}  for the expanding background are,
\begin{align}
\frac{d r_k}{d a} &= -\frac{1}{a} \cos(2\phi_k)\,,
\label{rkdiffeq}\\
\frac{d \phi_k}{d a} &= k \frac{|\eta_0|}{|\beta|} \frac{1}{a^{1+1/\beta}} + \frac{1}{a} \coth(2r_k) \sin(2\phi_k)\,.
\label{pkdiffeq}
\end{align}
\begin{figure}[h!]
    \centering
    \includegraphics[width=1\linewidth]{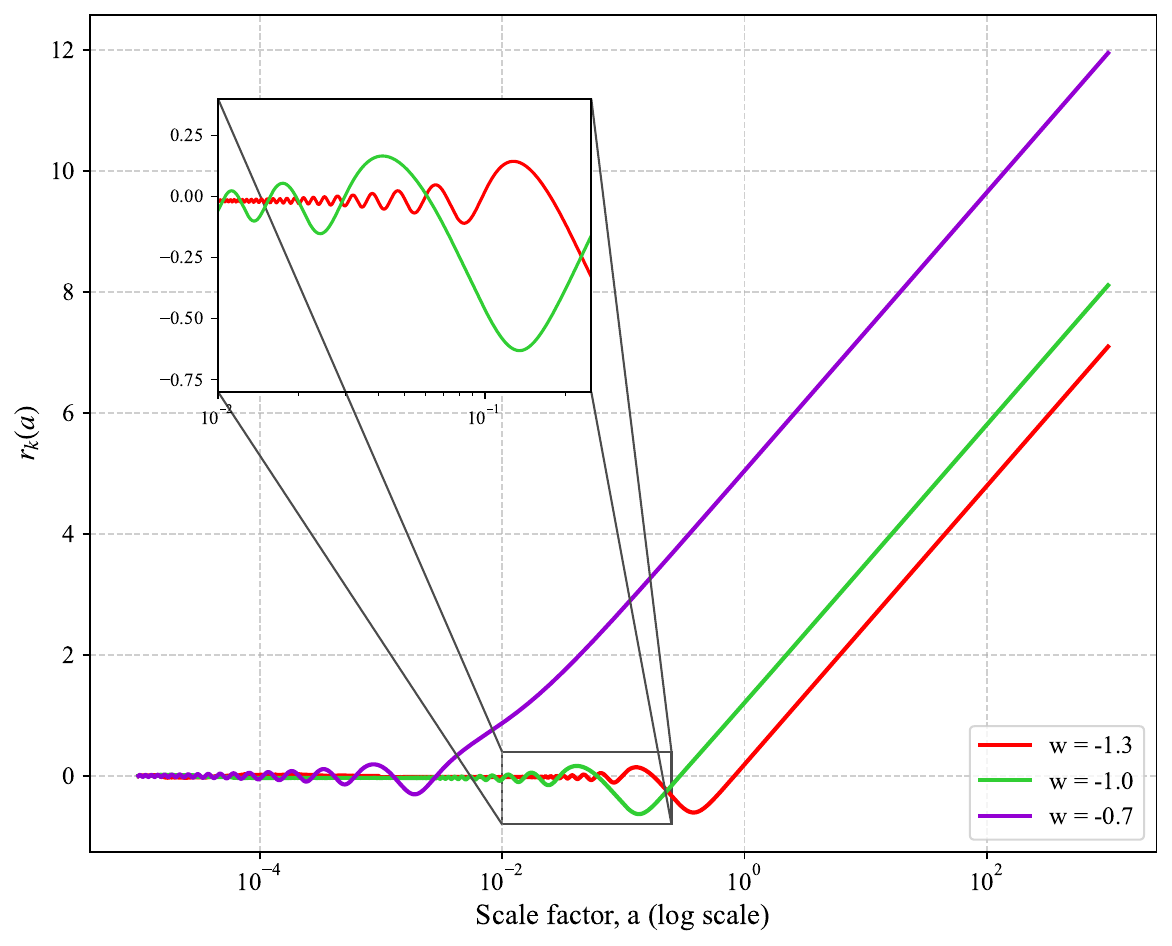}
    \caption{Solutions for $r_k$ versus the scale factor $a$. All curves show decaying oscillations at early times before transitioning to a period of growth. This growth phase begins earlier for less negative values of $w$. Consequently, at any given late time, a less negative $w$ (e.g., $w=-0.7$) results in a significantly larger value for $r_k$ compared to more negative values.}
    \label{fig:1a}
\end{figure}
The squeezing parameter $r_k$ is plotted against the scale factor $a$ on a log scale in Figure \ref{fig:1a}. Each colored line on this graph represents a different constant equation of state parameter $w$. At early times ($a \ll 1$), $r_k$ shows small oscillations around zero, as seen in the inset. As the scale factor $a$ gets larger, these oscillations become slower. When the modes begin to cross the Hubble horizon, $r_k$ stops oscillating and starts to grow. The time when this growth starts depends on the value of $w$. The modes in universes with larger $w$ values ($w=-0.7$) exit the horizon earlier than modes with smaller $w$ values ($w=-1.3$). At late times ($a \gtrsim 1$), the straight lines on the plot show that $r_k$ grows in proportion to $\log a$. The slope of this growth is the same for all values of $w$. However, the final value of $r_k$ is very different for each case. The plot shows that larger values of $w$ lead to a larger final value for $r_k$.

\begin{figure}[h!]
    \centering
    \includegraphics[width=1\linewidth]{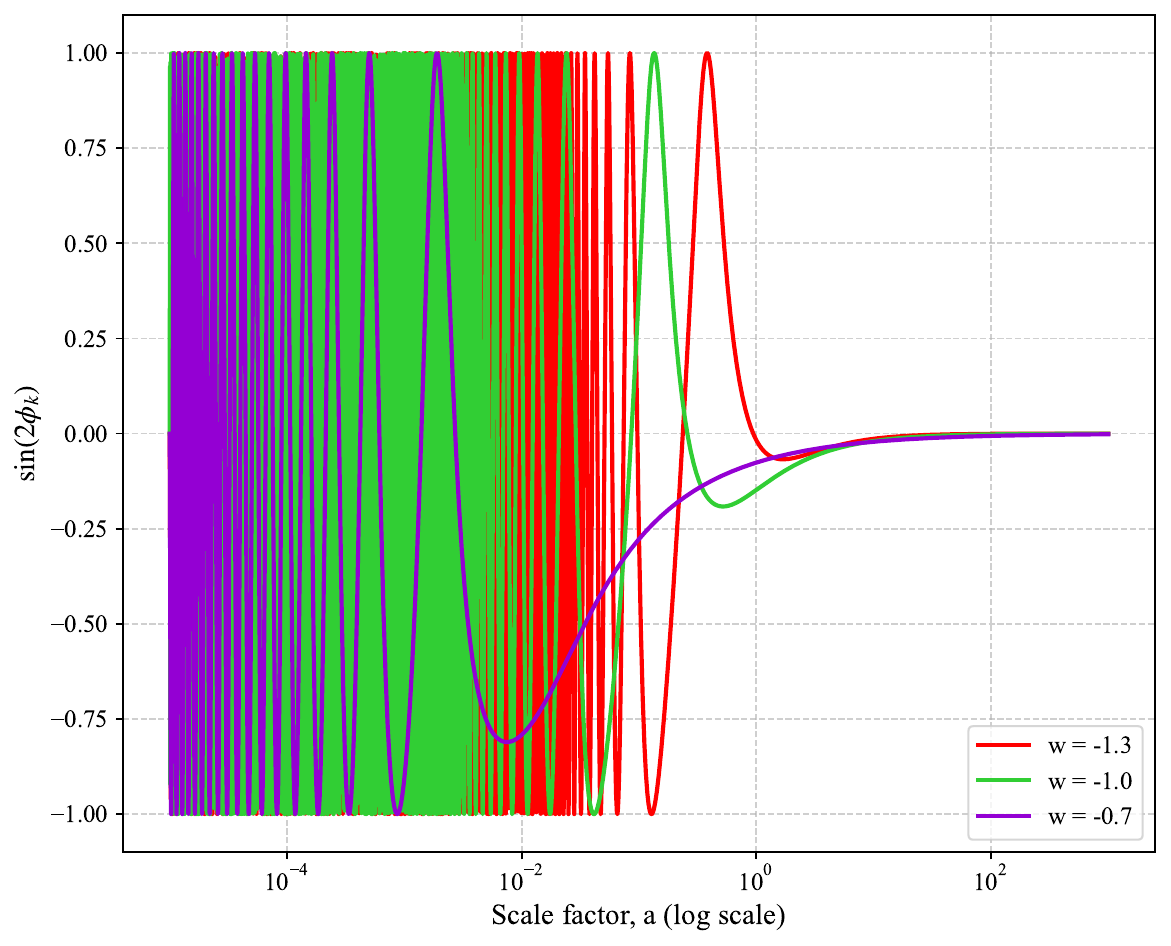}
    \caption{Numerical solutions for $sin(2\phi_k)$ as a function of scale factor $a$ for expanding accelerating background. The solution oscillates rapidly initially and then dampens depending on the value of $w$. For higher values ($w=-0.7$) transition occurs earlier, but for lower values ($w=-1.3$) the oscillations occur until much later and then finally settles at zero.}
    \label{fig:1b}
\end{figure}
In Figure \ref{fig:1b}, the plot shows the value of $\sin(2\phi_k)$ as a function of the scale factor $a$. At early times, when the modes are sub-Hubble, $\sin(2\phi_k)$ oscillates very quickly between $+1$ and $-1$. This happens at the same time that the squeezing parameter $r_k$ from Figure \ref{fig:1a} was showing small oscillations around zero. As the universe expands and the modes begin to cross the horizon, the oscillations stop, and the value of $\sin(2\phi_k)$ goes towards zero. At late times, after the modes are super-Hubble, $\sin(2\phi_k)$ stays at zero. This means the squeezing angle $\phi_k$ freezes to a constant value. This freezing of the angle happens exactly when the squeezing parameter $r_k$ starts to grow. Once the expansion starts to dominate the mode's evolution, it causes the amount of squeezing ($r_k$) to increase continuously while locking the squeezing direction ($\phi_k$) in place.

\begin{figure}[h!]
    \centering
    \includegraphics[width=1\linewidth]{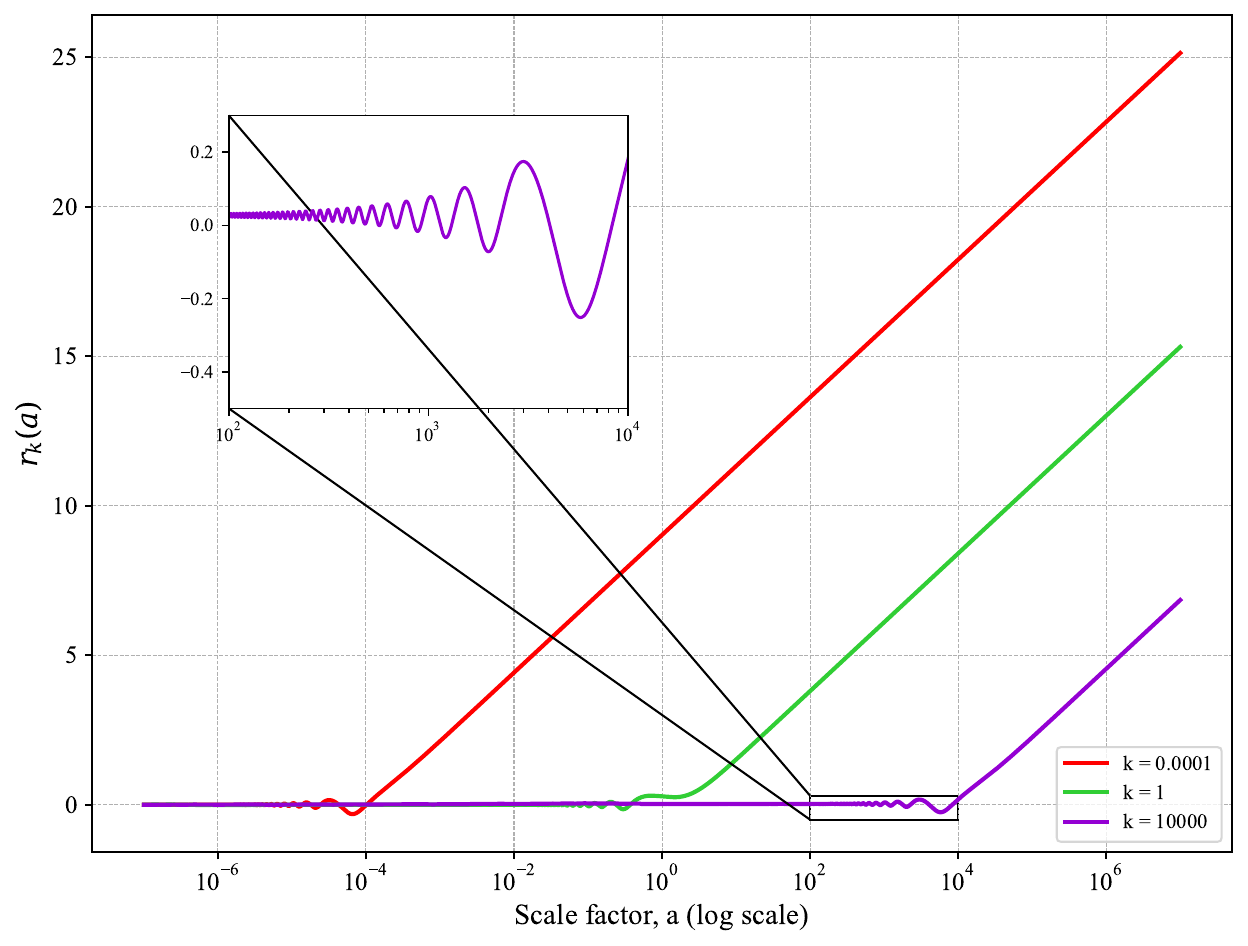}
    \caption{Numerical solutions for $r_k$ as a function of scale factor $a$ for expanding accelerating background ($w= -1$), with varying k. Modes with smaller k exit the horizon earlier, which allows them more time to grow and reach a larger final value. The inset shows the oscillations of the modes before the horizon exit.}
    \label{fig:6}
\end{figure}
In Figure \ref{fig:6}, we have plotted the squeezing parameter $r_k$ as a function of the scale factor $a$, but for this plot, the equation of state $w$ is kept constant, and we change the wavenumber $k$'s values. Here as well, all modes start with small oscillations around $r_k = 0$ when they are sub-Hubble. The inset shows that the modes oscillate around 0. The main difference in this plot is the time of horizon exit. Modes with a very small $k$ (e.g., the red line, $k=0.0001$) are large-scale modes and exit the horizon very early. In comparison, the modes with a large $k$ (e.g., the purple line, $k=10000$) are small-scale modes and exit the horizon much later. After exiting the horizon, $r_k$ grows proportionally to $\log a$ for all modes. We can see that the lines are parallel, which means the growth rate is the same for every value of $k$. However, because the small-$k$ modes exit earlier, they have more time to grow and reach a much larger final value of $r_k$. This is different from Figure \ref{fig:1a}, where larger values of $w$ led to a larger final $r_k$; here, it is the smaller values of $k$ that lead to a larger final $r_k$.

\begin{figure}[h!]
    \centering
    \includegraphics[width=0.9\linewidth]{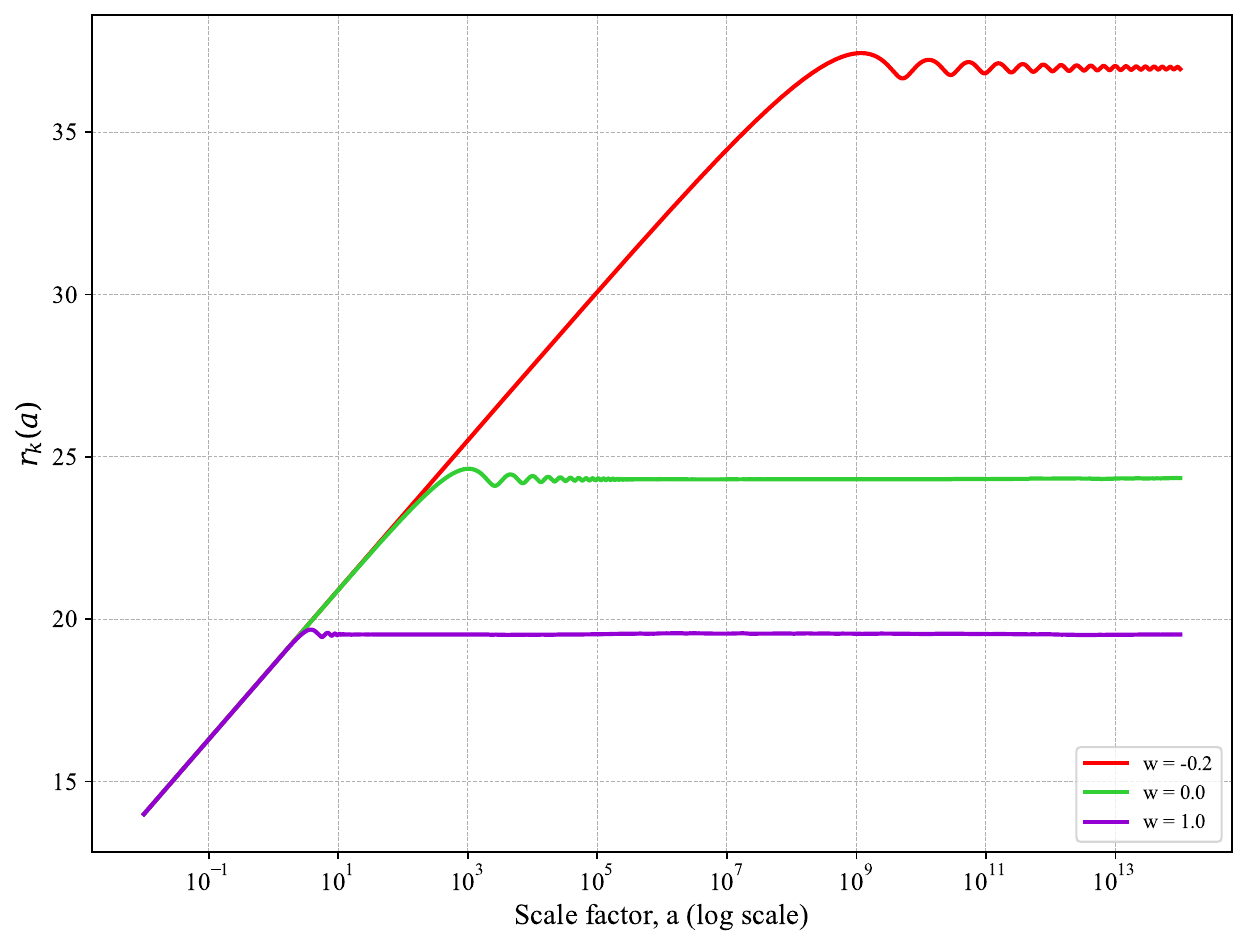}
    \caption{Numerical solutions for $r_k$ as a function of scale factor $a$ for expanding decelerating background. For smaller $w$, the squeezing occurs for a much longer time, and it reaches a higher final value compared to higher $w$ values. Oscillations can be seen after the squeezing stops.} 
    \label{fig:2a}
\end{figure}

In Figure \ref{fig:2a}, we show the evolution of $r_k$ in a decelerating universe, where the equation of state parameter $w$ is varied. The modes start in a super-hubble state with a large initial value of $r_k \approx 14$. Because the modes are super-Hubble, $r_k$ first grows in proportion to $\log a$. As the universe continues to expand, these modes eventually re-enter the horizon. After a mode re-enters the horizon, the growth of its squeezing parameter stops. The value of $r_k$ then freezes to a large constant value, with small oscillations on top. The squeezing that was generated while the mode was super-Hubble is preserved. Therefore, $r_k$ does not return to zero. The plot shows that modes in universes with a smaller $w$ ($w=-0.2$) re-enter the horizon later and freeze at a much higher value of $r_k$.
\begin{equation}
    r_k(a) \approx r_0 + \ln \left( \frac{a}{a_0} \right)\,,
    \label{eq:squeezing_amplitude}
\end{equation}

\begin{figure}[h!]
    \centering
    \includegraphics[width=1\linewidth]{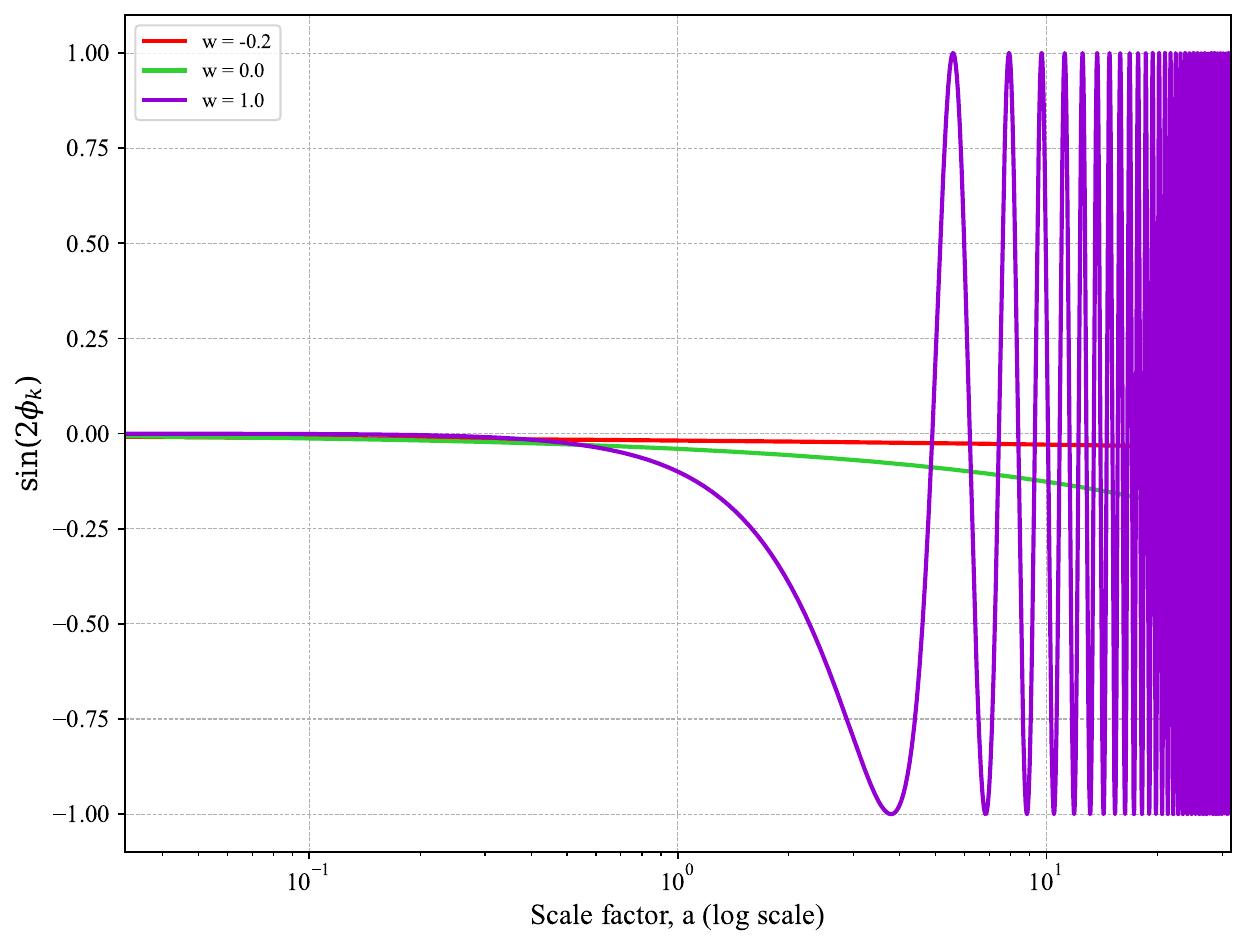}
    \caption{Numerical solutions for $sin(2\phi_k)$ as a function of scale factor $a$ for expanding decelerating background. All curves begin at $0$ and start oscillating. For higher $w$, the oscillation begins much earlier compared to lower values. For $w = -0.2$ and $0.0$, the transition to oscillation occurs much later, which isn't visible in the graph.}
    \label{fig:2b}
\end{figure}

Figure \ref{fig:2b} shows the behavior of the squeezing angle, represented by $\sin(2\phi_k)$, for the same decelerating universes as above. We begin with the modes already super-Hubble, so the squeezing angle $\phi_k$ is initially fixed at $-\pi/2$. This makes $\sin(2\phi_k)$ start at zero. During this time, the squeezing parameter $r_k$ grows steadily. As the modes start to re-enter the horizon, the angle is no longer frozen. The plot shows that $\sin(2\phi_k)$ first moves towards -1 and then, once the modes are fully inside the horizon, it begins to oscillate rapidly between $+1$ and $-1$. This transition from a fixed angle to an oscillating one happens at the same time that the squeezing parameter $r_k$ stops growing and freezes. It shows that when modes re-enter the horizon, their squeezing properties stop evolving with the expansion and start oscillating
\begin{equation}
    \phi_k(a) \approx -\frac{\pi}{2} + \frac{k \eta_0}{|\beta| (|\beta| + 2)} a^{1/|\beta|}\,.
    \label{eq:squeezing_phase}
\end{equation}

\subsubsection{Contracting background}
\begin{figure}[h!]
    \centering
    \includegraphics[width=1\linewidth]{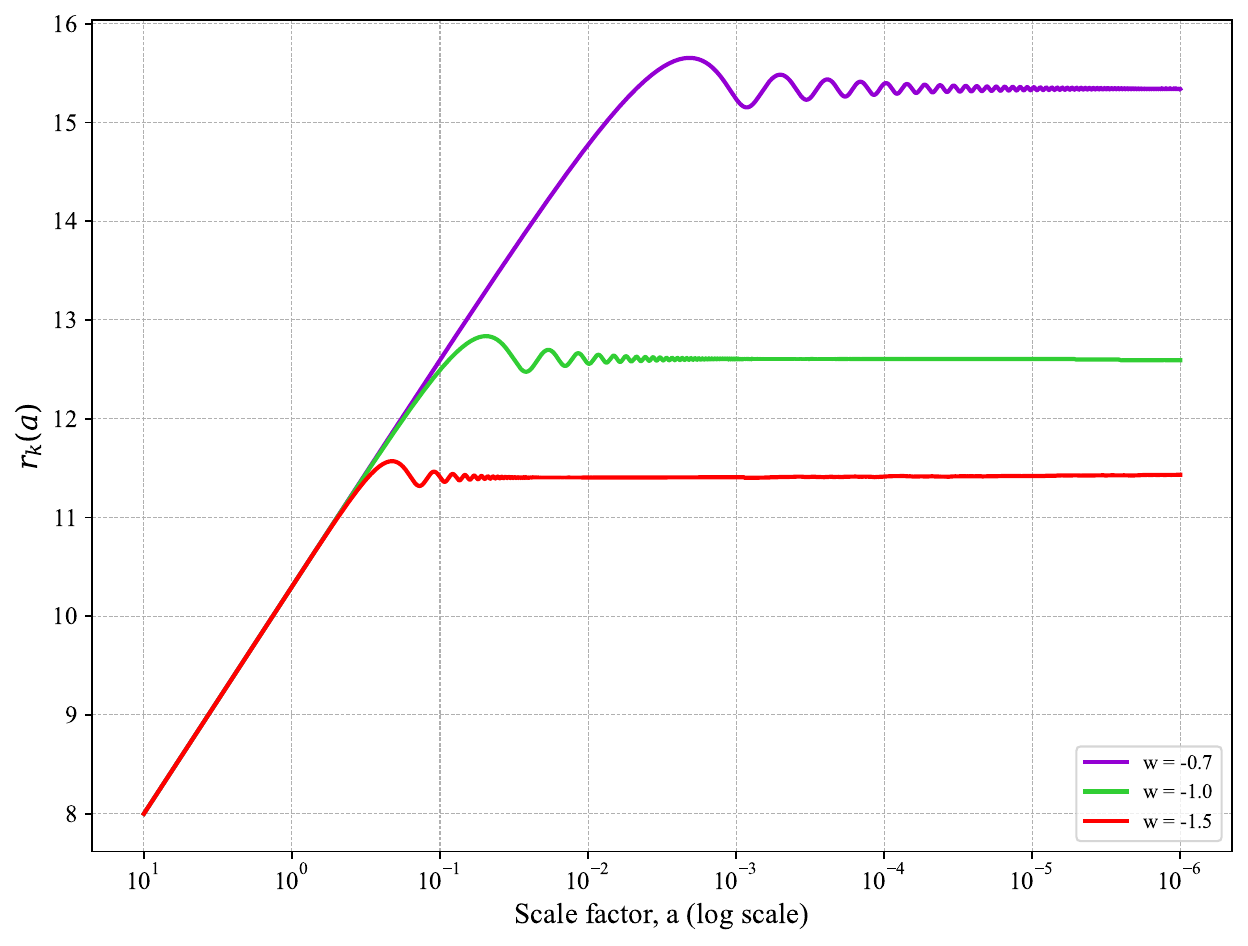}
    \caption{Numerical solutions for $r_k$ as a function of scale factor $a$ for contracting accelerating background. The plateauing of the values of $r_k$ depends on $w$. For lower values of $w$, the growth stops early, and for higher values, $r_k$ reaches a higher value and then plateaus at a later time.}
    \label{fig:3a}
\end{figure}

\begin{figure}[h!]
    \centering
    \includegraphics[width=1\linewidth]{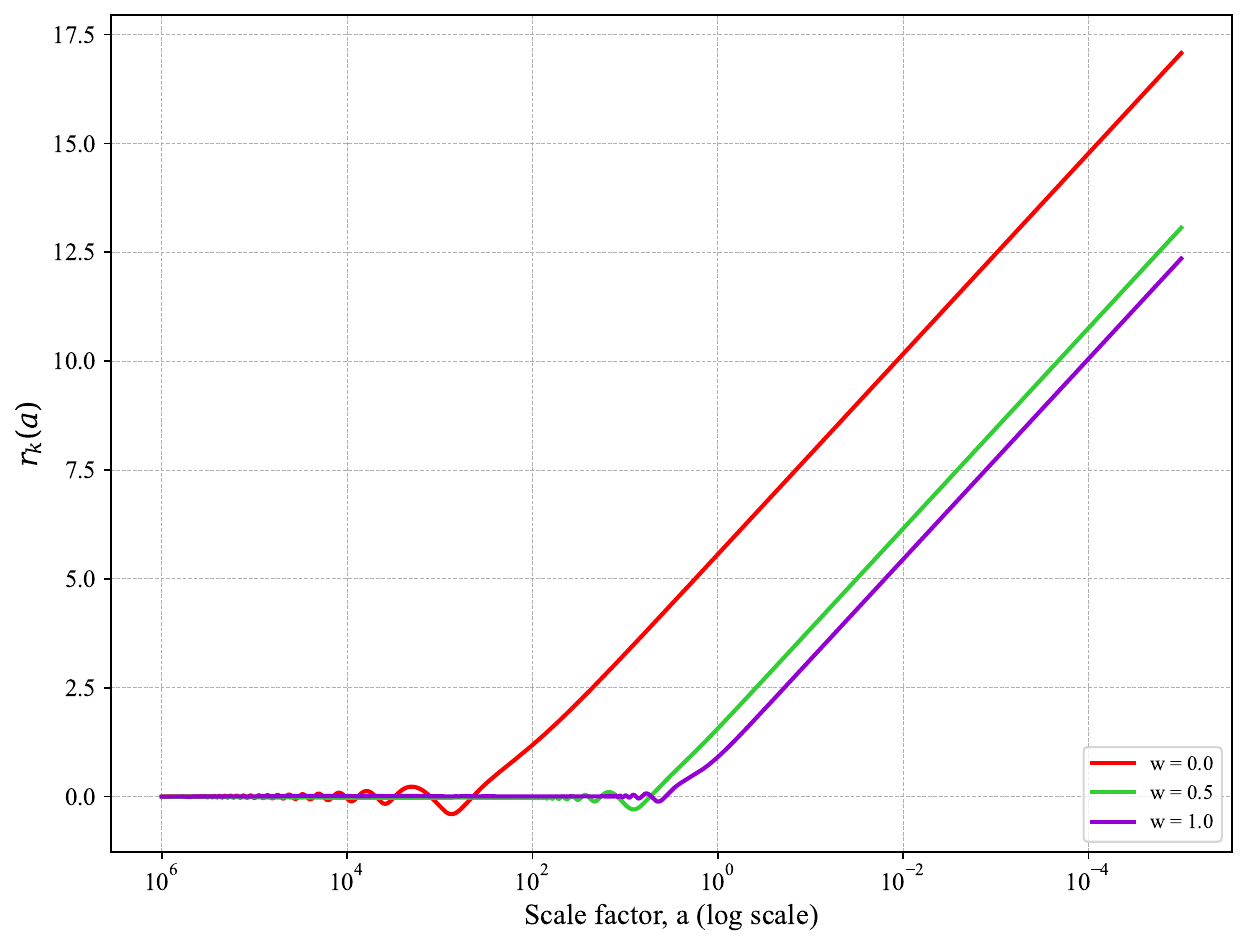}
    \caption{Numerical solutions for $r_k$ as a function of scale factor $a$ for contracting decelerating background. The plots start at $0$, while oscillating and rising steadily based on the value of $w$ at different times. Lower $w$ values ($w=0.0$) start the rise much earlier compared to the higher values.}
    \label{fig:3b}
\end{figure}

The expression for the scale factor $a(\eta)$ for the contracting background is given by~\cite{Bhattacharyya:2020kgu}:
\begin{equation}
a(\eta)=
\begin{cases}
\displaystyle\left(\frac{\eta_{0}}{\eta}\right)^{\beta},
& 0<\eta<\infty,\ \eta_{0}>0,\ \beta>0\;(w<-1/3)\quad\text{(accelerating)}\,,\\[8pt]
\displaystyle\left(\frac{\eta}{\eta_{0}}\right)^{|\beta|},
& -\infty<\eta<0,\ \eta_{0}<0,\ \beta<0\;(w>-1/3)\quad\text{(decelerating)}\,,
\end{cases}
\label{eq:scale_factor_contr}
\end{equation}
where,  $\beta = -\frac{2}{1+3w}$. Then, the equations of motion, Eqs. \ref{eq:squeezing_parameter} and \ref{eq:squeezing_angle}  for the contracting background are,
\begin{align}
\frac{d r_k}{d a} &= -\frac{1}{a} \cos(2\phi_k)\,,
\label{rkdiffeq2}
\\
\frac{d \phi_k}{d a} &=- k \frac{|\eta_0|}{|\beta|} \frac{1}{a^{1+1/\beta}} + \frac{1}{a} \coth(2r_k) \sin(2\phi_k)\,.
\label{pkdiffeq2}
\end{align}
To generate Figures \ref{fig:3a} and \ref{fig:3b}, we fixed the wavenumber at $k = 0.1$ and set the initial squeezing angle to $\phi_k(a_0) = -\pi/2$. For Figure \ref{fig:3a}, the scale factor starts at $a_0 = 10^{1}$ and evolves down to $a_{\text{end}} = 10^{-6}$, representing a contracting background. The squeezing parameter was initialized at $r_k(a_0) = 8$, with conformal time set to $\eta = 1.0$. For Figure \ref{fig:3b}, the evolution begins from $a_0 = 10^{6}$ and ends at $a_{\text{end}} = 10^{-5}$. We used an initial squeezing parameter of $r_k(a_0) = 1$ and the axis rescaled to zero, and started the evolution at $\eta = -1.0$.

Figure \ref{fig:3a} shows the evolution of the squeezing parameter $r_k$ in a contracting, accelerating universe. In this case, the scale factor $a$ starts at a large value and becomes smaller over time. We start with a large value of $r_k \approx 8$ when the modes are super-Hubble. As the universe contracts (as $a$ decreases), the squeezing parameter $r_k$ grows. On the log plot, this growth appears as a straight line. As the contraction continues, the modes eventually become sub-Hubble. When this happens, the growth of $r_k$ stops, and it freezes to a constant value, with small oscillations on top. The plot shows that the final amount of squeezing depends on the equation of state. Universes with a larger $w$ end up with a much larger final value for $r_k$.
Figure \ref{fig:3b} shows the squeezing parameter $r_k$ for a contracting, decelerating universe. The x-axis shows the scale factor $a$ decreasing, so time moves from left to right. At the start of the contraction (large $a$), the modes are sub-Hubble, and $r_k$ shows small oscillations around zero. As the universe contracts, the modes become super-Hubble, and at this point, $r_k$ begins to grow. In the super-Hubble regime (small $a$), the plot shows that $r_k$ grows as a straight line, which means its growth is proportional to $-\log a$. 

\section{Continuous variable quantum computing}
\label{section 3}
The concept of continuous variable quantum computing (CVQC) is based on infinite-dimensional quantum systems called qumodes. A quantum mode or qumode is an elementary unit of quantum information processing in the CVQC model whose Hilbert space corresponds to that of a quantum harmonic oscillator. The unitary operators acting on the infinite-dimensional Hilbert space are equivalent to the logic gates in this model~\cite{Cochran:2024cri,Braunstein:2005zz}. This concept of using quadrature operators to form logic gates has been discussed by Lloyd and Braunstein~\cite{Lloyd:1998jk}. The notion of universality in CVQC can be defined for transformations corresponding to Hamiltonians that are polynomial in the operators of continuous variables. A set of quantum gates is referred to as universal if any arbitrary unitary transformation can be expressed as a finite product of these gates to an arbitrary accurate approximation. So, the techniques to decompose any given unitary in terms of a universal gate set are necessary to program a quantum computer. Such methods for CVQC have been discussed in~\cite{Kalajdzievski:2021axm}. In the case of modes of the electromagnetic field, linear operations such as translations, phase shifts, squeezers, and beam splitters, in addition to a nonlinear operation such as a Kerr nonlinearity, make it possible to perform arbitrary polynomial transformations on those modes~\cite{Braunstein:2005zz}.

A CV system~\cite{Ferraro:2005hen,Serafini:2017rrn,Brask:2021kvs,Weedbrook:2011wxo} of $n$ modes is described by n bosons with corresponding creation and annihilation operators $\{\hat c^{\dagger}_j, \hat c_j\}$ with $j=1,2,...,n$. These operators satisfy the canonical commutation relations, 
$[\hat{c}_j, \hat{c}_k] = [\hat{c}_j^\dagger, \hat{c}_k^\dagger] = 0$,and 
$[\hat{c}_j, \hat{c}_k^\dagger] = \delta_{jk}$, where $\delta_{jk}$ is the Kronecker delta.
We now define the dimensionless quadrature operators ($\hbar$ = $1$),
\begin{align}
\hat{q}_j &= \frac{1}{\sqrt{2}} (\hat{c}_j + \hat{c}_j^\dagger),
\label{3}
\\
\hat{p}_j &= \frac{1}{i\sqrt{2}} (\hat{c}_j - \hat{c}_j^\dagger)\,,
\label{4}
\end{align}
with the commutation relation $[\hat{q}_j, \hat{p}_k] = i\delta_{jk}$. We now group the quadrature operators for the $n$ modes in a column vector as follows,
\begin{equation}
\hat{\mathbf{r}} = 
(\hat{q}_1,\hat{p}_1,\dots,\hat{q}_n,\hat{p}_n)^T\,.
\end{equation}
The canonical commutation relations can then be written $[\hat{r}_j, \hat{r}_k] = i \Omega_{jk}$, where $\Omega$ is the n-mode symplectic form defined as,
\begin{equation}
\label{eq:Symplectic_form}
\Omega = \bigoplus_{j=1}^{2n}
\begin{pmatrix}
0 & 1 \\
-1 & 0
\end{pmatrix}\,.
\end{equation}
The continuous variable approach provides an equivalent reformulation of the density operator in terms of a quasi-probability distribution referred to as the Wigner function, defined on the phase space~\cite{Weedbrook:2011wxo}.  An arbitrary quantum state $\hat{\rho}$ of an n-mode bosonic system is equivalent to a Wigner function W(r) defined over a $2n$-dimensional phase space $(\mathbb{R}^{2n}, \Omega)$. The characteristic function of a quantum state $\hat{\rho}$ is defined as,
\begin{equation}
    \chi(\zeta) = \Tr{[\hat{\rho}D(\zeta)]}\,,
\end{equation}
where $D(\zeta) = \exp{(i\hat{r}^T\Omega\zeta)}$ is the Weyl operator. Now, the Fourier transform of this characteristic function leads to the corresponding Wigner function,
\begin{equation}
\label{eq:Wigner_def}
    W(r) = \int_{\mathbb{R}^{2n}} \frac{d^{2n}\zeta}{(2\pi)^{2n}} exp(-ir^T\Omega\zeta) \chi(\zeta)\,.
\end{equation}
Here, the Wigner function is normalized to one.
\begin{equation}
    \int_{\mathbb{R}^{2n}} W(r) dr = 1\,.
\end{equation}
An interesting property of the Wigner function is that it can be negative, and hence it is referred to as a quasi-probability distribution. In Eq. \ref{eq:Wigner_def} the continuous variables $r \in \mathbb{R}^{2r}$ are the eigenvalues of quadrature operators $\hat
r$. These variables span a real symplectic space $(\mathbb{R}^{2n}, \Omega)$ called the phase space. The Wigner representation provides a complete description through the statistical moments of the quantum state. The first and second moment are called mean ($\mu$) and covariance ($V$) respectively, and defined by,
\begin{align}
    \mu_j &= \langle \hat r_j \rangle\,,\\
V_{jk} &= \langle \{\hat{r}_j - \mu_j, \hat{r}_k - \mu_k\} \rangle \,,
\end{align}
where $\{\cdot, \cdot\}$ represents the anti-commutator, and $\langle\cdot\rangle$ gives the expectation value with respect to the state. The covariance matrix is a $2n \times 2n$ matrix which is real, symmetric, and positive definite. The variances of the quadrature operators are stored in the diagonal entries, and the off-diagonal entries represent correlations between different quadratures. The covariance matrix obeys the following inequality,
\begin{equation}
V + i\Omega \geq 0\,.
\end{equation}
For a particular class of quantum states, the first two moments provide a complete description, i.e., $\hat{\rho} = \hat{\rho}(\mu,V)$. Such bosonic states are called Gaussian states and whose Wigner representation is Gaussian,
\begin{equation}
    W(r) = \frac{1}{(2\pi)^n \sqrt{det(V)}} \exp\left[-\frac{1}{2}(r-\mu)^TV^{-1}(r-\mu)\right]\,.
\end{equation}
We note that for a Gaussian state, the non-negative nature of its Wigner function implies that the state is pure. Some important examples of Gaussian states are vacuum state ($\mu = 0, V = \mathbf{I}_{2n}$), coherent state ($\mu \neq 0, V = \mathbf{I}_{2n}$), squeezed state ($V \neq \mathbf{I}_{2n}$) and thermal state ($\mu = 0, V = \bigoplus_i (1+2n_i)\mathbf{I}_{2n}$) with $i= 1,2,...,n$, where $\mathbf{I}_{2n}$ is $2n \times 2n$ identity matrix and $n_i$ is the mean number of noise quanta. In the Gaussian formalism, the evolution of the quantum state is described by the symplectic transformations. The output state can be obtained from the input states using the symplectic matrix $S$ as follows,
\begin{align}
    \label{eq:mean_S}
    \mu_{out} &= S \mu_{in}\,,\\
    \label{eq:cov_S}
    V_{out} &= S V_{in} S^T\,.
\end{align}
\subsection{CV Gates}

The unitary transformations in the Gaussian formalism must preserve the Gaussian nature of the given quantum state. This is true for the Hamiltonian, which is quadratic in the creation and annihilation operators. We consider a general quadratic Hamiltonian with complex coefficients~\cite{Adesso:2014npz},
\begin{equation}
    \hat H = P_{ij} \hat{c}_i \hat{c}_j + P^*_{ij} \hat{c}^\dagger_i \hat{c}^\dagger_j + Q_{ij} \hat{c}^\dagger_i \hat{c}_j + Q^*_{ij} \hat{c}_i \hat{c}^\dagger_j \,.
\end{equation}
In the Gaussian formalism, such a quadratic Hamiltonian corresponding to a unitary transformation $\hat{U} = \exp{-i\hat{H}}$ is mapped by symplectic transformations on the mean and covariance matrix as given by Eqs. \ref{eq:mean_S} and \ref{eq:cov_S}. Indeed, the cosmological Hamiltonian in Eq. \ref{eq:Hamiltonian} is quadratic in the mode operators and can thus be studied using symplectic transformations like squeezing and phase rotation from linear optics. 

In the CV framework, we can work with Gaussian and non-Gaussian CV gates. In this paper, we only use Gaussian gates as the Hamiltonian we model is quadratic in nature, but the higher-order dynamics can also be studied using the non-Gaussian gates. The combination of Gaussian and non-Gaussian gates can be used to form a universal gate set that allows us to simulate any arbitrary Hamiltonian to desired precision. Now we present the symplectic transformations which correspond to some frequently used Gaussian unitary operations. We also describe the Kerr gate, which is a non-Gaussian gate, to provide an example of a universal gate set~\cite{Kalajdzievski:2021axm,Weedbrook:2011wxo,Brask:2021kvs,Serafini:2017rrn}. 
\subsubsection*{Displacement}
The displacement operator is defined by,
\begin{equation}
    \hat D(\alpha) = \exp{(\alpha \hat{c}^\dagger - \alpha^* \hat{c})}\,,
\end{equation}
where $\alpha_j = \frac{q_j + ip_j}{2}$ is a complex number. It transforms the mode operator as,
\begin{equation}
    \hat D^\dagger(\alpha)\hat{c} \hat D(\alpha) = \hat{c} + \alpha\,.
\end{equation}
\begin{figure}[hbt!]
    \centering
    \includegraphics[width=0.25\linewidth]{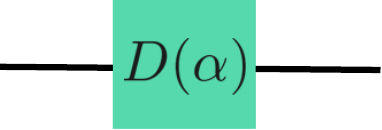}
    \caption{Circuit diagram of displacement operator}
    \label{fig:displacement}
\end{figure}
\subsubsection*{Phase rotation}
\begin{figure}[hbt!]
    \centering
    \includegraphics[width=0.25\linewidth]{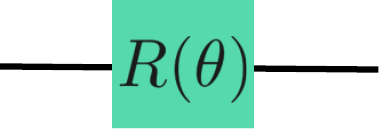}
    \caption{Circuit diagram of phase rotation operator}
    \label{fig:phase_rotation}
\end{figure}

The phase rotation operator is defined by,
\begin{equation}
   \hat R(\theta) = \exp{(-i\theta\hat{c}^\dagger \hat{c})}\,.
\end{equation}
Here $\theta$ is the rotation angle. The circuit diagram is shown in \ref{fig:phase_rotation}, and the corresponding symplectic matrix is,
\begin{equation}
    R(\theta) = \begin{pmatrix}
        \cos{\theta} & \sin{\theta}\\
        -\sin{\theta} & \cos{\theta}
    \end{pmatrix}.
\end{equation}

\subsubsection*{Beam Splitter}
A beam splitter for two bosonic modes is defined by,
\begin{equation}
    \hat{BS}(\delta) = exp[\delta(\hat{c_1}^\dagger \hat{c_2} - \hat{c_1} \hat{c_2}^\dagger)]\,.
\end{equation}
Here $\tau = \cos^2\delta$ is the transmissivity of the beam splitter. A beam splitter modifies the annihilation operators $\hat{c}_1$ and $\hat{c}_2$ as follows,
\begin{align}
\hat{c}_1 &\to \sqrt{\tau}\,\hat{c}_1 + \sqrt{1-\tau}\,\hat{c}_2\,, \\
\hat{c}_2 &\to -\sqrt{1-\tau}\,\hat{c}_1 + \sqrt{\tau}\,\hat{c}_2\,.
\end{align}
The symplectic matrix for quadrature transformation is:
\begin{equation}
BS(\delta) = \begin{pmatrix}
\cos{\delta}\mathbf{I}_2& \sin{\delta}\mathbf{I}_2\\
-\sin{\delta}\mathbf{I}_2& \cos{\delta}\mathbf{I}_2
\end{pmatrix}\,.
\end{equation}
Here $\mathbf{I}_2$ is the $2 \times 2$ identity matrix.

\begin{figure}[hbt!]
        \centering
        \includegraphics[width=0.25\linewidth]{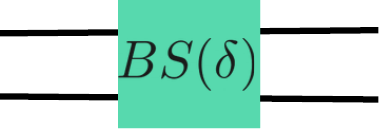}
        \caption{Circuit diagram for beam splitter}
        \label{fig:beam_splitter}
\end{figure}

\subsubsection*{Squeezing operators}

The squeezing operator acts on a mode by squeezing one canonical variable and expanding the conjugate variable. The action of the single-mode squeezing operator is given by,
\begin{equation}
\hat{S}_1(\xi) = \exp \{\xi (\hat{c}^\dagger)^2 - \xi^* \hat{c}^2\},
\end{equation}
where $\xi = r e^{i2\phi}$. This operator transforms the annihilation operators as follows:
\begin{align}
\hat{S}_1(\xi)^\dagger \hat{c} \hat{S}_1(\xi) &= \cosh(r)\hat{c} + e^{i2\phi}\sinh(r)\hat{c}^\dagger,  \\
\hat{S}_1(\xi)^\dagger \hat{c}^\dagger \hat{S}_1(\xi) &= \cosh(r)\hat{c}^\dagger + e^{-i2\phi}\sinh(r)\hat{c}\,.
\end{align}
Similarly, the two-mode squeezing operator acting on modes 1 and 2 is given by,
\begin{equation}
\hat{S}_2(\xi) = \exp \{\xi \hat{c}_1^\dagger \hat{c}_2^\dagger - \xi^* \hat{c}_1 \hat{c}_2\},
\end{equation}
where $\xi = r e^{i2\phi}$. This operator transforms the annihilation operators as follows:
\begin{align}
\hat{S}_2(\xi)^\dagger \hat{c}_1 \hat{S}_2(\xi) &= \cosh(r)\hat{c}_1 + e^{i2\phi}\sinh(r)\hat{c}_2^\dagger,  \\
\hat{S}_2(\xi)^\dagger \hat{c}_2 \hat{S}_2(\xi) &= \cosh(r)\hat{c}_2 + e^{i2\phi}\sinh(r)\hat{c}_1^\dagger\,.
\end{align}
A symplectic transformation on two modes is,
\begin{equation}
S_2(r,\phi) = 
\begin{pmatrix}
\cosh(r)\mathbf{I}_2 & -\sinh(r)S_{\phi} \\
-\sinh(r)S_{\phi} & \cosh(r)\mathbf{I}_2
\end{pmatrix}\,,
\end{equation}
where,
\begin{equation}
S_{\phi} =
\begin{pmatrix}
\cos(2\phi) & \sin(2\phi) \\
\sin(2\phi) & -\cos(2\phi)
\end{pmatrix}\,.
\end{equation}
A two-mode squeezed state can also be generated by using two single-mode squeezed states mixed by a symmetric beam splitter~\cite{Wolf:2002bhs}. Consider a beam splitter with transmissivity, $\delta = 0.5$ then 
\begin{equation}
    BS^\dagger(0.5) S_2(\xi) BS(0.5) = S_1(\xi) \otimes S_1(-\xi)\,.
\end{equation}
Since the action of the beam splitter is reversible, we can generate a two-mode squeezed state by combining two single-mode squeezed states using a beam splitter~\cite{Ferraro:2005hen,Braunstein:2005zz}.
\begin{figure}[hbt!]
        \centering
        \includegraphics[width=0.8\linewidth]{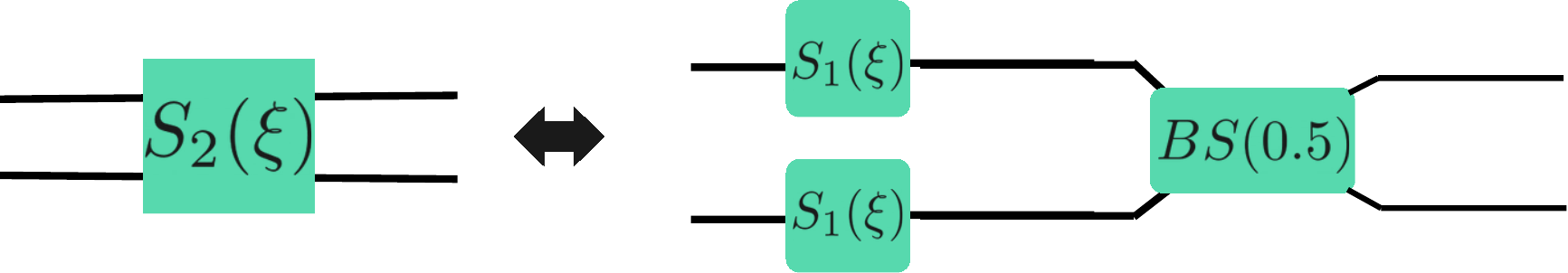}
        \caption{Circuit diagram showing the equivalence of two-mode squeezing to two single-mode squeezings followed by a beam splitter}
        \label{fig:Wolf:2002bhs}
\end{figure}
\subsubsection*{Kerr gate}
Kerr gate is a non-Gaussian gate which is given by the Hamiltonian,
\begin{equation}
    \hat{H} = (\hat{c}^\dagger \hat{c})^2 = \hat{n}^2\,.
\end{equation}
The Kerr gate is defined with parameter $\epsilon$ as,
\begin{equation}
    \hat K(\epsilon) = exp(i\epsilon\hat{n}^2)\,.
    \label{kerr Hamiltonian}
\end{equation}
The transformation of mode operators by the Kerr gate is as follows,
\begin{align}
    \hat{K}^\dagger(\epsilon)\hat{c}\hat{K}(\epsilon) = \hat{c}e^{i\epsilon(2\hat{c}^\dagger\hat{c} + 1)}\,,\\    \hat{K}^\dagger(\epsilon)\hat{c}^\dagger\hat{K}(\epsilon) = \hat{c}^\dagger e^{-i\epsilon(2\hat{c}^\dagger\hat{c} + 1)}\,.
\end{align}

\begin{figure}[hbt!]
    \centering
    \includegraphics[width=0.25\linewidth]{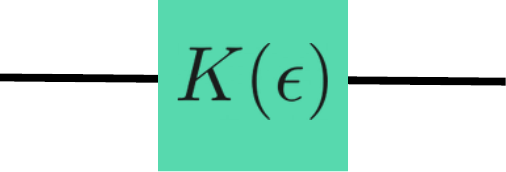}
    \caption{Circuit diagram of kerr gate}
    \label{fig:kerr-gate}
\end{figure}
\subsection{Entanglement measures for gaussian states}

As we have already discussed, a Gaussian state is completely characterized by its first and second moments. Consequently, the entanglement measures in the Gaussian formalism can be obtained from the covariance matrix of the given quantum state. In this section, we describe various entanglement measures such as von Neumann entropy and logarithmic negativity, along with purity and R\'enyi entropy for the Gaussian states. First, we briefly discuss some important concepts such as symplectic eigenvalues and partial transposition, which are essential for the calculation of the entanglement measures discussed below. For more details, refer to~\cite{Plenio:2007zz,Serafini:2017rrn,Plenio:2005cwa,Vidal:2002zz}.

A covariance matrix, $V$, can be transformed into its normal form by a symplectic matrix $S$ as,
\begin{equation}
    SVS^T = \bigoplus_{i=1}^n \begin{pmatrix}
        v_i & 0\\
        0 & v_i
    \end{pmatrix}\,.
\end{equation}
Here, the set of diagonal entries is referred to as the symplectic spectrum of the covariance matrix $V$. The symplectic eigenvalues are the absolute eigenvalues of the matrix $V\Omega$. This matrix gives $2n$ eigenvalues which are purely imaginary and come in pairs $\pm iv_i$ where $v_i$ are real with $i=1, 2,..., n$. Here we note that the symplectic eigenvalues of a covariance matrix have two-fold degeneracy. The symplectic spectrum of a covariance matrix encodes the information about the entanglement of the quantum state~\cite{Brady:2022ffk,Kranas:2023aph}.

The entanglement of Gaussian states can also be quantified by using the property of partial transposition of the covariance matrix. The positivity of partial transposition (PPT) is the basis of entanglement measures like logarithmic negativity, which is described in detail below~\cite{Plenio:2007zz}. Here we discuss the concept of partial transposition and the corresponding covariance matrix, particularly for a bipartite Gaussian state. The partially transposed density matrix ($\rho^{pt}$) is the matrix in which one subsystem is transposed while the other subsystem remains unchanged. The partial transpose of a bipartite quantum state with density matrix $\rho = \sum P_{kl,mn} \ket{k}\bra{l}_A \otimes \ket{m}\bra{n}_B$, with respect to subsystem B, is,
\begin{align}
\label{eq:rho_pt}
    \rho^{pt}_B = \sum P_{kl,mn} \ket{k}\bra{l}_A \otimes \ket{m}\bra{n}_B\,.
\end{align}
The spectrum of the partially transposed matrix in Eq. \ref{eq:rho_pt} is independent of the local orthonormal basis and also the choice of subsystem over which the partial transposition is performed~\cite{Plenio:2007zz}. The covariance matrix of the partially transposed density matrix is obtained as,
\begin{equation}
V^{pt} = T V T\,,
\end{equation}
where $T = \mathbb{I_{2n_1}} \bigoplus \sigma$, with $\sigma = \bigoplus_{i=1}^{n_2}\sigma_z$, where $\sigma_z$ is the z-Pauli matrix. For a two-mode Gaussian state, the matrix $V^{pt}$ is obtained from the covariance matrix, $V$, by reversing the sign of all components involving momenta $p_i$ of the subsystem. This amounts to changing the sign of the fourth row and fourth column of the covariance matrix~\cite{Brady:2022ffk}.

Now, using the concepts described above, we provide the descriptions of purity, R\'enyi entropy, von Neumann entropy, and logarithmic negativity in the context of the Gaussian formalism.
\subsubsection{Purity}
In the context of Gaussian systems—especially in the presence of decoherence or thermal noise—tracking purity and related entropic measures offer a practical and informative way to assess coherence and entanglement degradation~\cite{Serafini:2017rrn}.

A Gaussian quantum state is pure if and only if the determinant of its covariance matrix equals one, i.e., $\det(V) = 1$, which corresponds to all symplectic eigenvalues being unity. This condition ensures that the state exhibits minimal uncertainty and no classical statistical mixing. In general, the purity of a quantum state $\rho$ is defined as
\[
\mu = \mathrm{Tr}(\rho^2),
\]
and for Gaussian states with vanishing first moments, it is given by
\begin{equation}
\mu = \frac{1}{\sqrt{\det(V)}}.
\end{equation}
This makes purity a direct function of the phase-space fluctuations encoded in the covariance matrix. 
\subsubsection{von Neumann and  R\'enyi  entropy}

For a quantum state described by density matrix $\rho$, the von Neumann entropy is defined as,
\begin{equation}
S(\rho)=-\Tr{( \rho \log_2( \rho)}\,.
\end{equation}
For a pure quantum state, only one of the eigenvalues of $\rho$ is 1, and the rest are zero; hence, the entropy of a pure state is $0$, and $S( \rho) > 0$ for a mixed quantum state. For a Gaussian state of $n$ modes, the von Neumann entropy is calculated as,
\begin{equation}
\label{eq:Entropy_cov}
S(V) = \sum_{j=1}^n \frac{v_j + 1}{2} \log_2 \left( \frac{v_j + 1}{2} \right) - \frac{v_j - 1}{2} \log_2 \left( \frac{v_j - 1}{2} \right)\,,
\end{equation}
where $v_j$ are the symplectic eigenvalues of the covariance matrix, $V$. The symplectic eigenvalues of a covariance matrix are the absolute eigenvalues of the matrix $  V\Omega$ as discussed above.

The von Neumann entropy is a reliable measure of entanglement only if the total quantum state is pure~\cite{Kranas:2023aph,Brady:2022ffk}. To illustrate this, consider a two-mode Gaussian state. If the total state is pure and  the subsystems are mixed, then von Neumann entropy quantifies the entanglement entropy of the state. But if the total state is mixed, the von Neumann entropy is no longer a reliable entanglement measure, as the system may or may not be entangled.

The generalization of von Neumann entropy is the quantum R\'enyi entropy defined as~\cite{Li:2022ktm},
\begin{equation}
    S_\mu(\rho) = \frac{\ln \Tr{\rho^\mu}}{1-\mu} = \frac{\ln \left( \Sigma_{i=1}^n \lambda^{\mu}_{i} \right)} {1-\mu}\,,
\end{equation}
where $\lambda_i$ are the eigenvalues of the density matrix and $\mu$ is the order of the R\'enyi entropy. The R\'enyi entropy becomes von Neumann entropy as $\mu$ approaches $1$. Similarly, it can be used to bound the von Neumann entropy as,
\begin{equation}
    S_2 < S_1 < S_{1/2}\,.
\end{equation}

\subsubsection{Logarithmic negativity}

 The positivity of partial transposition (PPT) criterion states that a quantum state $\rho$ is separable if $\rho^{pt} \geq 0$ (i.e.,$\lambda_j^{pt} \geq 0 \quad \forall \, j$ is a necessary condition.
where $\lambda_j^{pt}$ are the eigenvalues of $\rho^{pt}$).
Hence, a sufficient condition for entanglement is that if $\exists$ j such that $\lambda_j^{pt} < 0 $.
In conclusion, a quantum state is entangled if it violates the PPT criterion~\cite{Serafini:2017rrn}. Logarithmic negativity (LN) is an entanglement measure based on the PPT criterion and defined as~\cite{Plenio:2005cwa,Vidal:2002zz},
\begin{equation}
LN = \log_2 ||\rho^{T_B}||\,,
\end{equation}
where $||\rho|| = Tr\sqrt{\rho^\dagger \rho}$ is the trace norm, and since $\rho$ is Hermitian, 
\begin{equation}
LN = \log_2\left(\sum_i |\lambda_i^{pt}|\right)\,,
\end{equation}
where $\lambda_i$'s are the eigenvalues of state $\rho^{pt}$.

The logarithmic negativity is easy to compute, and on Gaussian states it has an interpretation as a special type of entanglement cost. It is a reliable entanglement measure even if the total quantum state is mixed. It is important to note that $LN > 0$ indicates the existence of entanglement, but $LN = 0$ does not ensure separability, i.e., in general, $LN>0$ is a sufficient but not a necessary condition for entanglement. But LN does not reduce to the entropy of entanglement on pure states~\cite{Plenio:2005cwa,Plenio:2007zz}.

Logarithmic negativity can be computed for a bipartite Gaussian state described with a covariance matrix $V$ as~\cite{Serafini:2017rrn,Kranas:2023aph},
\begin{equation}
LN = \sum_j^{n_1 + n_2} \text{max}\{0,-\log_2(v_j^{PT})\}\,,
\end{equation}
where $v_j^{pt}$ are the symplectic eigenvalues of the covariance matrix, $V^{pt}$, of the partially transposed density matrix $\rho^{pt}$. The method to obtaine the covariance matrix $V^{pt}$ has been described above. For a two-mode system, there is only one symplectic eigenvalue that can be less than one, and we have~\cite{Serafini:2017rrn},
\begin{equation}
\label{eq:LN_two_mode}
LN = \text{max}\{0,-\log_2 v_{min}^{pt}\}\,.
\end{equation}
Consequently, for the two-mode Gaussian state, $LN>0$ is both a necessary and a sufficient condition for entanglement~\cite{Kranas:2023aph}.

\section{Quantum information theoretic quantities for cosmological perturbations}
\label{section 4}
In this section, we analyze the quantum information-theoretic structure that is inherent in cosmological perturbations. Here, we particularly focus on entanglement generated during inflation. We know that the quantum fluctuations in the early universe evolve into highly entangled two-mode squeezed states correlating modes as momenta $k$ and $-k$. So, we explore several fundamental entanglement measures to quantitatively characterize this structure. We first analytically compute the von Neumann entropy by deriving the reduced density matrix through a partial trace over one of the modes and then simultaneously applying a coarse-graining procedure. Parallely, the logarithmic negativity is calculated via partial transposition of the two-mode density matrix. Furthermore, the R\'enyi entropy is introduced as a generalized entropy. Purity is also evaluated as a function of the covariance matrix determinant, which thereby offers insight into the degree of mixedness of the reduced states.

To calculate these quantities in a clear and organized way, we use the Gaussian formalism and describe the evolution of cosmological modes using a quantum circuit made up of two-mode squeezing operations. This circuit representation enables efficient tracking of the first and second moments of the state, thereby simplifying the computation of subsystem entropy and correlations. The formalism is extended to include thermal noise by modifying the initial covariance matrix, allowing us to examine how thermal effects degrade entanglement. Finally, we address numerical instabilities inherent in the entropy expressions, particularly at the large squeezing values, and demonstrate this through explicit evaluation and comparative plots of divergent terms.

\subsection{Entanglement entropy}
\label{sec error}
 The entanglement entropy quantifies the degree of quantum entanglement between these modes, which arises due to the squeezing operation that correlates the particle number states of modes $k$ and $-k$. The density matrix in terms of the two-mode occupation number basis using Eq. \ref{eq:Squeezed state}  is: 
\begin{equation}
\label{density}
\rho = \prod_k \prod_p \sum_{n=0}^\infty \sum_{m=0}^\infty 
\frac{1}{\cosh r_k \cosh r_p} 
e^{-2i\phi_k(n - m)} \tanh^n r_k \tanh^m r_p 
\, |n_k, n_{-k} \rangle \langle m_p, m_{-p}|\,.
\end{equation}
To compute the entanglement entropy between modes $k$ and $-k$, we trace out one of the modes $-k$ to obtain the reduced density matrix for mode $k$ as
\begin{equation}
\rho_k = \Tr_{-k }{ \rho} = \prod_k \sum_{n=0}^{\infty}\frac{\tanh^{2n}(r_k)}{\cosh^2(r_k)}|n_k,n_{-k}\rangle \langle n_k,n_{-k}|\,.
\end{equation}
The von Neumann entropy of $\rho_k$ becomes,
\begin{align}
\label{entropyeq}
    S(\rho_k) &= -\mathrm{Tr}(\rho_k \ln \rho_k) \notag \\
    &=  -\sum_{n=0}^\infty \frac{\tanh^{2n} r_k}{\cosh^2 r_k} \ln \left( \frac{\tanh^{2n} r_k}{\cosh^2 r_k} \right) \notag  \\
    &= -2 \ln (\sinh r_k) \sum_{n=0}^\infty n \frac{\tanh^{2n} r_k}{\cosh^2 r_k} + 2 \ln (\cosh r_k) \sum_{n=0}^\infty (n + 1) \frac{\tanh^{2n} r_k}{\cosh^2 r_k} \notag \\
    &= \left( 1 + \sinh^2 r_k \right) \ln \left( 1 + \sinh^2 r_k \right) - \sinh^2 r_k \ln \left( \sinh^2 r_k \right)\,,
\end{align}
where we used the identity
\begin{align}
    \ln \left( \frac{\tanh^{2n} r_k}{\cosh^2 r_k} \right) &= 2n \ln (\sinh r_k) - (2n + 2) \ln (\cosh r_k)\,.
\end{align}
\begin{figure}[h!]
    \centering
    \includegraphics[width=1\linewidth]{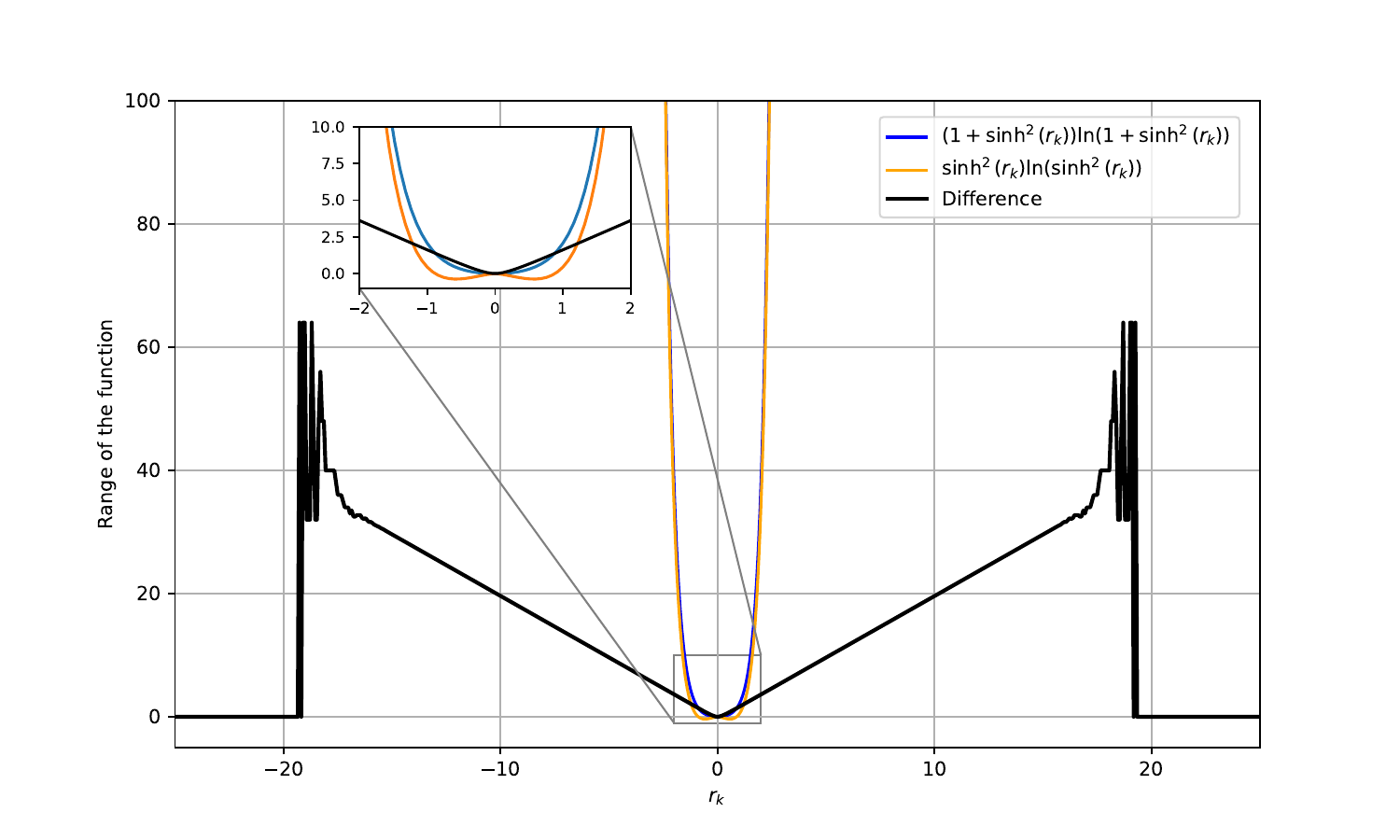}
    \caption{ Comparison of two terms of the entanglement entropy equation. The two terms grow exponentially while their difference grows linearly before numerical instability.}
    \label{terms}
\end{figure}

Figure \ref{terms} displays the plots of the two constituent terms of the entanglement entropy expression in Eq. \ref{entropyeq}. The blue curve corresponds to the first term, $\left( 1 + \sinh^2(r_k) \right) \ln \left( 1 + \sinh^2(r_k) \right)$  while the orange curve represents the second term $\sinh^2(r_k) \ln \left( \sinh^2(r_k) \right).$ The difference between these two terms yields the entanglement entropy for a single mode $k$, shown by the black curve.
Although the entropy is theoretically well-defined for all values of $r_k$, numerical evaluation becomes increasingly unstable at large $r_k$ due to the subtraction of nearly equal, exponentially large numbers. For example, at $r_k = 10$, the first term evaluates to approximately $2.2577 \times 10^9$, while the second term is approximately $2.2577 \times 10^9 - 19.6$. Their difference, the entropy, is thus $\sim 19.6$, a relatively small number obtained from subtracting two values of similar magnitude.

As $r_k$ increases, both terms grow rapidly and approach each other exponentially. However, their difference, being small compared to the absolute scale of the terms, becomes sensitive to machine precision. Once the difference falls below the numerical resolution (typically around $10^{-15}$ for double-precision floating point), the result is truncated to zero or becomes dominated by rounding errors. Consequently, the computed entropy appears to vanish or fluctuate erratically for large $r_k$, which is an artifact of numerical instability, not a physical feature of the entropy.

\begin{figure}[h!]
    \centering
    \includegraphics[width=1\linewidth]{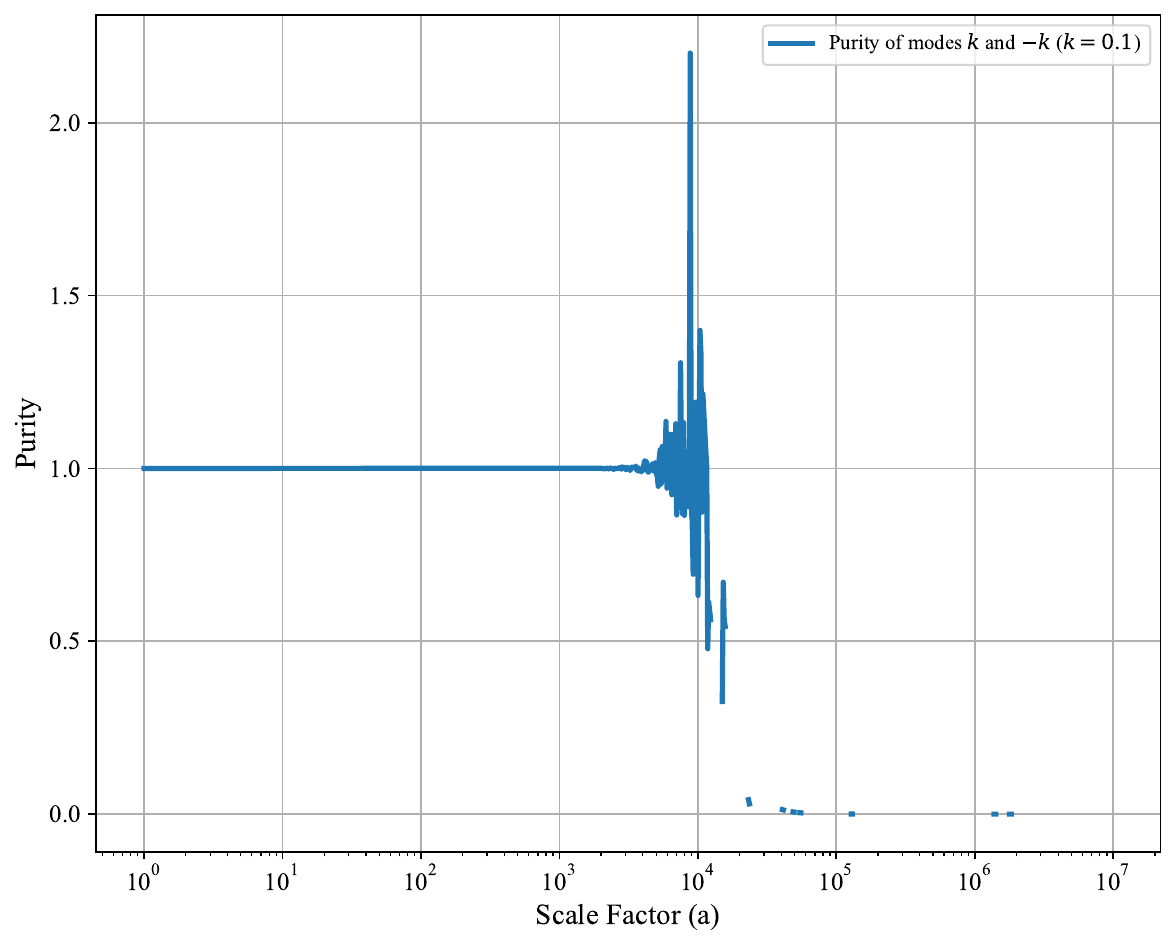}
    \caption{Purity of the two-mode squeezed vacuum state for modes $k$ and $-k$ with $k = 0.1$, plotted as a function of the scale factor $a$. The state remains pure (purity = 1) until around $a \sim 10^4 $, beyond which deviations and fluctuations appear due to numerical instability.}
    \label{fig:purity_check}
\end{figure}
When two vacuum modes are transformed into a two-mode squeezed vacuum using a two-mode squeezer, the resulting state becomes entangled. Each subsystem shows increasing von Neumann entropy due to this entanglement, while the full system remains pure, with zero entropy and a purity of one. In numerical simulations, the purity is calculated for each value of $k$, where the scale factor $(a)$ depends on $k$. For a large value of the scale factor $(a)$, the purity begins to fluctuate and deviates from the ideal value of 1, which indicates that there are errors. For example, at a scale factor of $10^4$, the purity begins to deviate from unity for $k = 0.1$, as shown in Figure \ref{fig:purity_check}. These deviations are not physical, but rather signs of numerical instability in the simulation.

\begin{figure}[h!]
    \centering
    \includegraphics[width=1\linewidth]{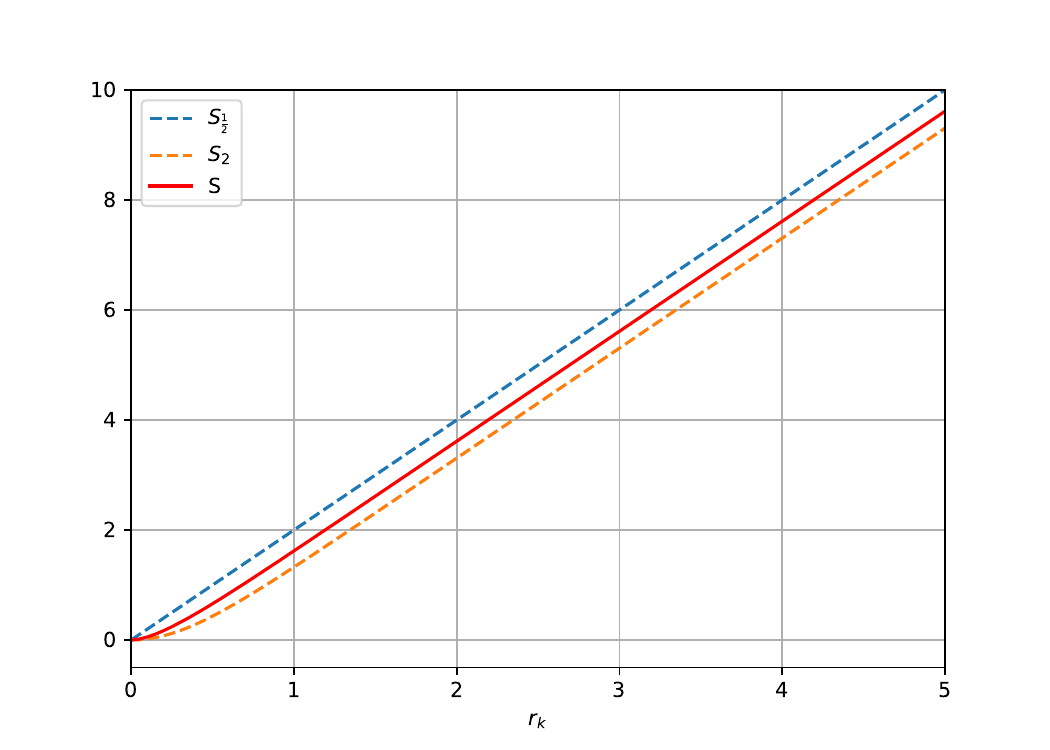}
    \caption{Comparison between von Neumann Entropy and Rényi Entropy. For lower values of $r_k$, von Neumann Entropy stays within the bounds of the first and the second Rényi entropies.}
    \label{simryn}
\end{figure}

In order to tackle these numerical instabilities, we will rely on R\'enyi Entropy, which, for suitable R\'enyi parameters, can be more stable than von Neumann entropy. Since $S_1$ corresponds to the usual von Neumann entropy, finding stable R\'enyi parameters can be used to bound von Neumann entropy, as for any two Rényi entropies $S_n$ and $S_m$, $S_n \geq S_m$ for $n < m$. The R\'enyi entropy associated with cosmological perturbations is given by~\cite{Seroje:2015tsa}, 
\begin{equation}
\label{R\'enyi_entropy}
    S_\mu (r_k) = \frac{1}{1-\mu} \ln \sum_{n=1}^d P_n = \frac{2\mu \ln \cosh r_k + \ln (1 - \tanh^{2\mu}r_k)}{\mu - 1}\,,
\end{equation}
where $\mu \geq 0$ is the R\'enyi parameter and $P_n$ are the eigen values of reduced density matrix of squuezed vacumm state. The second R\'enyi entropy $S_2(r_k)$ is then given by:
\begin{equation}
\label{S_2}
    S_2(r_k) = \ln \cosh 2r_k\,,
\end{equation} 
which provides a lower bound on the von Neumann entropy. For $\mu=\frac{1}{2}$, the R\'enyi entropy simplifies to: 
\begin{equation}
\label{S_1/2}
S_{1/2}(r_k) = \ln(1 - \tanh^2 r_k) - 2 \ln(1 - \tanh r_k) = 2r_k\,,
 \end{equation}
which provides an upper bound on the von Neumann entropy as shown in figure \ref{simryn}. Since $S_2(r_K)$ and $S_{1/2}(r_k)$ are numerically stable for large $r_k$, we can confidently use them to provide a lower and upper bound for von Neumann entropy for large squeezing parameters.

\subsection{Logarithmic negativity}
In order to compute the logarithmic negativity, we perform the partial transpose of the density operator \ref{density} with respect to mode $-k$,
\begin{equation}
\rho^{T_{-k}} = \frac{1}{\cosh^2 r_k} \sum_{n,m=0}^\infty e^{i\phi_k(m-n)} (\tanh r_k)^{n+m} |n_k, m_{-k}\rangle \langle m_k, n_{-k}|.
\label{eq:rhoT}
\end{equation}
Diagonal elements yield eigenvalues,
\begin{equation}
\lambda_n = \frac{(\tanh r_k)^{2n}}{\cosh^2 r_k},
\end{equation}
which are all positive. When $n \neq m$, each pair $(a, b)$ with $a \neq b$ contributes a $2 \times 2$ block:
\begin{equation}
\frac{(\tanh r_k)^{a+b}}{\cosh^2 r_k}
\begin{pmatrix}
0 & e^{i\phi_k(b-a)} \\
e^{i\phi_k(a-b)} & 0
\end{pmatrix},
\end{equation}
with eigenvalues
\[
\lambda_\pm = \pm \frac{(\tanh r_k)^{a+b}}{\cosh^2 r_k}.
\]
Trace norm of a partially transposed density matrix is given by the sum of its absolute eigenvalues:

\begin{align}
    \|\rho^{T_{-k}}\|_1 &= \sum_n |\lambda_n| + \sum_{a < b} 2|\lambda_\pm|  \notag \\
    &= \frac{1}{\cosh^2 r_k} \left[ \sum_{n=0}^{\infty} (\tanh r_k)^{2n} + 2 \sum_{a < b} (\tanh r_k)^{a+b} \right] \notag \\
    & =  \frac{1}{\cosh^2 r_k} \left[ \cosh^2 r_k + \cosh^2 r_k (e^{2r_k} - 1) \right] \notag \\
    & = e^{2r_k} 
\end{align}
where we used:
\begin{equation}
\sum_{n=0}^\infty (\tanh^2 r_k)^n = \frac{1}{1 - \tanh^2 r_k} = \cosh^2 r_k \text{   ,   }   2 \sum_{a < b} (\tanh r_k)^{a+b} = \cosh^2 r_k (e^{2r_k} - 1).
\end{equation}
Finally, the logarithmic negativity, which quantifies entanglement between the two modes, is
\begin{equation}
LN = \log_2 \|\rho^{T_{-k}}\|_1 = \log_2 (e^{2r_k}) = \frac{2r_k}{\ln 2}.
\end{equation}

\subsection{Symplectic formalism}
\begin{figure}[h!]
    \centering
    \includegraphics[width=0.8\linewidth]{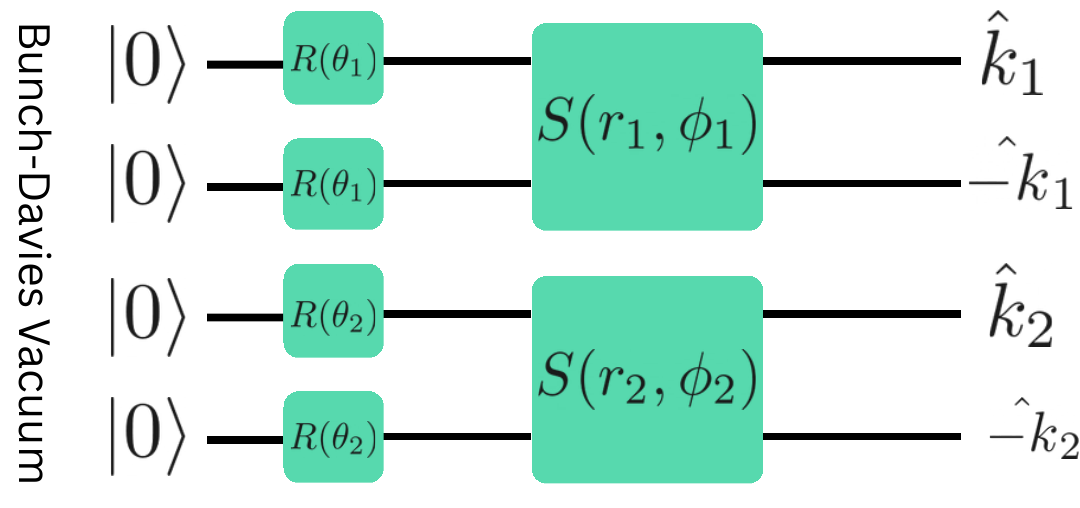}
    \caption{Circuit diagram illustrating the generation of entangled and squeezed modes. Two input Bunch-Davies vacuum modes, $\ket{0}$ and $\ket{0}$, interact individually through a rotation operator before entering a two-mode squeezer. This process results in entangled and squeezed outgoing modes, $\hat{k}_1$ and $-\hat{k}_1$ (upper circuit), and similarly $\hat{k}_2$ and $-\hat{k}_2$ (lower circuit). The rotation operators are included for completeness but can be neglected as they have no effect in this configuration.}
    \label{fig:Symplectic_circuit}
\end{figure}

In this section, we reformulate the calculations for entropies using the covariance formalism of Gaussian states and provide a useful diagrammatic representation referred to as the symplectic circuit. This approach is simple  and allows calculation of the relevant physical quantities like entanglement measures in an efficient way. The dynamics of the cosmological perturbations model is described by the Hamiltonian given in Eq. \ref{eq:Hamiltonian}. The second term of this Hamiltonian is a two-mode squeezer as described in Eq. \ref{eq:squeezing_operator}. We simulate the dynamics of the squeezer in Eq. \ref{eq:squeezing_operator} using the two-mode squeezer in quantum optics. The physics of cosmological perturbations corresponding to the Hamiltonian in Eq. \ref{eq:Hamiltonian} can be described by the symplectic circuit in Figure \ref{fig:Symplectic_circuit}. In the circuit, two input vacuum modes interact individually through a rotation operator and then pass through a two-mode squeezing gate. This interaction leads to two outgoing modes, $\hat{k}_1$ and $-\hat{k}_1$, which are entangled. The rotation operator has no effect in this configuration and can therefore be neglected. Similarly, consider the case where the two-mode squeezer leads to two outgoing modes, $\hat{k}_2$ and $-\hat{k}_2$, which are also entangled. This circuit also allows the study of interaction between $\hat{k_1}$ and $\hat{k_2}$ modes through a CV gate, such as a beam splitter. However, in this paper, we focus on one pair of modes and study their dynamics. We also analyze the effect of thermal noise on the entanglement measures. The use of a symplectic circuit is a powerful tool to simulate the dynamics of not only quadratic Hamiltonians but also Hamiltonians of higher degree. The use of CV gates and the circuit representation can make it possible to study not only Gaussian dynamics but also the non-Gaussian interactions. Here we show, in a simple manner, the results obtained in section $4.1$. 
We consider two modes $k$ and $-k$ described by the annihilation and creation operator $\{\hat{c}_i, \hat{c}^\dagger_{i}\}$ for $i = k, -k$, respectively. Then the two-mode squeezing operator is defined as,
\begin{equation}
   \hat S_2(\xi) = \exp\left(\xi \hat{c}^{\dagger}_k
\hat{c}^{\dagger}_{-k} - \xi^* \hat{c}_k
\hat{c}_{-k}\right) \,,
\end{equation}
where $\xi = r_k e^{i2\phi_k}$, $r_k$ is the squeezing intensity, and $2\phi_k$ is the squeezing phase. The symplectic matrix form of $S_2$ is given as,
\begin{equation}
    S_2(r_k,\phi_k) = 
\begin{pmatrix}
\cosh r_k & 0 & \sinh r_k \cos2\phi_k & \sinh r_k \sin2\phi_k \\
0 & \cosh r_k & \sinh r_k \sin2\phi_k & -\sinh r_k \cos2\phi_k \\
\sinh r_k \cos2\phi_k & \sinh r_k \sin2\phi_k & \cosh r_k & 0 \\
\sinh r_k \sin2\phi_k & -\sinh r_k \cos2\phi_k & 0 & \cosh r_k
\end{pmatrix}.
\end{equation}
Now we consider two cases for the initial state: the vacuum state and the thermal state. For each case, we use the symplectic matrix $S_2$ to calculate the von Neumann entropy and logarithmic negativity.

 The mean and covariance matrix for the input state, as the vacuum stat,e are,
\begin{equation}
\mu_{in} = 0, \quad V_{in} =
\begin{pmatrix}
1 & 0 & 0 & 0\\
0 & 1 & 0 & 0\\
0 & 0 & 1 & 0\\
0 & 0 & 0 & 1
\end{pmatrix}\,.
\end{equation}
The two-mode squeezer squeezes these vacuum modes and returns two output modes. The output mean and covariance matrix are given by,
\begin{align}
&\mu_{out} = S_2 \cdot \mu^{(in)} = 0, \\
&V_{out} = S_2 \cdot V_{in} \cdot S_2^T =
\begin{pmatrix}
\label{V_out_vacuum}
\cosh2r_k & 0 & \sinh2r_k \cos2\phi_k & \sinh2r_k \sin 2\phi_k\\
0 & \cosh2r_k & \sinh2r_k \sin 2\phi_k & -\sinh2r_k \cos2\phi_k\\
\sinh2r_k \cos2\phi_k & \sinh2r_k \sin 2\phi_k & \cosh2r_k & 0\\
\sinh2r_k \sin 2\phi_k & -\sinh2r_k \cos2\phi_k & 0 & \cosh2r_k
\end{pmatrix}
\end{align}
Now we extract the covariance matrix for each mode from Eq. \ref{V_out_vacuum} as a block matrix from the diagonal, and we get,
\begin{equation}
\label{V_k_V_-k}
    V_k = V_{-k} = 
\begin{pmatrix}
\cosh2r_k & 0\\
0 & \cosh2r_k
\end{pmatrix}\,.
\end{equation}
Now we can use the covariance matrices from Eqs. \ref{V_out_vacuum} and \ref{V_k_V_-k} to compute the entanglement measures. Since the total output state $V_{out}$ is a pure state and the subsystems are mixed, the von Neumann entropy of each output mode gives the entanglement entropy for the corresponding mode. For each output mode, the eigenvalues are $\{\cosh2r, \cosh2r\}$, and the entanglement entropy given by Eq. \ref{eq:Entropy_cov} is,
\begin{equation}
    S_k = S_{-k} = \left( 1 + \sinh^2r_k \right) \log_2 \left( 1 + \sinh^2r_k \right) - \sinh^2r_k \log_2 \left( \sinh^2r_k \right)\,.
\end{equation}
For the calculation of logarithmic negativity for the two-mode Gaussian state $V_{out}$, we use Eq. \ref{eq:LN_two_mode}. Now we calculate $v_{min}^{pt}$, the minimum symplectic eigenvalue of the covariance matrix $V^{pt}$ corresponding to the partially transposed state $\rho^{pt}$. The matrix $V^{pt}$ is obtained from $V_{out}$ by reversing the sign of all matrix elements involving momenta $p_i$ corresponding to mode $-k$. Thus,
\begin{equation}
    V^{pt} =
\begin{pmatrix}
\cosh2r_k & 0 & \sinh2r_k \cos2\phi_k & - \sinh2r_k \sin 2\phi_k\\
0 & \cosh2r_k & \sinh2r_k \sin 2\phi_k & \sinh2r_k \cos2\phi_k\\
\sinh2r_k \cos2\phi_k & \sinh2r_k \sin 2\phi_k & \cosh2r_k & 0\\
- \sinh2r_k \sin 2\phi_k & \sinh2r_k \cos2\phi_k & 0 & \cosh2r_k
\end{pmatrix}\,.
\end{equation}
The symplectic spectrum of $V^{pt}$ are the absolute eigenvalues of $V^{pt}\Omega$, where $\Omega = \bigoplus_N \begin{pmatrix}
0 & 1\\
-1 & 0
\end{pmatrix}$ is the symplectic form. 
Solving for eigenvalues, we find $v^{pt} = \{e^{2r_k}, e^{2r_k}, e^{-2r_k}, e^{-2r_k}\}$. Here we see that symplectic eigenvalues show two-fold degeneracy and the minimum eigenvalue is $v_{min}^{pt} = e^{-2r_k}$. Finally, we get the logarithmic negativity for the system as,
\begin{equation}
    LN = \frac{2r_k}{\ln 2}\,.
\end{equation}
\begin{figure}[h!]
    \centering
    \includegraphics[width=1\linewidth]{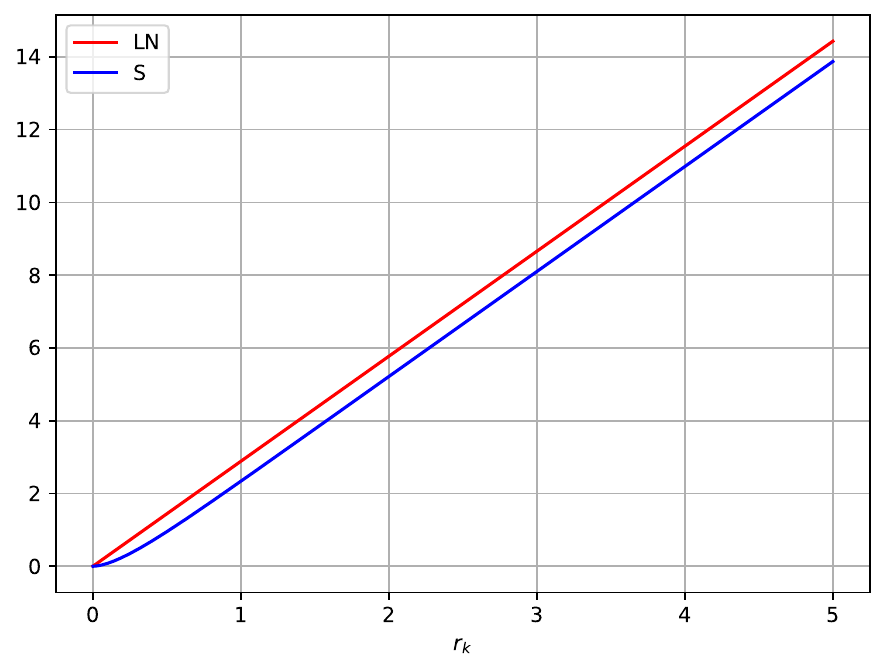}
    \caption{Logarithmic Negativity (LN) and von Neumann entropy (S) for vacuum state plotted against squeezing parameter ($r_k$).} 
    \label{fig:LN_S_vacuum}
\end{figure}
In Figure \ref{fig:LN_S_vacuum}, we cannot compare LN and S quantitatively as they have different definitions. But we can compare them qualitatively, and we find that both quantities increase as the squeezing intensity ($r_k$) is increased, which indicates an increase in entanglement. Since the total quantum state is pure, the von Neumann entropy gives the entanglement entropy of the corresponding mode.

The input vacuum state with the addition of thermal noise is referred to as the thermal state and is characterized by,
\begin{equation}
    \mu_{in} = 0 \quad \text{and} \quad V_{in} = (1 + 2 \bar n)\mathbf{I}_4\,,
\end{equation}
where $\mathbf{I}_4$ is $4 \times 4$ identity matrix and  $\bar n$ is the mean number of noise quanta. This system is similar to the two-mode squeezed vacuum state and only differs by the factor $(1 + 2 \bar n)$. Therefore, the logarithmic negativity is given by,
\begin{equation}
\label{eq:LN_thermal}
    LN_{thermal} = \max\{0, -\log_2[(1 + 2 \bar n)e^{-2r_k}]\}\,.
\end{equation}
The condition for the system to be entangled is $(1 + 2 \bar n)e^{-2r_k} < 1$ and the number of mean quanta beyond which the state is not entangled is,
\begin{equation}
   \bar n_{critical} = \frac{1}{2}(e^{2r_k} - 1)\,.
\end{equation}
Similarly, the von Neumann entropy is,
\begin{equation}
\label{eq:S_thermal}
    S_{thermal} = \frac{v + 1}{2} \log_2 \left( \frac{v + 1}{2} \right) - \frac{v - 1}{2} \log_2 \left( \frac{v - 1}{2} \right)\,,
\end{equation}
where $v = (1 + 2 \bar n) \cosh 2r_k$.
Note that, since in this case the total quantum state is mixed, the von Neumann entropy does not qualify as a reliable entanglement measure but only quantifies the entropy of the corresponding mode.
\begin{figure}[h!]
    \centering
    \includegraphics[width=1\linewidth]{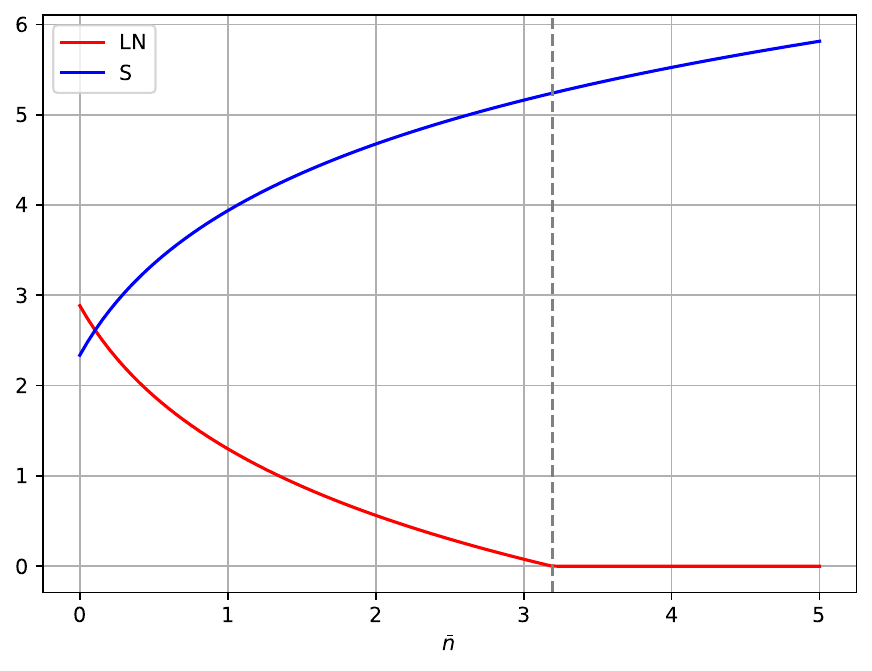}
    \caption{Logarithmic Negativity (LN) and von Neumann entropy ($S$) versus mean thermal noise quanta ($\bar{n}$) for a thermal state. $S$ increases while LN decreases with $\bar{n}$, vanishing beyond a specific value represented by the vertical dashed line.}
    \label{fig:LN_S_thermal}
\end{figure}
From Figure \ref{fig:LN_S_thermal}, we find that von Neumann entropy increases and logarithmic negativity decreases as we increase the mean number of noise quanta. This indicates that in this case, von Neumann entropy is not a reliable entanglement measure but only gives the entropy of the particular mode and not the entanglement entropy. This is because the total quantum state is mixed due to the presence of thermal noise. The logarithmic negativity decreases, indicating a decrease in entanglement of the system as the noise increases. Beyond a certain critical value of $\bar n$ (shown by the gray dotted line), the entanglement of the system is completely lost due to excessive noise, and the logarithmic negativity drops to zero. 

The above analysis shows the application of the Gaussian formalism and symplectic circuit to describe the dynamics of cosmological perturbations corresponding to a quadratic Hamiltonian. This approach is also a powerful tool to include non-linearities and apply non-Gaussian states to study such interactions. This can be done by the use of Gaussian and non-Gaussian CV gates. The circuit provides a comprehensive visual representation of the dynamics of phenomena like cosmological perturbations, which can then be simulated using a Python library like QuGIT~\cite{Brandao:2022voc}. This allows us to study the dynamics in an efficient way, which would be very complex to perform analytically. The symplectic circuit helps us visualize the symplectic matrices applied to construct the working of the Hamiltonian. Hence, the use of techniques from continuous variable quantum optics provides a powerful framework to study analog gravity models.

\section{Simulation for expanding and contracting background }
\label{section 5}
In this section, we explore the simulation of von Neumann entropy between the mode pairs $k$ and $-k$ in both expanding and contracting backgrounds, with fixed equation of state $w$. We also analyze the logarithmic negativity of the cosmological squeezed modes in the quantum optics formalism and explore the behavior of these entropy measures in the presence of thermal noise. 

\subsection{Entanglement entropy}
In this subsection, we investigate the behavior of the entanglement entropy of cosmological perturbations that evolve in both expanding and contracting FLRW backgrounds. We compute three key entropy measures: von Neumann entropy $S_1$, and Rényi entropies $S_{1/2}$, and $S_2$. The analytical expressions for Rényi entropies serve as theoretical bounds.

We analyze how these entropies evolve with the scale factor $a$, under different equations of state: an accelerating background with $w = -1$, and a decelerating one with $w = 1/3$. In each case, we examine whether the simulated von Neumann entropy lies within the expected analytical bounds. And we also interpret deviations as signatures of numerical instabilities. Finally, we present the total entanglement entropy accumulated over a wide range of wavenumbers, providing knowledge about the whole entanglement dynamics of the quantum field.

\subsubsection*{Expanding background}
We consider an expanding cosmological background with a fixed equation of state of $w=-1$ for the accelerating case and $w=1/3$ for the decelerating case. For an expanding, accelerating background, modes are inside the horizon at sufficiently early times $a \ll 1$, and they cross the horizon at $k \approx a$ as shown in Eq. \ref{k_horizon}. We take two values of k, $k=10^{-4}$ and $k=10^4$, and plot the analytic upper and lower Rényi‑entropy bounds, $S_{1/2}$ and $S_{2}$, respectively, against $\ln a$, while the von Neumann entropy $S_{1}$ is obtained via a quantum optics simulation using QuGIT~\cite{Brandao:2022voc}.
\begin{figure}[h!]
    \centering
    \includegraphics[width=1\linewidth]{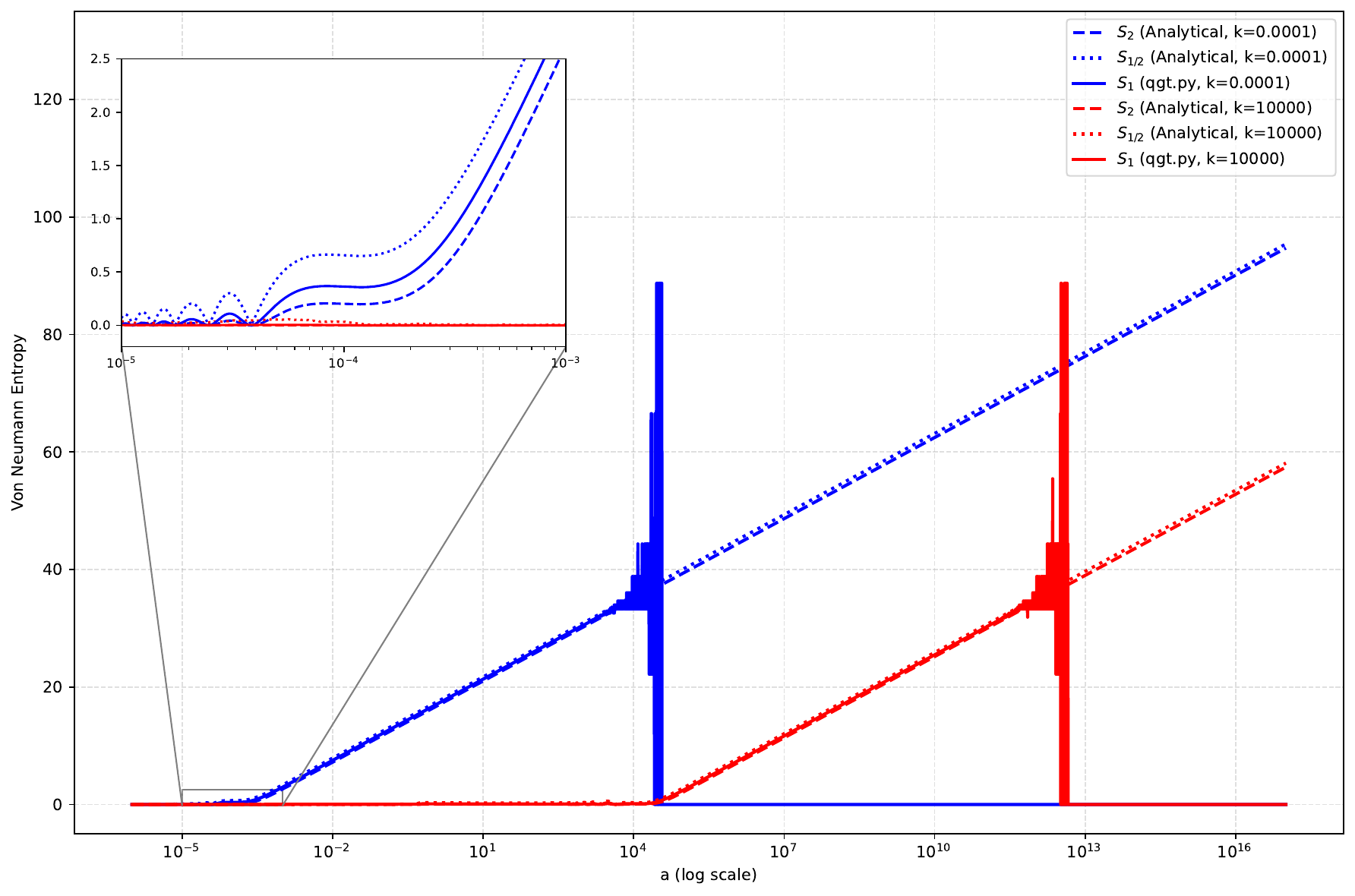}
    \caption{
    von Neumann entropy ($S$) for two different modes in an expanding accelerating background ($w = -1$). Rényi entropies ($S_{\frac{1}{2}}$ and $S_2$) for each mode are also plotted to validate the behaviour of $S$. The von Neumann entropy remains bounded between the two Rényi entropies until numerical instabilities appear.}
    \label{fig:entropy_exp_acc}
\end{figure}
After each mode cross the horizon, the analytic  Rényi entropies rise linearly with $\ln(a)$, and the simulated $S_1$ lies between $S_{1/2}$ and $S_{2}$ at early as well as late times, implying the circuit simulates cosmological squeezing Hamiltonian effectively as shown in Figure \ref{fig:entropy_exp_acc}. Zooming in near the early regime of a, simulated entropy and the analytic ones all hover near zero, with small oscillations, showing that before they exit the horizon, the state remains nearly pure and unsqueezed. However, for both values of $k$, we observe unusual spikes in $S_1$ around $ln(a)\approx10^{4}$ and $\ln(a)\approx10^{12}$ respectively, followed by abrupt drops back to zero. These transient violations of the Rényi bounds originate from the numerical instabilities discussed in the section \ref{sec error}. The important thing to see is that before this breakdown, the von Neumann entropy $S_1$ correctly stays between the two Rényi entropies, $S_2 \le S_1 \le S_{1/2}$.

\begin{figure} [h!]
    \centering
    \includegraphics[width=1\linewidth]{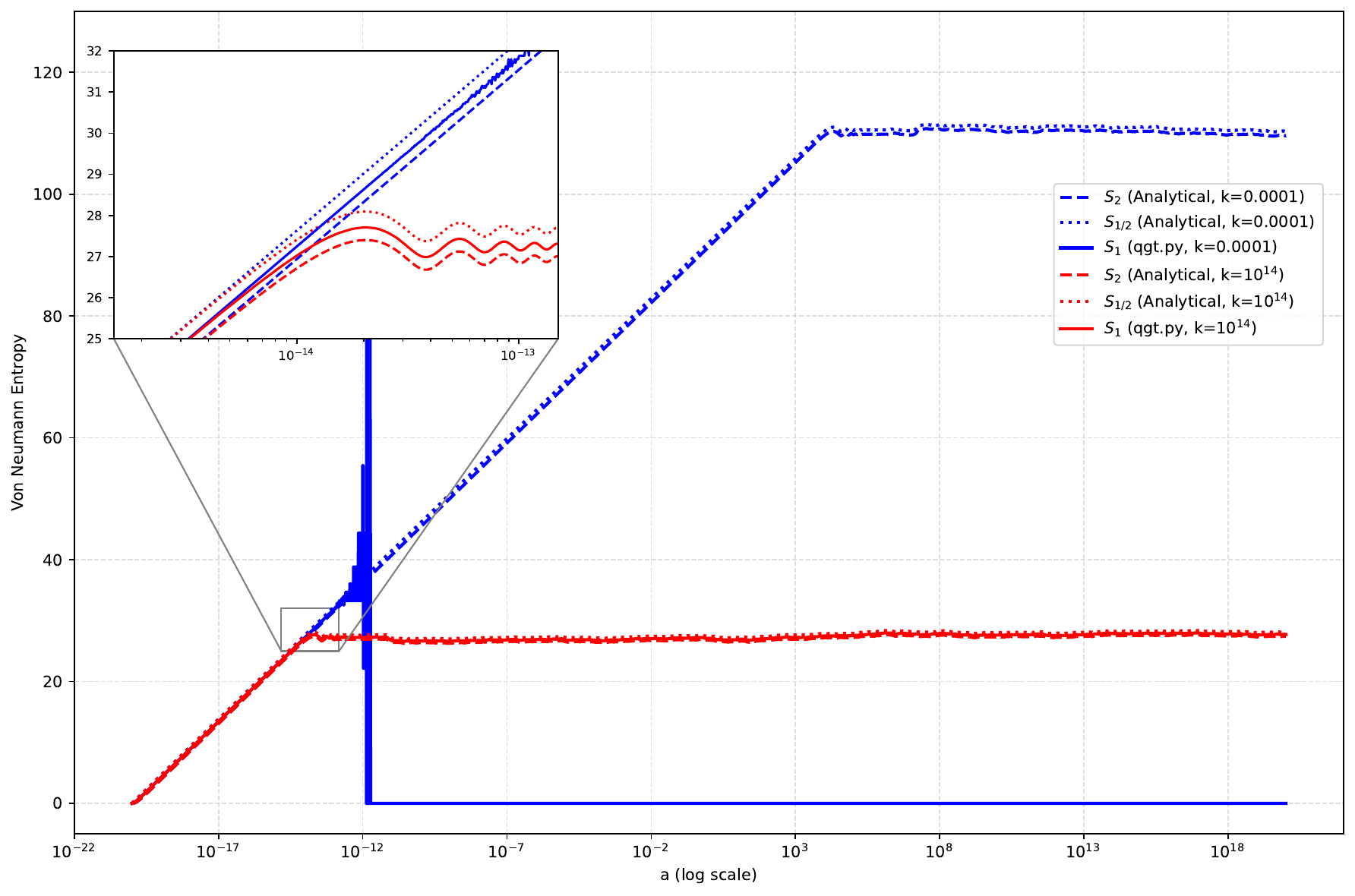}
    \caption{von Neumann entropy ($S$) for two different modes in an expanding decelerating background ($w = 1/3$). Rényi entropies ($S_{\frac{1}{2}}$ and $S_2$) for each mode are also plotted to validate the behaviour of $S$. The entropy values remain within the bounds of the two Rényi entropies until numerical instabilities occur. For $k = 10^{14}$, the entropy never reaches the point of numerical instability.}
    \label{fig:entropy_exp_dec}
\end{figure}

Perturbations for decelerating solutions $w > -1/3$ begin outside the horizon for sufficiently early times $a \ll 1$. Since the entropy is a function of $r_k$, both the analytic  Rényi entropies and the simulated $S_1$ grow when a mode is superhorizon. At late times, after the modes re-enter the horizon, the entropy measures freeze in, and similar to the accelerating case, simulated $S_1$ lies between $S_{1/2}$ and $S_{2}$. However, in this case, the von Neumann entropy freezes at higher values for smaller values of k, that is, after the onset of the entropy instability. For sufficiently large values of $k$ (we have chosen $k=10^{13}$), the von Neumann entropy saturates earlier, before reaching the instability point, as shown in Figure \ref{fig:entropy_exp_dec}. 

\subsubsection*{Contracting background}
 \begin{figure}[h!]
    \centering
    \includegraphics[width=1\linewidth]{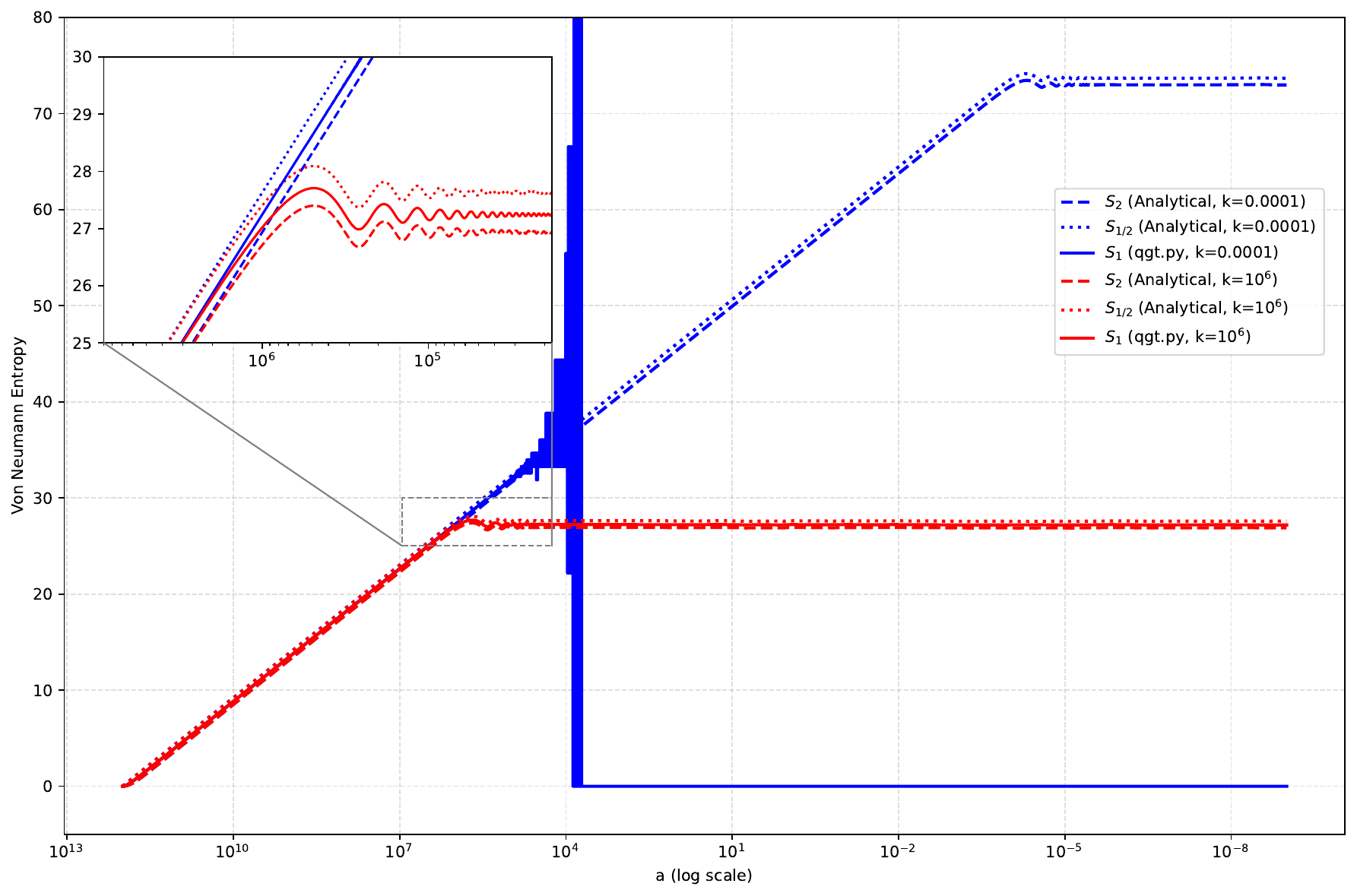}
    \caption{von Neumann entropy ($S$) for two different modes in a contracting accelerating background ($w = -1$). Rényi entropies ($S_{\frac{1}{2}}$ and $S_2$) for each mode are also plotted to validate the behaviour of $S$. The entropy values remain within the bounds of the two Rényi entropies until numerical instabilities occur. For $k = 10^{6}$, the entropy never reaches the point of numerical instability.}
    \label{fig:entropy_con_acc}
\end{figure}
Now we consider a contracting cosmological background. For contracting acceleration, the modes begin outside the horizon for early times $a \gg 1$. Given that a contracting universe is simply the time‐reverse of an expanding one, the entanglement measures in a contracting accelerating background must mirror those in an expanding decelerating background, and hence their plots coincide. In particular, at early time $a \gg 1$,  both the analytic Rényi entropy and the simulated von Neumann entropy are sufficiently large, and they freeze in when the modes re-enter the horizon, as they did for expanding decelerating solutions. To avoid the numerical instability of von Neumann entropy at low $k$, we chose one of our values of $k$, a sufficiently large value, $k = 10^{6}$, that eliminates the instability. With this choice, the simulated von Neumann entropy in Figure \ref{fig:entropy_con_acc} produces a smooth curve that remains bounded by the corresponding Rényi entropies. The sub-graph magnifies the small region where the entropy measures for $k=10^{14}$ freeze in, as they enter the horizon, and have small oscillations before settling into their plateau.

\begin{figure}[h!]
    \centering
    \includegraphics[width=1\linewidth]{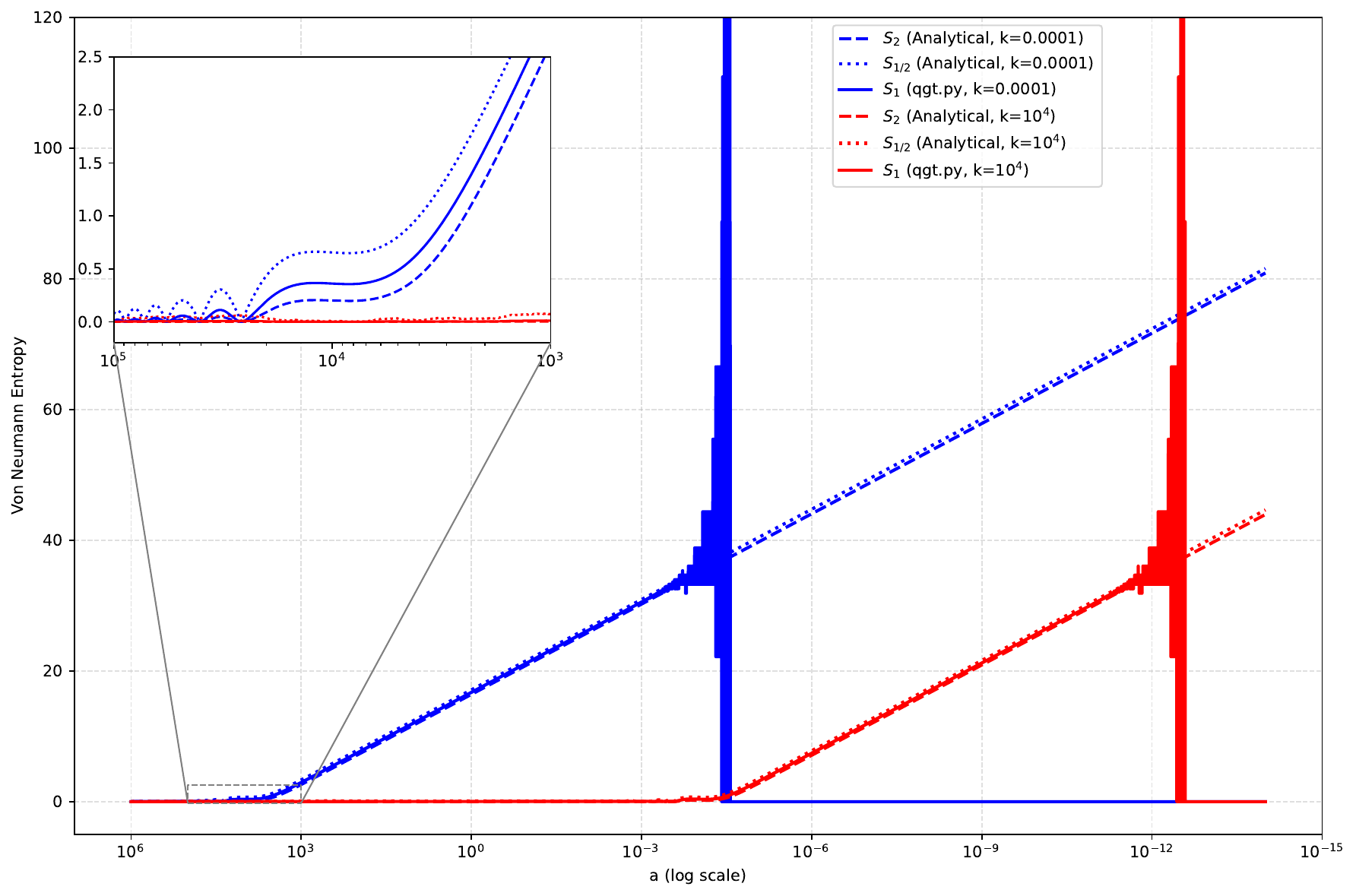}
    \caption{von Neumann entropy ($S$) for two different modes in a contracting decelerating background ($w = 1/3$). Rényi entropies ($S_{\frac{1}{2}}$ and $S_2$) for each mode are also plotted to validate the behaviour of $S$. The von Neumann entropy remains bounded between the two Rényi entropies until numerical instabilities appear.}
    \label{fig:entropy_con_dec}
\end{figure}

For the contracting decelerating case, modes are inside the horizon at early times $a \gg 1$, and exit the horizon as the universe evolves. Due to time reversal symmetry, it must resemble the squeezing and thus the entropy parameters of expanding accelerating, as this can be seen comparing Figures \ref{fig:entropy_con_dec} and \ref{fig:entropy_exp_acc}. In the early time $a \gg 1$ regime, both the simulated and analytical entropies stay close to zero with slight oscillation, indicating that the state remains almost pure and unsqueezed, before they exit the horizon. In the late-time regime, $a \ll 1$, the analytical Rényi entropies exhibit a clear linear growth with $\ln(a)$. Throughout both early and late times, simulated von Neumann entropy $S_1$ remains confined between $S_{1/2}$ and $S_2$, reflecting the expected behavior, imposed by the squeezing Hamiltonian realized in the circuit.

\begin{figure}[h!]
    \centering
    \includegraphics[width=1\linewidth]{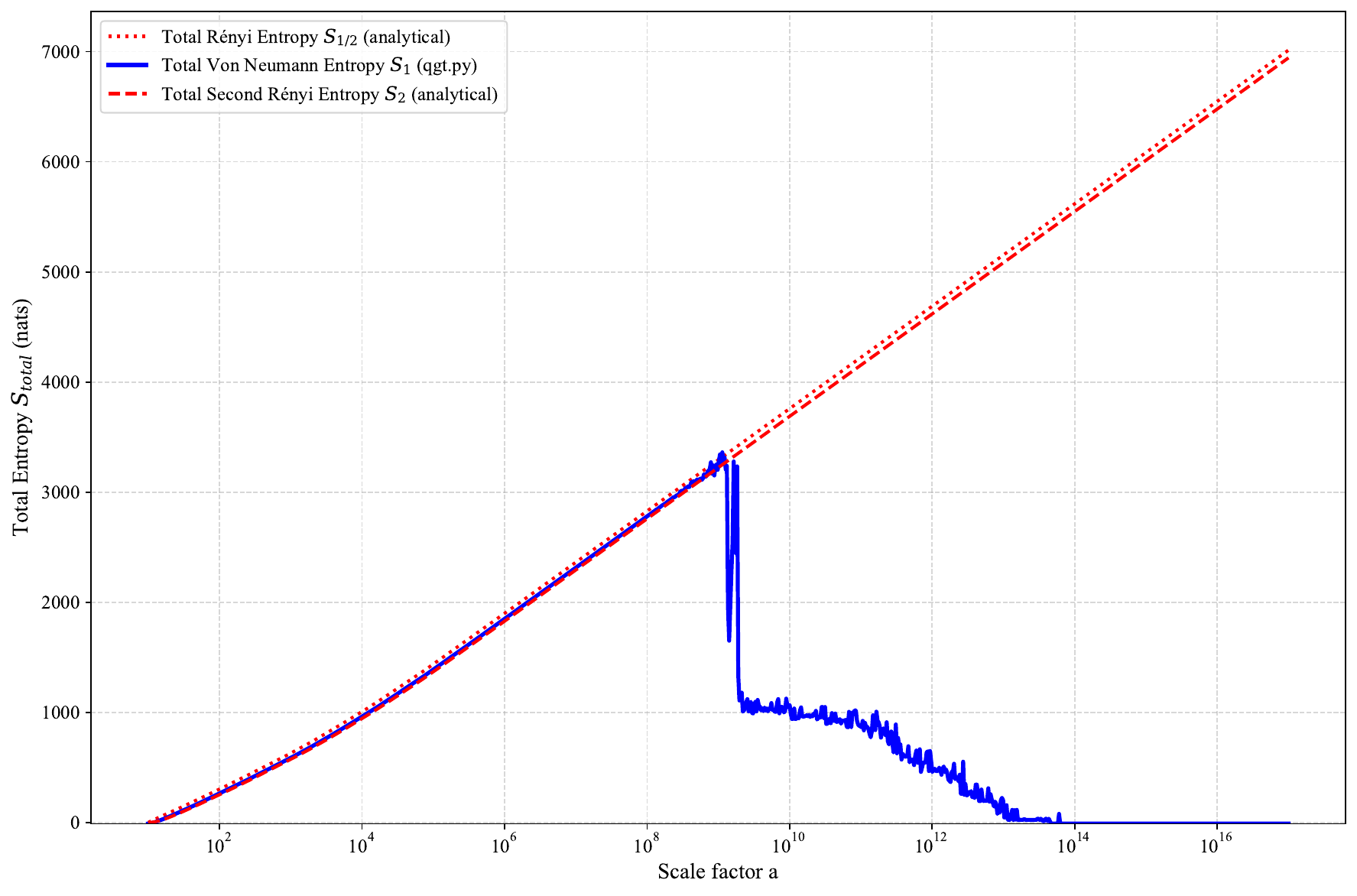}
    \caption{Total von Neumann Entropy for a hundred different modes between $k=10^{-5}$ to $k=10^{5}$ for expanding accelerating background ($w=-1$). Total entropy remains bounded between the two Rényi entropies until numerical instabilities appear.}
    \label{fig:total_entanglement}
\end{figure}

In Figure \ref{fig:total_entanglement}, we plot the total entropy against the scale factor $a$. To get this total entropy, we calculated the entropy for 100 different wavenumbers, from $k=10^{-5}$ to $k=10^{5}$, and then added them all up at each point in time. The plot shows three curves: the total von Neumann entropy ($S_1$, solid blue line), and the total Rényi 1/2 and Second entropies ($S_{1/2}$ and $S_2$, red lines). We can see that the two red lines for the Rényi entropies are almost identical and show a stable, straight-line growth. The blue line for the von Neumann entropy follows the red lines perfectly at early times. However, at late times (around $a = 10^9$), the numerical calculation for the von Neumann entropy becomes unstable. This causes the entropy to show a large, noisy spike and then drop to zero. 
\subsection{Logarithmic negativity}
 The degree of entanglement is quantified by logarithmic negativity, a computable measure of bipartite entanglement between quantum modes. Here, we simulate the behavior of Logarithmic negativity across four distinct cosmological scenarios, expanding accelerating ($ w = -1 $), expanding decelerating ($ w = 1/3 $), contracting accelerating ($ w = -1 $), and contracting decelerating ($ w = 1/3 $) which is presented in the Figure \ref{logn}.

\subsubsection*{Expanding background}

\begin{figure} [h!]
    \centering
    \includegraphics[width=1\linewidth]{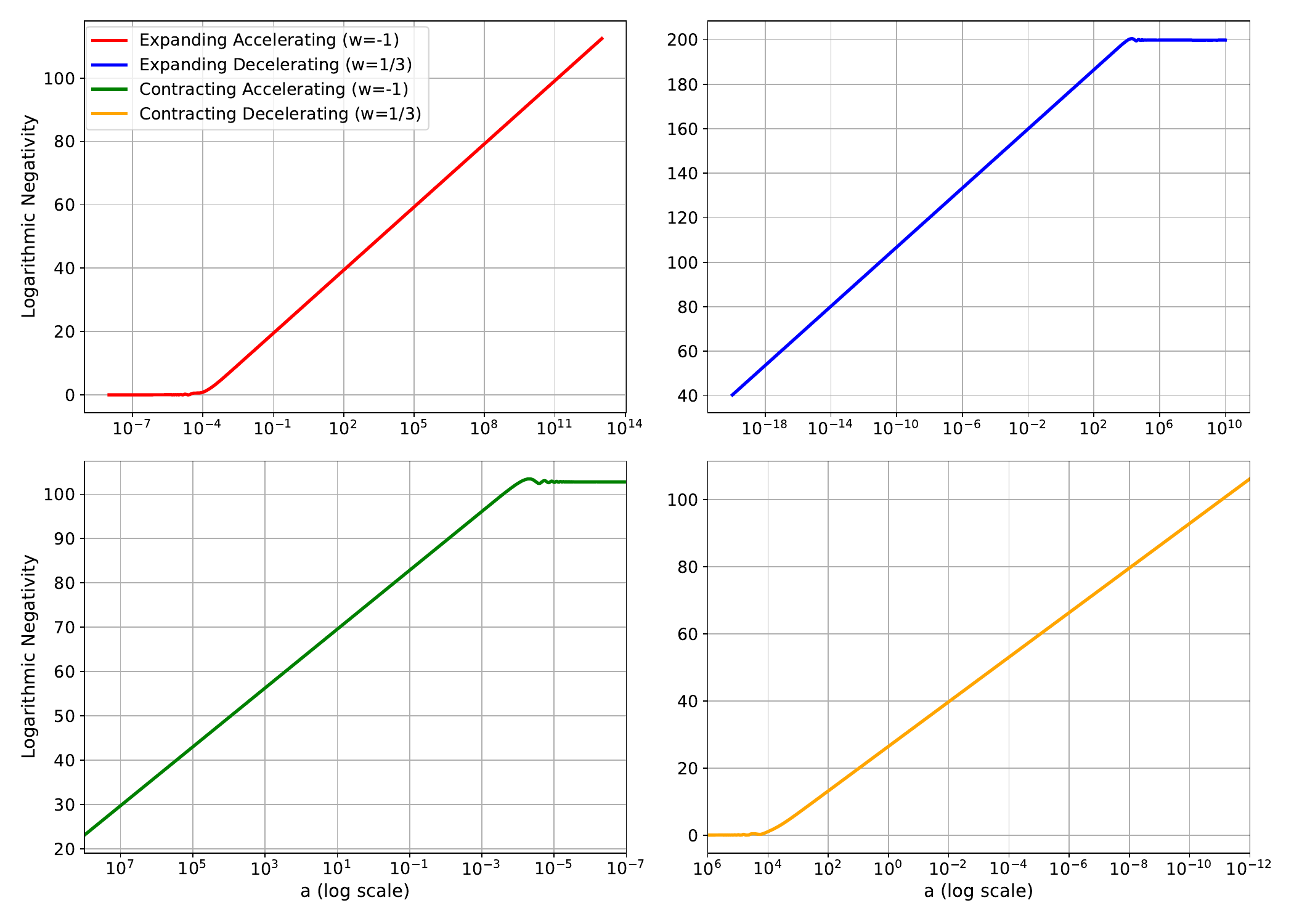}
    \caption{Evolution of Logarithmic Negativity as a function of scale factor (a) for different cosmological backgrounds.}
    \label{logn}
\end{figure}
The top-left plot in Figure \ref{logn} illustrates the evolution of logarithmic negativity in an expanding, accelerating universe. Here, logarithmic negativity begins to grow rapidly at small values of the scale factor $ a $, indicating that entanglement starts developing early during the expansion phase. As $ a $ increases from $ 10^{-7} $ to $ 10^{14} $, the logarithmic negativity increases monotonically without reaching a saturation point. This suggests that the accelerated expansion continuously enhances mode mixing, leading to unbounded growth in entanglement. Notably, the onset of growth occurs almost immediately as the system evolves.

In the top-right panel, we observe the behavior of logarithmic negativity in an expanding decelerating universe. Unlike the accelerating case, the growth of logarithmic negativity is significantly slower and more gradual. The entanglement begins to increase at moderate values of $ a $, and the curve exhibits a smoother slope over time. As the universe decelerates, the LN becomes flat. This is due to the fact that as the  scale factor $a$ grows, the spatial separation between quantum modes increases. As the modes are relatively close together, this gives rise to an increase in entanglement, as evidenced by the growth of logarithmic negativity. However, as the expansion progresses, the distance between modes becomes increasingly large, leading to a reduction of their effective coupling. When the spatial separation reaches a critical threshold, the interactions between the modes may stabilize, resulting in a steady state where further expansion no longer significantly enhances entanglement. Thus, dynamical saturation is observed as a plateau in the logarithmic negativity, indicating that the system has reached a maximal or equilibrium level of entanglement under the given cosmological conditions.

\subsubsection*{Contracting background}

The bottom-left plot  in Figure \ref{logn}  illustrates the entanglement dynamics in a contracting accelerating universe, an intriguing time-reversed analogue of the inflationary expansion. In this case, the scale factor decreases rapidly, indicating accelerated contraction. Initially, logarithmic negativity increases significantly  as the scale factor decreases from $ 10^7 $ to $ 10^{-7} $, indicating a strong squeezing of modes due to the growing curvature and energy density during contraction. However, unlike the expanding case, the LN curve eventually flattens, approaching a plateau. Thus, flattening implies that although the squeezing parameter increases initially, it becomes bounded as contraction continues, and further enhancement of entanglement is suppressed.

Finally, the bottom-right panel presents the result for a contracting, decelerating universe.
As the universe evolves toward smaller scale factors, LN increases steadily without evident saturation within the simulated domain. This implies that the squeezing parameter $r_k$ continues to grow consistently as contraction proceeds, enhancing quantum entanglement between the modes. The decelerating contraction allows sustained squeezing because modes are increasingly compressed within a shrinking causal horizon, leading to a persistent growth in quantum correlations. Unlike in the expanding decelerating scenario, where LN saturates due to limited horizon stretching, the causal compression in this contracting case maintains efficient entanglement generation.

In this scenario, logarithmic negativity increases continuously as the scale factor decreases from $ 10^6 $ to $ 10^{-12} $, with no signs of saturation within the simulated range. The onset of entanglement growth occurs gradually, reflecting the slower pace of contraction compared to the accelerating case.

\subsection{Thermal noise}
In realistic quantum systems, perfect isolation from the environment is rarely achievable. Therefore, it is essential to consider the effects of thermal noise, which introduces classical uncertainty into the pure quantum state. In this work, we extend our analysis of cosmological perturbations by introducing thermal noise at the input stage of our quantum simulation, modeled as a two-mode thermal Gaussian state~\cite{Brady:2022ffk}.

From our analysis of thermal state as input in section $4.2$, Eqs. \ref{eq:S_thermal} and \ref{eq:LN_thermal} give the value of von Neumann entropy and logarithmic negativity, respectively. We set the value of $\bar{n}$ to the critical value beyond which LN is zero for the system. This is an indication of loss of entanglement due to the addition of thermal noise in the system. The value of $\bar{n}_{\text{critical}}$ is given by,
\begin{equation}
\bar{n}_{\text{critical}} = \frac{1}{2}(e^{2r_k} - 1).
\end{equation}
On the other hand, von Neumann entropy increases with the addition of thermal noise, indicating that it does not quantify entanglement if the total quantum state is mixed.

In this section, we analyze the effect of thermal noise on the entanglement measures under different cosmological backgrounds.
\begin{figure} [h]
    \centering
    \includegraphics[width=1\linewidth]{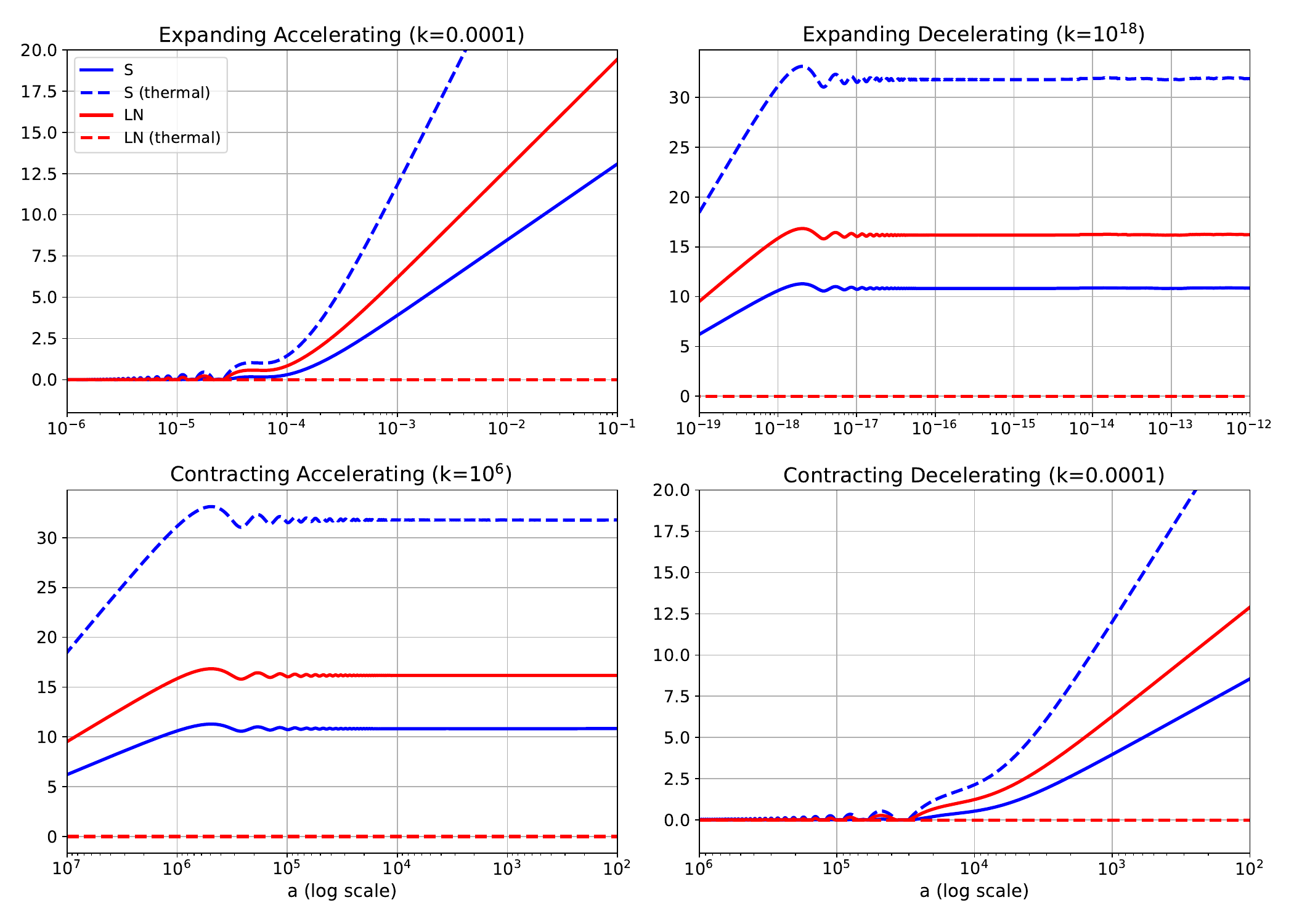}
    \caption{Comparison of von Neumann Entropy and Logarithmic Negativity for pure states and thermal mixed states (after adding thermal noise) for different cosmological backgrounds. Thermal noise, introduced as a two-mode thermal state representing environmental effects, reduces entanglement (LN) while increasing overall entropy, highlighting the impact of decoherence on quantum correlations.}
    \label{noise}
\end{figure}

\subsubsection*{Expanding background}
In the Figure \ref{noise}, the top most (both left and right) sub plots shows the evolution of von Neumann entropy $S$ and logarithmic negativity $LN$ for expanding accelerating $(w=-1, k=10^{-4})$ and expanding decelerating $(w=\frac{1}{3}, k=10^{18})$ case of different modes with and without thermal noise. We plot both the von Neumann entropy and logarithmic negativity against the scale factor. Here, we take different modes to avoid numerical errors.

From the topmost left subplot, it can be interpreted that both entropy $S$ and $S$(thermal) increase as the scale factor grows. And the logarithmic negativity also similarly increases with the scale factor. The thermal noise (dashed lines) strongly suppresses both measures, which indicates a significant decrease in entanglement. It can also be observed from the plots that there is flat logarithmic entropy in the presence of noise, and there is an increment in $S$(thermal), which suggests that the increase in entropy corresponds to a decrease in entanglement. So in both cases, there is degradation of entanglement. Similarly, for expanding decelerating case, we see some gradual increase in $LN$ and $S$ for a period of time, which later plateaus for higher values of scale factor.

\subsubsection*{Contracting background}
We can see from the left bottom subplot that both $S$ and ${LN}$ decrease as $a$ decreases in the case of a contracting background. We find that the thermal noise still suppresses these measures, but the effect is mild. As the modes re-enter the horizon, it leads to a reduction in entanglement. For the contracting accelerating case $(k=10^6)$, both the von Neumann $S$ and logarithmic negativity $LN$ decrease with the scale factor, and there is also the presence of suppression from the thermal noise, but it is relatively low compared to the expanding part. In the contracting decelerating part $(k=10^{-4})$, the von Neumann entropy $S$ and logarithmic negativity $LN$ decrease as the scale factor decreases, similar to the accelerating part. And entanglement continues to decrease as it re-enters the horizon.

\section{Conclusion}
\label{section 6}
In this paper, we make use of concepts from quantum optics to study the entanglement entropy of cosmological perturbations. We focus on simulating various entanglement entropy measures of cosmological perturbations for different backgrounds using the quantum optics tool. The paper makes extensive use of the covariance formalism of Gaussian states using continuous variables, and the diagrammatic approach corresponding to the symplectic circuit. The use of Gaussian states and Gaussian unitaries provides important methods to study the evolution of bosonic systems under a quadratic Hamiltonian. \\
This formalism is useful to study different physical quantities like entanglement entropy, logarithmic negativity, purity, particle number, etc. In particular, we examine how von Neumann entropy and logarithmic negativity evolve in both expanding (accelerating and decelerating) and contracting (accelerating and decelerating) cosmologies. By comparing analytic R\'enyi‑entropy bounds with von Neumann entropy obtained via QuGIT simulations in section \ref{section 5}, we validate that our circuit faithfully implements the cosmological squeezing Hamiltonian across accelerating and decelerating scenarios. Finally, we introduce thermal noise at the input stage to demonstrate that environmental mixedness increases the von Neumann entropy and sets logarithmic negativity to zero.

The techniques used here, like the two-mode squeezer, are ubiquitous in experimental setups. We want to emphasize the simplicity and compactness of the Gaussian formalism in quantifying entanglement between any desired modes of the system. This formalism allows the inclusion and study of multi-mode entanglement generated in cosmological perturbations with simplicity. This paper deals with a two-mode system, and in the future, we aim to use these techniques from quantum optics to study interactions between sub- and super-Hubble modes and include non-linearities, which would otherwise be a complex task. The symplectic circuit provides a powerful visual tool to understand the dynamics underlying the phenomena of cosmological perturbations. Also, the circuit helps to approximate the dynamics of the system, and we must check its accuracy with the actual dynamics of the system. Finally, we have studied how thermal noise contributes to the loss of entanglement of the output state, and such noise must be controlled to study entanglement in any experiment aiming to observe quantum properties of the cosmological perturbations.

\section{Outlook}
\label{sec:outlook}
While $\pm k$ entanglement via two‑mode squeezing is a fundamental channel, there is an additional processes that also generate entanglement of modes, and hence entanglement entropy. In~\cite{Bhattacharyya:2020rpy,Brahma:2021mng}, the coupling between super-Hubble modes and the sub-Hubble modes due to the presence of gravitational nonlinearities is studied, where super-Hubble modes are taken as the system and the sub-Hubble modes as the bath. Tracing over the sub-Hubble modes leads to decoherence and entropy growth for super-Hubble modes. Their result suggests that entanglement entropy due to cubic gravitational nonlinearities is larger than due to vacuum squeezing. However, analytical calculations quickly become daunting once nonlinear effects are included. In contrast, by mapping the problem onto quantum optics circuits, we can leverage well‐controlled squeezing and beam-splitting operations to compute these otherwise tedious quantities with ease. In our future work, we aim to include the nonlinearities in our circuit analysis and study the effect of thermal noise in such cases. Furthermore, we plan to extend our analysis to the multi-field inflation involving two scalar fields instead of a a single one which corresponds to four modes in the CV gate framework. We also aim to extend our analysis to an initial state different from the Bunch-Davies vacuum and study the effect on the entanglement measures for such a system.

Cosmological perturbations are imprinted as the anisotropies in the cosmic microwave background (CMB), which serve as the seeds for the formation of galaxies and large-scale structures. The statistical properties of these perturbations are quantified by their power spectrum, which describes how the amplitude of fluctuations varies with the spatial scale~\cite{Mukhanov:1981xt}. The spectrum of curvature fluctuations in inflationary cosmology is defined as:

\begin{equation}
{\cal P}_{\cal R}(k,t) \equiv k^{3}{\cal R}_{k}^{2}(t)\,,
\end{equation}
where $k$ is the magnitude of ${\bf k}$. There is also a primordial power spectrum, which is generated from co-moving curvature perturbations, which mainly connect the early universe to the large-scale structure formation of today. The observed CMBR data from the recombination era are mainly calculated on the basis of the primordial power spectrum. Hence, the power spectrum plays a central role in connecting theoretical models of the early universe with observational data. We aim to extend the circuit analysis to study the power spectrum of fluctuations and carry out a comprehensive analysis of the theoretical and symplectic methods. We can also compare the results between the observed power spectrum from WMAP, CMBR, or any observatories to our theoretically calculated power spectrum. We will extend our analysis to four‑mode squeezed states, simulating the entanglement entropy across different momentum modes $k$. In addition, using quantum computation methods, we plan to study the quantum–classical transitions in a cosmological setting. To bridge theory and experiment, we also plan to study how to emulate the cosmological perturbations of order ($r\gtrsim40$) using the standard laboratory squeezing parameters of order  ($r \sim 1$).
\acknowledgments
This research is part of the Abdus Salam International Centre for Theoretical Physics (ICTP): Physics Without Frontiers (PWF) initiative, and we acknowledge support from the PWF program of the ICTP, Italy. R.S. is supported partly by a postdoctoral fellowship from the West University of Timișoara, Romania and acknowledge the kind hospitality of CERN Theory Department where part of this work has been carried out. K.A. is supported by the Munich Quantum Valley, which is supported by the Bavarian state government with funds from the Hightech Agenda Bayern Plus. 
\bibliographystyle{JHEP}
\bibliography{ref3.bib}

\providecommand{\href}[2]{#2}\begingroup\raggedright\begin{thebibliography}{10}

\bibitem{Ferraro:2005hen}
A.~Ferraro, S.~Olivares and M.G.A.~Paris, \emph{{Gaussian states in continuous variable quantum information}} (3, 2005), [\href{https://arxiv.org/abs/quant-ph/0503237}{{\ttfamily quant-ph/0503237}}].

\bibitem{Weedbrook:2011wxo}
C.~Weedbrook, S.~Pirandola, R.~Garc{\'\i}a-Patr{\'o}n, N.J.~Cerf, T.C.~Ralph, J.H.~Shapiro et~al., \emph{{Gaussian quantum information}}, \href{https://doi.org/10.1103/RevModPhys.84.621}{\emph{Rev. Mod. Phys.} {\bfseries 84} (2012) 621} [\href{https://arxiv.org/abs/1110.3234}{{\ttfamily 1110.3234}}].

\bibitem{Serafini:2017rrn}
A.~Serafini, \emph{{Quantum Continuous Variables}}, CRC Press (10, 2017), \href{https://doi.org/10.1201/9781315118727}{10.1201/9781315118727}.

\bibitem{Adhikari:2022oxr}
K.~Adhikari and S.~Choudhury, \emph{{Cosmological Krylov Complexity}}, \href{https://doi.org/10.1002/prop.202200126}{\emph{Fortsch. Phys.} {\bfseries 70} (2022) 2200126} [\href{https://arxiv.org/abs/2203.14330}{{\ttfamily 2203.14330}}].

\bibitem{Adhikari:2022whf}
K.~Adhikari, S.~Choudhury and A.~Roy, \emph{{Krylov Complexity in Quantum Field Theory}}, \href{https://doi.org/10.1016/j.nuclphysb.2023.116263}{\emph{Nucl. Phys. B} {\bfseries 993} (2023) 116263} [\href{https://arxiv.org/abs/2204.02250}{{\ttfamily 2204.02250}}].

\bibitem{Adhikari:2023evu}
K.~Adhikari, A.~Rijal, A.K.~Aryal, M.~Ghimire, R.~Singh and C.~Deppe, \emph{{Krylov Complexity of Fermionic and Bosonic Gaussian States}}, \href{https://doi.org/10.1002/prop.202400014}{\emph{Fortsch. Phys.} {\bfseries 72} (2024) 2400014} [\href{https://arxiv.org/abs/2309.10382}{{\ttfamily 2309.10382}}].

\bibitem{Li:2023ekd}
T.~Li and L.-H.~Liu, \emph{{Cosmological complexity of the modified dispersion relation}}, \href{https://doi.org/10.1016/j.physletb.2024.138728}{\emph{Phys. Lett. B} {\bfseries 854} (2024) 138728} [\href{https://arxiv.org/abs/2309.01595}{{\ttfamily 2309.01595}}].

\bibitem{Li:2024kfm}
T.~Li and L.-H.~Liu, \emph{{Inflationary Krylov complexity}}, \href{https://doi.org/10.1007/JHEP04(2024)123}{\emph{JHEP} {\bfseries 04} (2024) 123} [\href{https://arxiv.org/abs/2401.09307}{{\ttfamily 2401.09307}}].

\bibitem{Rabinovici:2025otw}
E.~Rabinovici, A.~S{\'a}nchez-Garrido, R.~Shir and J.~Sonner, \emph{{Krylov Complexity}},  \href{https://arxiv.org/abs/2507.06286}{{\ttfamily 2507.06286}}.

\bibitem{Plenio:2007zz}
M.B.~Plenio and S.S.~Virmani, \emph{{An Introduction to Entanglement Theory}}, \href{https://doi.org/10.1007/978-3-319-04063-9_8}{\emph{Quant. Inf. Comput.} {\bfseries 7} (2007) 001} [\href{https://arxiv.org/abs/quant-ph/0504163}{{\ttfamily quant-ph/0504163}}].

\bibitem{Brahma:2020zpk}
S.~Brahma, O.~Alaryani and R.~Brandenberger, \emph{{Entanglement entropy of cosmological perturbations}}, \href{https://doi.org/10.1103/PhysRevD.102.043529}{\emph{Phys. Rev. D} {\bfseries 102} (2020) 043529} [\href{https://arxiv.org/abs/2005.09688}{{\ttfamily 2005.09688}}].

\bibitem{Guth:1980zm}
A.H.~Guth, \emph{{The Inflationary Universe: A Possible Solution to the Horizon and Flatness Problems}}, \href{https://doi.org/10.1103/PhysRevD.23.347}{\emph{Phys. Rev. D} {\bfseries 23} (1981) 347}.

\bibitem{Mukhanov:1982nu}
V.F.~Mukhanov and G.V.~Chibisov, \emph{{The Vacuum energy and large scale structure of the universe}}, {\emph{Sov. Phys. JETP} {\bfseries 56} (1982) 258}.

\bibitem{Mukhanov:1990me}
V.F.~Mukhanov, H.A.~Feldman and R.H.~Brandenberger, \emph{{Theory of cosmological perturbations. Part 1. Classical perturbations. Part 2. Quantum theory of perturbations. Part 3. Extensions}}, \href{https://doi.org/10.1016/0370-1573(92)90044-Z}{\emph{Phys. Rept.} {\bfseries 215} (1992) 203}.

\bibitem{Dodelson:2003ft}
S.~Dodelson, \emph{{Modern Cosmology}}, Academic Press, Amsterdam (2003).

\bibitem{Straumann:2005mz}
N.~Straumann, \emph{{From primordial quantum fluctuations to the anisotropies of the cosmic microwave background radiation}}, \href{https://doi.org/10.1002/andp.200610212}{\emph{Annalen Phys.} {\bfseries 15} (2006) 701} [\href{https://arxiv.org/abs/hep-ph/0505249}{{\ttfamily hep-ph/0505249}}].

\bibitem{Mukhanov:1985rz}
V.F.~Mukhanov, \emph{{Gravitational Instability of the Universe Filled with a Scalar Field}}, {\emph{JETP Lett.} {\bfseries 41} (1985) 493}.

\bibitem{Sasaki:1986hm}
M.~Sasaki, \emph{{Large Scale Quantum Fluctuations in the Inflationary Universe}}, \href{https://doi.org/10.1143/PTP.76.1036}{\emph{Prog. Theor. Phys.} {\bfseries 76} (1986) 1036}.

\bibitem{Parker:1969au}
L.~Parker, \emph{{Quantized fields and particle creation in expanding universes. 1.}}, \href{https://doi.org/10.1103/PhysRev.183.1057}{\emph{Phys. Rev.} {\bfseries 183} (1969) 1057}.

\bibitem{Jiang:2025ktt}
X.-Y.~Jiang, X.-L.~Huang and S.-M.~Wu, \emph{{Cosmological entanglement of initial multipartite states}}, \href{https://doi.org/10.1140/epjc/s10052-025-14605-z}{\emph{Eur. Phys. J. C} {\bfseries 85} (2025) 851}.

\bibitem{Albrecht:1992kf}
A.~Albrecht, P.~Ferreira, M.~Joyce and T.~Prokopec, \emph{{Inflation and squeezed quantum states}}, \href{https://doi.org/10.1103/PhysRevD.50.4807}{\emph{Phys. Rev. D} {\bfseries 50} (1994) 4807} [\href{https://arxiv.org/abs/astro-ph/9303001}{{\ttfamily astro-ph/9303001}}].

\bibitem{Grishchuk:1990bj}
L.P.~Grishchuk and Y.V.~Sidorov, \emph{{Squeezed quantum states of relic gravitons and primordial density fluctuations}}, \href{https://doi.org/10.1103/PhysRevD.42.3413}{\emph{Phys. Rev. D} {\bfseries 42} (1990) 3413}.

\bibitem{Gasperini:1992xv}
M.~Gasperini and M.~Giovannini, \emph{{Entropy production in the cosmological amplification of the vacuum fluctuations}}, \href{https://doi.org/10.1016/0370-2693(93)91159-K}{\emph{Phys. Lett. B} {\bfseries 301} (1993) 334} [\href{https://arxiv.org/abs/gr-qc/9301010}{{\ttfamily gr-qc/9301010}}].

\bibitem{Martin:2021qkg}
J.{\'e}.~Martin and V.~Vennin, \emph{{Real-space entanglement in the Cosmic Microwave Background}}, \href{https://doi.org/10.1088/1475-7516/2021/10/036}{\emph{JCAP} {\bfseries 10} (2021) 036} [\href{https://arxiv.org/abs/2106.15100}{{\ttfamily 2106.15100}}].

\bibitem{Martin:2021xml}
J.~Martin and V.~Vennin, \emph{{Real-space entanglement of quantum fields}}, \href{https://doi.org/10.1103/PhysRevD.104.085012}{\emph{Phys. Rev. D} {\bfseries 104} (2021) 085012} [\href{https://arxiv.org/abs/2106.14575}{{\ttfamily 2106.14575}}].

\bibitem{Boutivas:2023ksg}
K.~Boutivas, G.~Pastras and N.~Tetradis, \emph{{Entanglement and expansion}}, \href{https://doi.org/10.1007/JHEP05(2023)199}{\emph{JHEP} {\bfseries 05} (2023) 199} [\href{https://arxiv.org/abs/2302.14666}{{\ttfamily 2302.14666}}].

\bibitem{Brandenberger:1990bx}
R.H.~Brandenberger, R.~Laflamme and M.~Mijic, \emph{{Classical Perturbations From Decoherence of Quantum Fluctuations in the Inflationary Universe}}, \href{https://doi.org/10.1142/S0217732390002651}{\emph{Mod. Phys. Lett. A} {\bfseries 5} (1990) 2311}.

\bibitem{Polarski:1995jg}
D.~Polarski and A.A.~Starobinsky, \emph{{Semiclassicality and decoherence of cosmological perturbations}}, \href{https://doi.org/10.1088/0264-9381/13/3/006}{\emph{Class. Quant. Grav.} {\bfseries 13} (1996) 377} [\href{https://arxiv.org/abs/gr-qc/9504030}{{\ttfamily gr-qc/9504030}}].

\bibitem{Lesgourgues:1996jc}
J.~Lesgourgues, D.~Polarski and A.A.~Starobinsky, \emph{{Quantum to classical transition of cosmological perturbations for nonvacuum initial states}}, \href{https://doi.org/10.1016/S0550-3213(97)00224-1}{\emph{Nucl. Phys. B} {\bfseries 497} (1997) 479} [\href{https://arxiv.org/abs/gr-qc/9611019}{{\ttfamily gr-qc/9611019}}].

\bibitem{Kiefer:1998qe}
C.~Kiefer, D.~Polarski and A.A.~Starobinsky, \emph{{Quantum to classical transition for fluctuations in the early universe}}, \href{https://doi.org/10.1142/S0218271898000292}{\emph{Int. J. Mod. Phys. D} {\bfseries 7} (1998) 455} [\href{https://arxiv.org/abs/gr-qc/9802003}{{\ttfamily gr-qc/9802003}}].

\bibitem{Calzetta:1995ys}
E.~Calzetta and B.L.~Hu, \emph{{Quantum fluctuations, decoherence of the mean field, and structure formation in the early universe}}, \href{https://doi.org/10.1103/PhysRevD.52.6770}{\emph{Phys. Rev. D} {\bfseries 52} (1995) 6770} [\href{https://arxiv.org/abs/gr-qc/9505046}{{\ttfamily gr-qc/9505046}}].

\bibitem{Perez:2005gh}
A.~Perez, H.~Sahlmann and D.~Sudarsky, \emph{{On the quantum origin of the seeds of cosmic structure}}, \href{https://doi.org/10.1088/0264-9381/23/7/008}{\emph{Class. Quant. Grav.} {\bfseries 23} (2006) 2317} [\href{https://arxiv.org/abs/gr-qc/0508100}{{\ttfamily gr-qc/0508100}}].

\bibitem{Kiefer:2008ku}
C.~Kiefer and D.~Polarski, \emph{{Why do cosmological perturbations look classical to us?}}, \href{https://doi.org/10.1166/asl.2009.1023}{\emph{Adv. Sci. Lett.} {\bfseries 2} (2009) 164} [\href{https://arxiv.org/abs/0810.0087}{{\ttfamily 0810.0087}}].

\bibitem{Colas:2021llj}
T.~Colas, J.~Grain and V.~Vennin, \emph{{Four-mode squeezed states: two-field quantum systems and the symplectic group $\mathrm {Sp}(4,{\mathbb {R}})$}}, \href{https://doi.org/10.1140/epjc/s10052-021-09922-y}{\emph{Eur. Phys. J. C} {\bfseries 82} (2022) 6} [\href{https://arxiv.org/abs/2104.14942}{{\ttfamily 2104.14942}}].

\bibitem{Li:2024ljz}
T.~Li and L.-H.~Liu, \emph{{Krylov complexity of thermal state in early universe}},  \href{https://arxiv.org/abs/2408.03293}{{\ttfamily 2408.03293}}.

\bibitem{Li:2024iji}
T.~Li and L.-H.~Liu, \emph{{Inflationary complexity of thermal state}},  \href{https://arxiv.org/abs/2405.01433}{{\ttfamily 2405.01433}}.

\bibitem{Liu:2020wtr}
J.~Liu and Y.-Z.~Li, \emph{{On Quantum Simulation Of Cosmic Inflation}}, \href{https://doi.org/10.1103/PhysRevD.104.086013}{\emph{Phys. Rev. D} {\bfseries 104} (2021) 086013} [\href{https://arxiv.org/abs/2009.10921}{{\ttfamily 2009.10921}}].

\bibitem{Boutivas:2023mfg}
K.~Boutivas, D.~Katsinis, G.~Pastras and N.~Tetradis, \emph{{Entanglement in cosmology}}, \href{https://doi.org/10.1088/1475-7516/2024/04/017}{\emph{JCAP} {\bfseries 04} (2024) 017} [\href{https://arxiv.org/abs/2310.17208}{{\ttfamily 2310.17208}}].

\bibitem{Lloyd:2013xba}
S.~Lloyd, \emph{{The universe as quantum computer}},  in \emph{{A Computable Universe}: {Understanding and Exploring Nature as Computation}}, pp.~567--582 (2013), \href{https://doi.org/10.1142/9789814374309_0029}{DOI} [\href{https://arxiv.org/abs/1312.4455}{{\ttfamily 1312.4455}}].

\bibitem{Bao:2017iye}
N.~Bao, C.~Cao, S.M.~Carroll and L.~McAllister, \emph{{Quantum Circuit Cosmology: The Expansion of the Universe Since the First Qubit}},  \href{https://arxiv.org/abs/1702.06959}{{\ttfamily 1702.06959}}.

\bibitem{Yang:2019kbb}
R.-Q.~Yang, H.~Liu, S.~Zhu, L.~Luo and R.-G.~Cai, \emph{{Simulating quantum field theory in curved spacetime with quantum many-body systems}}, \href{https://doi.org/10.1103/PhysRevResearch.2.023107}{\emph{Phys. Rev. Res.} {\bfseries 2} (2020) 023107} [\href{https://arxiv.org/abs/1906.01927}{{\ttfamily 1906.01927}}].

\bibitem{Maceda:2024rrd}
M.D.~Maceda and C.~Sab{\'\i}n, \emph{{Digital quantum simulation of cosmological particle creation with IBM quantum computers}}, \href{https://doi.org/10.1038/s41598-025-87015-6}{\emph{Sci. Rep.} {\bfseries 15} (2025) 3476} [\href{https://arxiv.org/abs/2410.02412}{{\ttfamily 2410.02412}}].

\bibitem{Piotrak:2025zhy}
M.~Piotrak, T.~Colas, A.~Alonso-Serrano and A.~Serafini, \emph{{Quantum estimation of cosmological parameters}},  \href{https://arxiv.org/abs/2507.12228}{{\ttfamily 2507.12228}}.

\bibitem{Li:2023gtf}
W.-M.~Li, R.-D.~Wang, H.-Y.~Wu, X.-L.~Huang, H.-S.~Zeng and S.-M.~Wu, \emph{{Quantum entanglement for continuous variables sharing in an expanding spacetime}}, \href{https://doi.org/10.1140/epjc/s10052-023-11344-x}{\emph{Eur. Phys. J. C} {\bfseries 83} (2023) 222} [\href{https://arxiv.org/abs/2303.09924}{{\ttfamily 2303.09924}}].

\bibitem{Grishchuk:1993ds}
L.P.~Grishchuk, \emph{{Quantum effects in cosmology}}, \href{https://doi.org/10.1088/0264-9381/10/12/006}{\emph{Class. Quant. Grav.} {\bfseries 10} (1993) 2449} [\href{https://arxiv.org/abs/gr-qc/9302036}{{\ttfamily gr-qc/9302036}}].

\bibitem{Marshall:2015mna}
K.~Marshall, R.~Pooser, G.~Siopsis and C.~Weedbrook, \emph{{Quantum simulation of quantum field theory using continuous variables}}, \href{https://doi.org/10.1103/PhysRevA.92.063825}{\emph{Phys. Rev. A} {\bfseries 92} (2015) 063825} [\href{https://arxiv.org/abs/1503.08121}{{\ttfamily 1503.08121}}].

\bibitem{Zhai:2024odw}
K.-H.~Zhai and L.-H.~Liu, \emph{{Krylov Complexity in early universe}},  \href{https://arxiv.org/abs/2411.18405}{{\ttfamily 2411.18405}}.

\bibitem{Fischer:2004bf}
U.R.~Fischer and R.~Schützhold, \emph{{Quantum simulation of cosmic inflation in two-component Bose-Einstein condensates}}, \href{https://doi.org/10.1103/PhysRevA.70.063615}{\emph{Phys. Rev. A} {\bfseries 70} (2004) 063615} [\href{https://arxiv.org/abs/cond-mat/0406470}{{\ttfamily cond-mat/0406470}}].

\bibitem{Wittemer:2019agm}
M.~Wittemer, F.~Hakelberg, P.~Kiefer, J.-P.~Schr{\"o}der, C.~Fey, R.~Sch{\"u}tzhold et~al., \emph{{Phonon Pair Creation by Inflating Quantum Fluctuations in an Ion Trap}}, \href{https://doi.org/10.1103/PhysRevLett.123.180502}{\emph{Phys. Rev. Lett.} {\bfseries 123} (2019) 180502} [\href{https://arxiv.org/abs/1903.05523}{{\ttfamily 1903.05523}}].

\bibitem{Bhardwaj:2020ndh}
A.~Bhardwaj, D.~Vaido and D.E.~Sheehy, \emph{{Inflationary Dynamics and Particle Production in a Toroidal Bose-Einstein Condensate}}, \href{https://doi.org/10.1103/PhysRevA.103.023322}{\emph{Phys. Rev. A} {\bfseries 103} (2021) 023322} [\href{https://arxiv.org/abs/2009.05611}{{\ttfamily 2009.05611}}].

\bibitem{Rhyno:2023kud}
B.~Rhyno, I.~Velkovsky, P.~Adshead, B.~Gadway and S.~Vishveshwara, \emph{{Mechanical cosmology: Simulating scalar fluctuations in expanding universes using synthetic mechanical lattices}}, \href{https://doi.org/10.1103/PhysRevResearch.7.L022004}{\emph{Phys. Rev. Res.} {\bfseries 7} (2025) L022004} [\href{https://arxiv.org/abs/2312.13467}{{\ttfamily 2312.13467}}].

\bibitem{Agullo:2024lry}
I.~Agullo, A.~Delhom and {\'A}.~Parra-L{\'o}pez, \emph{{Toward the observation of entangled pairs in BEC analog expanding universes}}, \href{https://doi.org/10.1103/PhysRevD.110.125023}{\emph{Phys. Rev. D} {\bfseries 110} (2024) 125023} [\href{https://arxiv.org/abs/2411.09596}{{\ttfamily 2411.09596}}].

\bibitem{Steinhauer:2021fhb}
J.~Steinhauer, M.~Abuzarli, T.~Aladjidi, T.~Bienaim{\'e}, C.~Piekarski, W.~Liu et~al., \emph{{Analogue cosmological particle creation in an ultracold quantum fluid of light}}, \href{https://doi.org/10.1038/s41467-022-30603-1}{\emph{Nature Commun.} {\bfseries 13} (2022) 2890} [\href{https://arxiv.org/abs/2102.08279}{{\ttfamily 2102.08279}}].

\bibitem{Pal:2024qno}
K.~Pal and U.R.~Fischer, \emph{{Quantum nonlinear effects in the number-conserving analog gravity of Bose-Einstein condensates}}, \href{https://doi.org/10.1103/PhysRevD.110.116022}{\emph{Phys. Rev. D} {\bfseries 110} (2024) 116022} [\href{https://arxiv.org/abs/2410.13596}{{\ttfamily 2410.13596}}].

\bibitem{Fischer:2001jz}
U.R.~Fischer and M.~Visser, \emph{{Riemannian geometry of irrotational vortex acoustics}}, \href{https://doi.org/10.1103/PhysRevLett.88.110201}{\emph{Phys. Rev. Lett.} {\bfseries 88} (2002) 110201} [\href{https://arxiv.org/abs/cond-mat/0110211}{{\ttfamily cond-mat/0110211}}].

\bibitem{Fedichev:2003id}
P.O.~Fedichev and U.R.~Fischer, \emph{{Gibbons-Hawking effect in the sonic de Sitter space-time of an expanding Bose-Einstein-condensed gas}}, \href{https://doi.org/10.1103/PhysRevLett.91.240407}{\emph{Phys. Rev. Lett.} {\bfseries 91} (2003) 240407} [\href{https://arxiv.org/abs/cond-mat/0304342}{{\ttfamily cond-mat/0304342}}].

\bibitem{Fedichev:2003bv}
P.O.~Fedichev and U.R.~Fischer, \emph{{'Cosmological' quasiparticle production in harmonically trapped superfluid gases}}, \href{https://doi.org/10.1103/PhysRevA.69.033602}{\emph{Phys. Rev. A} {\bfseries 69} (2004) 033602} [\href{https://arxiv.org/abs/cond-mat/0303063}{{\ttfamily cond-mat/0303063}}].

\bibitem{Fedichev:2003dj}
P.O.~Fedichev and U.R.~Fischer, \emph{{Observer dependence for the phonon content of the sound field living on the effective curved space-time background of a Bose-Einstein condensate}}, \href{https://doi.org/10.1103/PhysRevD.69.064021}{\emph{Phys. Rev. D} {\bfseries 69} (2004) 064021} [\href{https://arxiv.org/abs/cond-mat/0307200}{{\ttfamily cond-mat/0307200}}].

\bibitem{Cha:2016esj}
S.-Y.~Ch{\"a} and U.R.~Fischer, \emph{{Probing the scale invariance of the inflationary power spectrum in expanding quasi-two-dimensional dipolar condensates}}, \href{https://doi.org/10.1103/PhysRevLett.118.130404}{\emph{Phys. Rev. Lett.} {\bfseries 118} (2017) 130404} [\href{https://arxiv.org/abs/1609.06155}{{\ttfamily 1609.06155}}].

\bibitem{Tian:2020bze}
Z.~Tian, Y.~Lin, U.R.~Fischer and J.~Du, \emph{{Testing the upper bound on the speed of scrambling with an analogue of Hawking radiation using trapped ions}}, \href{https://doi.org/10.1140/epjc/s10052-022-10149-8}{\emph{Eur. Phys. J. C} {\bfseries 82} (2022) 212} [\href{https://arxiv.org/abs/2007.05949}{{\ttfamily 2007.05949}}].

\bibitem{Ribeiro:2021fpk}
C.C.H.~Ribeiro, S.-S.~Baak and U.R.~Fischer, \emph{{Existence of steady-state black hole analogs in finite quasi-one-dimensional Bose-Einstein condensates}}, \href{https://doi.org/10.1103/PhysRevD.105.124066}{\emph{Phys. Rev. D} {\bfseries 105} (2022) 124066} [\href{https://arxiv.org/abs/2103.05015}{{\ttfamily 2103.05015}}].

\bibitem{Ribeiro:2022gln}
C.C.H.~Ribeiro and U.R.~Fischer, \emph{{Impact of trans-Planckian excitations on black-hole radiation in dipolar condensates}}, \href{https://doi.org/10.1103/PhysRevD.107.L121502}{\emph{Phys. Rev. D} {\bfseries 107} (2023) L121502} [\href{https://arxiv.org/abs/2211.01243}{{\ttfamily 2211.01243}}].

\bibitem{Baak:2022hum}
S.-S.~Baak, C.C.H.~Ribeiro and U.R.~Fischer, \emph{{Number-conserving solution for dynamical quantum backreaction in a Bose-Einstein condensate}}, \href{https://doi.org/10.1103/PhysRevA.106.053319}{\emph{Phys. Rev. A} {\bfseries 106} (2022) 053319} [\href{https://arxiv.org/abs/2206.11317}{{\ttfamily 2206.11317}}].

\bibitem{Baak:2023zjf}
S.-S.~Baak, S.~Datta and U.R.~Fischer, \emph{{Petrov classification of analogue spacetimes}}, \href{https://doi.org/10.1088/1361-6382/acf08e}{\emph{Class. Quant. Grav.} {\bfseries 40} (2023) 215001} [\href{https://arxiv.org/abs/2305.12771}{{\ttfamily 2305.12771}}].

\bibitem{Chandran:2025azu}
S.M.~Chandran and U.R.~Fischer, \emph{{Expansion-contraction duality breaking in a Planck-scale sensitive cosmological quantum simulator}},  \href{https://arxiv.org/abs/2506.02719}{{\ttfamily 2506.02719}}.

\bibitem{Gerry_Knight_2004}
C.~Gerry and P.~Knight, \emph{Introductory Quantum Optics}, Cambridge University Press (2004).

\bibitem{Vidal:2002zz}
G.~Vidal and R.F.~Werner, \emph{{Computable measure of entanglement}}, \href{https://doi.org/10.1103/PhysRevA.65.032314}{\emph{Phys. Rev. A} {\bfseries 65} (2002) 032314} [\href{https://arxiv.org/abs/quant-ph/0102117}{{\ttfamily quant-ph/0102117}}].

\bibitem{Plenio:2005cwa}
M.B.~Plenio, \emph{{Logarithmic Negativity: A Full Entanglement Monotone That is not Convex}}, \href{https://doi.org/10.1103/PhysRevLett.95.090503}{\emph{Phys. Rev. Lett.} {\bfseries 95} (2005) 090503} [\href{https://arxiv.org/abs/quant-ph/0505071}{{\ttfamily quant-ph/0505071}}].

\bibitem{Choudhury:2016cso}
S.~Choudhury, S.~Panda and R.~Singh, \emph{{Bell violation in the Sky}}, \href{https://doi.org/10.1140/epjc/s10052-016-4553-3}{\emph{Eur. Phys. J. C} {\bfseries 77} (2017) 60} [\href{https://arxiv.org/abs/1607.00237}{{\ttfamily 1607.00237}}].

\bibitem{Choudhury:2016pfr}
S.~Choudhury, S.~Panda and R.~Singh, \emph{{Bell violation in primordial cosmology}}, \href{https://doi.org/10.3390/universe3010013}{\emph{Universe} {\bfseries 3} (2017) 13} [\href{https://arxiv.org/abs/1612.09445}{{\ttfamily 1612.09445}}].

\bibitem{Brady:2022ffk}
A.J.~Brady, I.~Agullo and D.~Kranas, \emph{{Symplectic circuits, entanglement, and stimulated Hawking radiation in analogue gravity}}, \href{https://doi.org/10.1103/PhysRevD.106.105021}{\emph{Phys. Rev. D} {\bfseries 106} (2022) 105021} [\href{https://arxiv.org/abs/2209.11317}{{\ttfamily 2209.11317}}].

\bibitem{Kranas:2023aph}
D.~Kranas, \emph{{Entanglement in the Hawking Effect: from Astrophysical to Optical Black Holes}}, Ph.D. thesis, Louisiana State U., Louisiana State U., 2023.

\bibitem{Brandenberger:2003vk}
R.H.~Brandenberger, \emph{{Lectures on the theory of cosmological perturbations}}, \href{https://doi.org/10.1007/978-3-540-40918-2_5}{\emph{Lect. Notes Phys.} {\bfseries 646} (2004) 127} [\href{https://arxiv.org/abs/hep-th/0306071}{{\ttfamily hep-th/0306071}}].

\bibitem{Brandenberger:1993zc}
R.H.~Brandenberger, H.~Feldman and V.F.~Mukhanov, \emph{{Classical and quantum theory of perturbations in inflationary universe models}},  in \emph{{37th Yamada Conference: Evolution of the Universe and its Observational Quest}}, pp.~19--30, 7, 1993 [\href{https://arxiv.org/abs/astro-ph/9307016}{{\ttfamily astro-ph/9307016}}].

\bibitem{Bhattacharyya:2020rpy}
A.~Bhattacharyya, S.~Das, S.~Shajidul~Haque and B.~Underwood, \emph{{Cosmological Complexity}}, \href{https://doi.org/10.1103/PhysRevD.101.106020}{\emph{Phys. Rev. D} {\bfseries 101} (2020) 106020} [\href{https://arxiv.org/abs/2001.08664}{{\ttfamily 2001.08664}}].

\bibitem{Bhattacharyya:2020kgu}
A.~Bhattacharyya, S.~Das, S.S.~Haque and B.~Underwood, \emph{{Rise of cosmological complexity: Saturation of growth and chaos}}, \href{https://doi.org/10.1103/PhysRevResearch.2.033273}{\emph{Phys. Rev. Res.} {\bfseries 2} (2020) 033273} [\href{https://arxiv.org/abs/2005.10854}{{\ttfamily 2005.10854}}].

\bibitem{Cochran:2024cri}
S.~Cochran, J.~Stokes, P.~Jayakumar and S.~Veerapaneni, \emph{{An application of continuous-variable gate synthesis to quantum simulation of classical dynamics}}, \href{https://doi.org/10.1116/5.0234007}{\emph{AVS Quantum Sci.} {\bfseries 7} (2025) 023801} [\href{https://arxiv.org/abs/2407.08006}{{\ttfamily 2407.08006}}].

\bibitem{Braunstein:2005zz}
S.L.~Braunstein and P.~van Loock, \emph{{Quantum information with continuous variables}}, \href{https://doi.org/10.1103/RevModPhys.77.513}{\emph{Rev. Mod. Phys.} {\bfseries 77} (2005) 513} [\href{https://arxiv.org/abs/quant-ph/0410100}{{\ttfamily quant-ph/0410100}}].

\bibitem{Lloyd:1998jk}
S.~Lloyd and S.L.~Braunstein, \emph{{Quantum computation over continuous variables}}, \href{https://doi.org/10.1103/PhysRevLett.82.1784}{\emph{Phys. Rev. Lett.} {\bfseries 82} (1999) 1784} [\href{https://arxiv.org/abs/quant-ph/9810082}{{\ttfamily quant-ph/9810082}}].

\bibitem{Kalajdzievski:2021axm}
T.~Kalajdzievski and N.~Quesada, \emph{{Exact and approximate continuous-variable gate decompositions}}, \href{https://doi.org/10.22331/q-2021-02-08-394}{\emph{Quantum} {\bfseries 5} (2021) 394} [\href{https://arxiv.org/abs/2010.07453}{{\ttfamily 2010.07453}}].

\bibitem{Brask:2021kvs}
J.B.~Brask, \emph{{Gaussian states and operations -- a quick reference}},  \href{https://arxiv.org/abs/2102.05748}{{\ttfamily 2102.05748}}.

\bibitem{Adesso:2014npz}
G.~Adesso, S.~Ragy and A.R.~Lee, \emph{{Continuous Variable Quantum Information: Gaussian States and Beyond}}, \href{https://doi.org/10.1142/s1230161214400010}{\emph{Open Syst. Info. Dyn.} {\bfseries 21} (2014) 1440001} [\href{https://arxiv.org/abs/1401.4679}{{\ttfamily 1401.4679}}].

\bibitem{Wolf:2002bhs}
M.M.~Wolf, J.~Eisert and M.B.~Plenio, \emph{{The entangling power of passive optical elements}}, \href{https://doi.org/10.1103/PhysRevLett.90.047904}{\emph{Phys. Rev. Lett.} {\bfseries 90} (2003) 047904} [\href{https://arxiv.org/abs/quant-ph/0206171}{{\ttfamily quant-ph/0206171}}].

\bibitem{Li:2022ktm}
D.~Li and C.~Zheng, \emph{{Non-Hermitian Generalization of R{\'e}nyi Entropy}}, \href{https://doi.org/10.3390/e24111563}{\emph{Entropy} {\bfseries 24} (2022) 1563}.

\bibitem{Seroje:2015tsa}
K.K.R.~Seroje, R.S.~dela Rosa and F.N.C.~Paraan, \emph{{Effective thermodynamics of isolated entangled squeezed and coherent states}}, \href{https://doi.org/10.1088/0143-0807/36/5/055051}{\emph{Eur. J. Phys.} {\bfseries 36} (2015) 055051} [\href{https://arxiv.org/abs/1507.00054}{{\ttfamily 1507.00054}}].

\bibitem{Brandao:2022voc}
I.~Brand{\~a}o, D.~Tandeitnik and T.~Guerreiro, \emph{{QuGIT: A numerical toolbox for Gaussian quantum states}}, \href{https://doi.org/10.1016/j.cpc.2022.108471}{\emph{Comput. Phys. Commun.} {\bfseries 280} (2022) 108471} [\href{https://arxiv.org/abs/2201.06368}{{\ttfamily 2201.06368}}].

\bibitem{Brahma:2021mng}
S.~Brahma, A.~Berera and J.~Calder{\'o}n-Figueroa, \emph{{Universal signature of quantum entanglement across cosmological distances}}, \href{https://doi.org/10.1088/1361-6382/aca066}{\emph{Class. Quant. Grav.} {\bfseries 39} (2022) 245002} [\href{https://arxiv.org/abs/2107.06910}{{\ttfamily 2107.06910}}].

\bibitem{Mukhanov:1981xt}
V.F.~Mukhanov and G.V.~Chibisov, \emph{{Quantum Fluctuations and a Nonsingular Universe}}, {\emph{JETP Lett.} {\bfseries 33} (1981) 532}.

\end{thebibliography}\endgroup
\end{document}